\documentclass[prd, twocolumn, nofootinbib]{revtex4-2}

\usepackage{amsmath}
\usepackage{amsfonts}
\usepackage{amssymb}

\usepackage{multirow}
\usepackage{makecell}

\usepackage{xcolor}
\usepackage[colorlinks=true, linkcolor=blue, citecolor=blue, urlcolor=blue]{hyperref}

\usepackage{graphbox}

\usepackage{comment}

\usepackage[normalem]{ulem}

\newcommand{\result}[1]{#1}

\newcommand{\BayesIsoAni}{\result{\ensuremath{\mathcal{B}^\mathrm{iso}_\mathrm{ani}}}}

\begin{document}

\title{
An Isotropy Measurement with Gravitational Wave Observations
}

\author{Reed Essick}
\email{reed.essick@gmail.com}
\affiliation{Perimeter Institute for Theoretical Physics \\ 31 Caroline Street North, Waterloo, ON N2L 2Y5}

\author{Will M. Farr}
\affiliation{Department of Physics and Astronomy, Stony Brook University, Stony Brook NY 11794, USA}
\affiliation{Center for Computational Astrophysics, Flatiron Institute, New York NY 10010, USA}

\author{Maya Fishbach}
\affiliation{NASA Hubble Fellowship Program Einstein Fellow}
\affiliation{Center for Interdisciplinary Exploration and Research in Astrophysics (CIERA) and Department of Physics and Astronomy, Northwestern University \\ 1800 Sherman Ave, Evanston, IL 60201, USA}

\author{Daniel E. Holz}
\affiliation{Department of Physics, Department of Astronomy \& Astrophysics, Enrico Fermi Institute, and Kavli Institute for Cosmological Physics, The University of Chicago, 5640 South Ellis Avenue, Chicago, Illinois 60637, USA}

\author{Erik Katsavounidis}
\affiliation{LIGO Laboratory and Kavli Institute for Astrophysics and Space Research, Massachusetts Institute of Technology, 185 Albany St, Cambridge, MA 02139, USA}

\begin{abstract}
    We constrain the distribution of merging compact binaries across the celestial sphere using the GWTC-3 catalog from the LIGO-Virgo-KAGRA Collaborations' (LVK) third observing run.
    With \result{63} confident detections from O3, we constrain the relative variability (standard deviation) of the rate density across the sky to be \result{$\lesssim 16\%$ at 90\% confidence} assuming the logarithm of the rate density is described by a Gaussian random field with correlation length \result{$\geq 10^\circ$}.
    This tightens to \result{$\lesssim 3.5\%$} when the correlation length is \result{$\geq 20^\circ$}.
    While the new O3 data provides the tightest constraints on anisotropies available to-date, we do not find overwhelming evidence in favor of isotropy, either.
    A simple counting experiment favors an isotropic distribution by a factor of \result{$\BayesIsoAni = 3.7$}, which is nonetheless an improvement of more than a factor of two compared to analogous analyses based on only the first and second observing runs of the LVK.
\end{abstract}

\maketitle


\section{Introduction}
\label{sec:introduction}

The observation of gravitational waves (GWs) from the coalescence of compact binaries provides a new way to study how these systems form, evolve, and are distributed throughout the universe (see Ref.~\cite{GWTC-3-RnP} and references therein).
In particular, the spatial distribution of GW sources can test the cosmological principle: is the universe statistically homogeneous and isotropic?
Deviations from perfect homogeneity have already been proposed as a way to infer cosmological parameters through cross-correlations of clustering within GW and electromagnetic observations (see, e.g.,~\cite{GWTC-3-Cosmo, Mukherjee:2021}).
However, these studies assume \textit{a priori} that GW events follow anisotropies measured from electromagnetic surveys.
That is, they do not directly measure anisotropies from the GW data.
Our goal in this paper is to constrain anisotropies in the population of merging binaries using only GW data.

Although large deviations from isotropy are not expected, it behooves us nevertheless to check this, similar to the motivation within Ref.~\cite{Vitale:2022}.
Directly constraining anisotropies with GW catalogs may be of interest in several astrophysical situations.
For example, resolving clustering scales from GW data alone may be used to test the assumption that GW sources are always associated with galaxies.
Along these lines, GW sources could be used to directly trace clustering scales; see, e.g., Refs.~\cite{Libanore:2021, Shao:2022} for discussion of this in the context of 3$^\mathrm{rd}$ generation detectors. Similarly, the identification of individual host galaxies for specific events and/or the statistical association between the full GW catalog and different types of galaxies may suggest, perhaps through the mass-dependent galaxy clustering scale, which types of galaxies most often host compact binary coalescences~\cite{Chen:2016, Singer:2016}.
This could be combined with knowledge of the star formation history to in turn constrain the delay time distribution between binary formation and coalescence~\cite{Fishbach:2021}.
See, e.g., Ref.~\cite{Zevin:2022} for a similar application to short gamma-ray bursts (GRBs).

In addition, the LIGO-Virgo-KAGRA (LVK) collaborations'~\cite{LIGO, Virgo} searches for unmodeled ``burst'' events in addition to compact binaries~\cite{O3-burst}.
Given that the source of such events will not be known \textit{a priori}, their spatial distribution will likely provide crucial clues as to their origins.
Indeed, determining whether burst events correlate with local structure will inform the distance to the sources and therefore their intrinsic energy scales, analogous to GRBs~\cite{Andrade:2019} and other high-energy astrophysical phenomena.

The detection of anisotropies within the distribution of merging binaries could be the signature of more exotic physics, such as wormholes that may effectively tunnel to larger volumes and therefore higher number of merging binaries~\cite{PhysRevD.104.044030} or lensed events, which appear as repeated signals from the same part of the sky~\cite{Xu:2021,Caliskan:2022}.
In particular, strong lensing may distort the shape of the waveform, particularly the relative phasing between different harmonics~\cite{Ezquiaga:2021}.
These effects may be difficult to distinguish from more general alternative theories of gravity~\cite{Ezquiaga:2022}, and the identification of anisotropies may be a cleaner signature of lensing than the waveform's phasing alone.
Indeed, many searches for lensed events begin with overlaps on the sky.

Several authors have already studied the distribution of merging binaries with the LIGO-Virgo Collaboration's~\cite{LIGO, Virgo} first catalog of 11 detections (GWTC-1~\cite{GWTC-1}).
Specifically, Ref.~\cite{Stiskalek:2020} modeled anisotropies with 12 pixels of equal area and a set of Euler angles that rotated the pixelization across the sky.
Using an approximation of the catalog's sensitivity that assumed constant and equal power spectral densities for both LIGO detectors throughout the run, neglecting the presence of Virgo, but accounting for the diurnal cycle and correlations in when the LIGO interferometers recorded science-quality data~\cite{Chen:2017}, they found weak evidence in favor of isotropy.
Similarly, Ref.~\cite{Payne:2020} used the same 11 events but a different estimate of survey sensitivity to constrain anisotropies with a model constructed from a low-order spherical harmonic expansion.
They considered several models with different numbers of harmonics up to $l_\mathrm{max} = 5$, finding equivalently weak evidence in favor of isotropy regardless of $l_\mathrm{max}$.
Finally, Ref.~\cite{Cavaglia:2020} attempted to measure the two-point correlation function of GW events with a spherical harmonic decomposition of the sum of individual event localizations while assuming the sensitivity of the detector network was uniform over the entire sky.
They also found no evidence for an excess of correlation at any angular scale.

Maps of upper limits on anisotropies in the stochastic GW background are routinely produced under various assumptions in either the pixel or spherical-harmonic domains.
Although no statistically significant detection has been made, these analyses typically make assumptions about the power spectrum of the stochastic GW background and produce maximum likelihood estimates of the angular distribution of the intensity.
See Ref.~\cite{Renzini:2022} for a review.
While there has been no unambiguous detection of the stochastic GW background to date, let alone the detection of anisotropies, there may still be information about the distribution of merging binaries at high redshift encoded in the nondetection (see, e.g., Ref.~\cite{Callister:2020}).

Additionally, anisotropies are of general interest in other high-energy astrophysical phenomena.
Analyses of GRBs show that they are consistent with isotropic distributions, regardless of how the catalog is subdivided~\cite{Andrade:2019}, the distribution of fast-radio bursts (FRBs) is an active area of research~\cite{Cordes:2019}, and multiple groups have claimed detections of anisotropies in the arrival directions of cosmic rays~\cite{Abeysekara:2014, Illuminati:2016, Aab:2018, McNally:2021}.

Therefore, it is of general interest to develop methods to constrain the rate of mergers as a function of their position on the celestial sphere.
We use hierarchical Bayesian inference to construct posterior processes for the distribution of merging compact binaries over the sky using \result{63} confidently detected binaries, including binary black hole (BBH), neutron star-black hole (NSBH), and binary neutron star (BNS) sources, from the the LVK's third observing run (O3, 1 April 2019--27 March 2022~\cite{GWTC-2, GWTC-2d1, GWTC-3}).
In addition to the nearly 6-fold increase in sample size from GWTC-1, our analysis benefits from estimates of survey sensitivity derived from simulated signals injected into real detector noise and processed directly with the searches used to construct the catalog~\cite{GWTC-3-injections}.
These injections implicitly account for variability in each detector's sensitivity and correlations between the times when detectors record data.\footnote{Appendix~\ref{sec:events and injections} shows that the O3 catalog's sensitivity is nearly uniform over the entire sky, although measurable deviations exist.}
This improves upon previous estimates of survey sensitivity, which depended on approximations with poorly quantified systematic uncertainties~\cite{Stiskalek:2020, Payne:2020}.
We also self-consistently incorporate realistic models of the masses, spins, and redshift distributions of merging binaries derived from GW observation~\cite{Farah:2021, GWTC-3-RnP}.

We find mild evidence in favor of isotropy.
This agrees with Refs.~\cite{Stiskalek:2020} and~\cite{Payne:2020}, but we place tighter constraints on anisotropies because of the larger sample size now available.
In fact, we find Bayes factors in favor of isotropy ($\BayesIsoAni$) similar to Refs.~\cite{Stiskalek:2020, Payne:2020} when we use only events from GWTC-1, and these increase by a factor of 2 when we use (only) the \result{63} events from O3.
Although there are a few persistent ``hot pixels'' from O3 in all our models on average, we cannot confidently bound the rate density in these directions to be inconsistent with isotropy.
Indeed, we bound the relative variability (standard deviation) in the rate density to \result{$\lesssim 16\%$} of the isotropic rate at \result{90\% credibility} if the correlation length scale in the rate density is \result{$\geq 10^\circ$}, and this is improved to \result{$\lesssim 3.5\%$} if the length scale is \result{$\geq 20^\circ$.} 

The rest of this paper is structured as follows.
In Sec.~\ref{sec:counting experiments}, we perform a simple counting experiment by dividing the sky into hemispheres, showing that most of the information about (an)isotropy comes from the best localized events (less than half our catalog).
Sec.~\ref{sec:hierarchical models} presents additional models of varying complexity, including pixelized representations like Ref.~\cite{Stiskalek:2020} (Sec.~\ref{sec:pixelized representations}) and representations based on low-order spherical harmonic expansions like Ref.~\cite{Payne:2020} (Sec.~\ref{sec:spherical harmonic representations}), culminating in a nonparametric description of the rate density as a Gaussian random field (Sec.~\ref{sec:gaussian process}).
We discuss implications of current constraints and conclude in Sec.~\ref{sec:discussion}.


\section{Counting Experiments}
\label{sec:counting experiments}

We begin with a simple counting experiment: divide the sky into two hemispheres and ``count'' the number of events that fall within each.\footnote{Because GW events often have very broad localizations, we always employ hierarchical Bayesian inference to account for measurement uncertainty. See Sec.~\ref{sec:formalism} for more details.}
In the context of GW catalogs, this simple model is useful because of the symmetry inherent in the sensitivity for current interferometers.
Each interferometer's sensitivity has even parity when reflected across the plane defined by its arms.
This means that the sensitivity to each hemisphere will be equal regardless of how many interferometers participate in the survey and exactly where hemispheres are drawn as long as they divide the sky in half equally.

To wit, we construct a model that divides the sky in half, assigning a different rate density to each hemisphere: the fraction of events coming from the ``northern'' hemisphere is $f$, and the corresponding fraction from the ``southern'' hemisphere is $1-f$.
We also consider all possible hemispheres by sampling over Euler angles that rotate the simple hemisphere model into a general partition of the sky.
This rotated model is similar to the approach in Ref.~\cite{Stiskalek:2020}.

\begin{figure}
    \centering
    \includegraphics[hsmash=c, width=1.0\columnwidth]{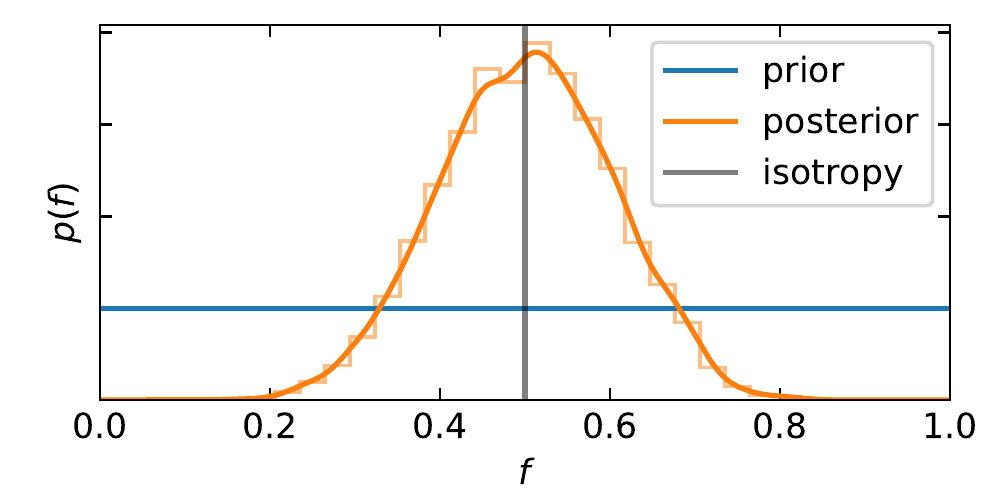}
    \caption{
        Marginal prior (\emph{blue}) and posterior (\emph{orange}) for the mixing fraction in the Rotated Hemisphere model (Table~\ref{tab:fitted population models}).
        The data favor isotropy (\emph{grey}, $f=0.5$) by a factor of \result{$\BayesIsoAni=3.7$} and rule out anisotropies bigger than \result{3:1}.
    }
    \label{fig:dirichlet posterior}
\end{figure}

Much of the information about isotropy comes from the best-localized events, and we find that the data prefer equal fractions of events from each hemisphere ($f=0.5$) by a factor of \result{$\BayesIsoAni=3.7$}, assuming uniform priors for $f$ and the Euler angles (Fig.~\ref{fig:dirichlet posterior}).
The model also finds no preference for specific rotations, which is expected if $f \sim 0.5$.
Furthermore, the number of events in one hemisphere is binomially distributed, and the uncertainty in the fraction of events will be $\sigma^2_f = f(1-f)/N$ with $N$ events.
With our selection of \result{63} events and assuming isotropy, we expect \result{$\sigma_f = 6.3\%$}.
However, this is significantly smaller than the actual standard deviation observed in Fig.~\ref{fig:dirichlet posterior}, which corresponds to only \result{$23.5$ effective events} (\result{$\sigma_f = 10.3\%$}).
While this could be due in part to the trials factor associated with sampling over possible rotations, it is likely because many of the events in our catalog have uninformative broad localization uncertainties.
Indeed, if we only use the \result{25} best-localized events from our catalog,\footnote{In general, selecting events in this way may significantly complicate our estimate the catalog's sensitivity. However, our Rotated Hemisphere model is immune to such considerations because of the symmetry of the interferometer antenna patterns.} we find \result{$\sigma_f = 12.6\%$} and \result{$\BayesIsoAni = 2.7$}, only slightly less constraining than the uncertainty obtained with the full catalog.
Similarly, if we only include the \result{10} best-localized events, we obtain \result{$\sigma_f = 17.5\%$} and \result{$\BayesIsoAni = 1.9$}, only slightly worse than expected from the binomial distribution ($\sigma_f = 15.8\%$ with 10 events).

This should be contrasted with the constraints obtained using only GWTC-1: \result{$\BayesIsoAni = 1.3$} and $\sigma_f = 18.9\%$.\footnote{Although our sensitivity estimates only cover O3, we can analyze GWTC-1 without accounting for selection effects because of the symmetry in this model.}
Ref.~\cite{Stiskalek:2020} found $\BayesIsoAni = 1.3$ and Ref.~\cite{Payne:2020} quote $\BayesIsoAni \sim 1.1$ - $1.6$ depending on how many spherical harmonics they include.
We see, then, that our larger sample size provides the tightest constraints to-date.

\section{Hierarchical Models}
\label{sec:hierarchical models}

We now consider several additional representations of the distribution of merging binaries and construct maps of the merger rate across the sky with each.
Section~\ref{sec:formalism} briefly reviews hierarchical Bayesian inference before Section~\ref{sec:results} presents our results.
Comparing different modeling choices allows us to examine, to some extent, which features are constrained by the data and which are dominated by our modeling choices.

\begin{table*}
    \caption{
        Fixed population models for the source-frame primary mass ($m_1$), secondary mass ($m_2 \leq m_1$), Cartesian spin vectors for each component ($\vec{s}_1$, $\vec{s}_2$), and redshift ($z$).
        We employ the \textit{maximum a posteriori} values for the Broken Power-Law + Dip model from Ref.~\cite{Farah:2021} as well as a flat $\Lambda$CDM cosmology with $H_0 = 67.32 \, \mathrm{km}/\mathrm{s}/\mathrm{Mpc}$, $\Omega_M = 0.3158$, and $\Omega_\Lambda=1-\Omega_M$ (first column of Table 1 in Ref.~\cite{Planck:2020}).
        We also assume events' orbital inclinations are isotropically distributed, events' phases at coalescence and polarization angles are uniformly distributed throughout their physical ranges, and that events' arrival times are uniformly distributed throughout the duration of the experiment.
    }
    \label{tab:fixed population models}
    {\renewcommand{\arraystretch}{1.5}
    \begin{tabular}{c c c p{4.5cm} p{4.5cm}}
        \hline
        \hline
            Variates & Name/Description & Functional Form & \multicolumn{2}{c}{Visualization} \\
        \hline
        \hline
            $m_1$, $m_2$
              & \thead{Broken Power-Law \\ + Dip \\ (BPL+Dip)}
                & \thead{$p(m_1,m_2) \quad\quad\quad\quad\quad\quad$ \\ $\quad\quad \propto p(m_1) p(m_2) \quad\quad$ \\ $\quad\quad\quad\quad \times \, p_\mathrm{pair}(m_2, m_2/m_1)$ \\ see Refs.~\cite{Farah:2021, GWTC-3-RnP}}
                & \includegraphics[hsmash=r, align=c, width=4.5cm, clip=True, trim=0.25cm 0.55cm 0.15cm 0.20cm]{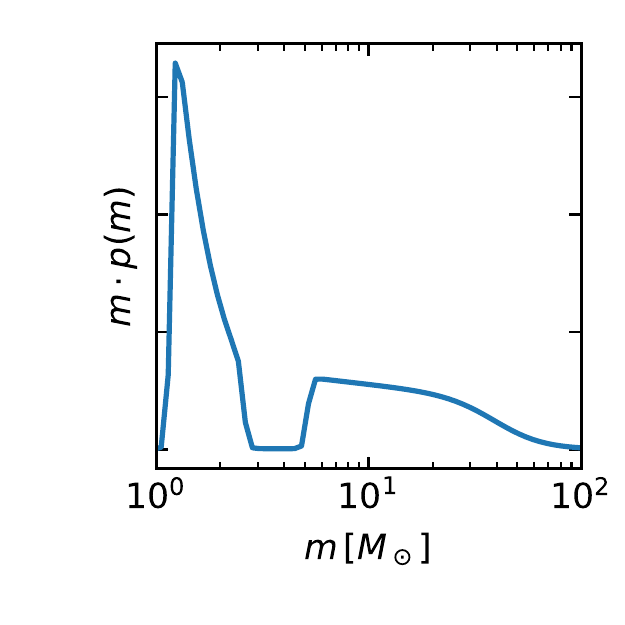}
                & \includegraphics[hsmash=r, align=c, width=4.5cm, clip=True, trim=0.25cm 0.55cm 0.15cm 0.20cm]{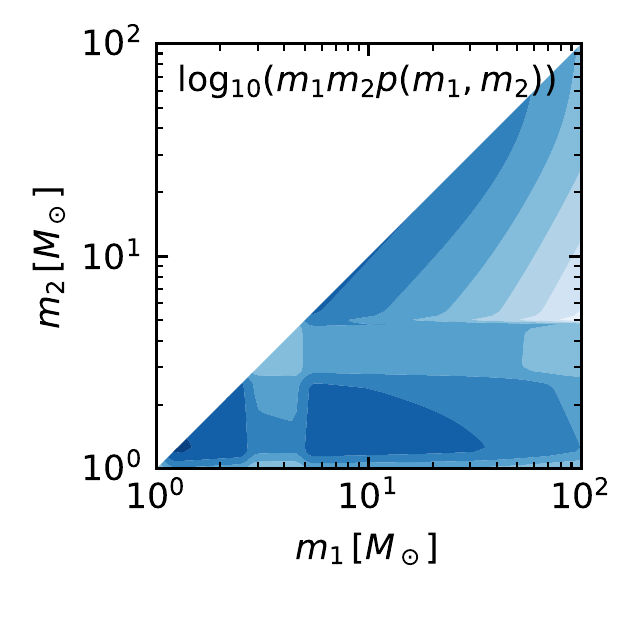} \\
            $z$
              & \thead{uniform in \\ comoving volume \\ \& source-frame time}
                & $p(z) \propto (dV_c/dz) / (1+z)$
                & \multicolumn{2}{c}{\includegraphics[hsmash=c, align=c, width=4.5cm, clip=True, trim=0.25cm 0.60cm 0.15cm 0.40cm]{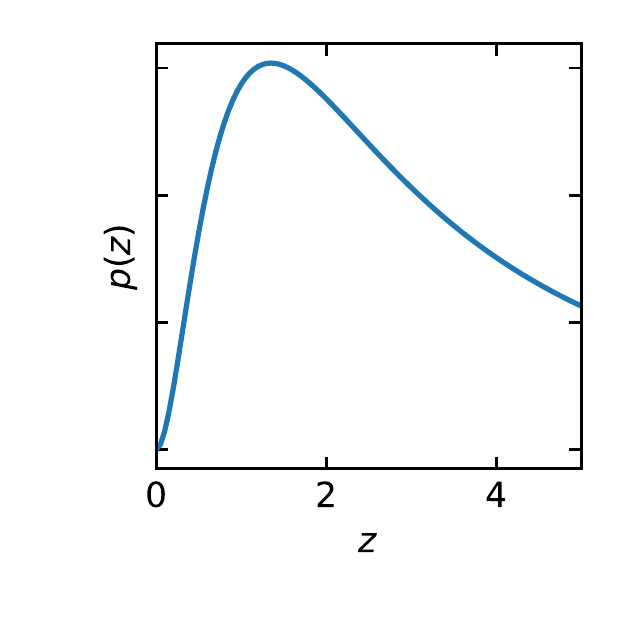}} \\
            $\vec{s}_1, \vec{s}_2$
              & \thead{uniform in magnitude \\ \& isotropic orientation}
                & $p(\vec{s}_i) = 1/4\pi|\vec{s}_i|^2$
                & \includegraphics[hsmash=r, align=c, width=4.5cm, clip=True, trim=0.25cm 0.55cm 0.15cm 0.30cm]{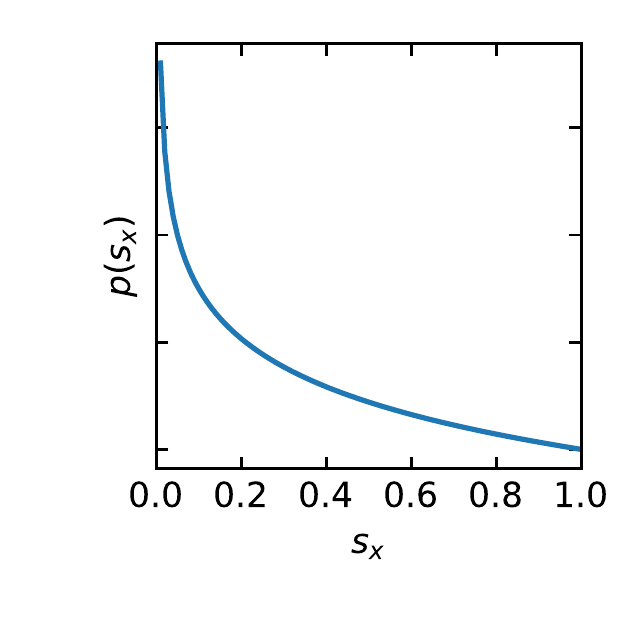}
                & \includegraphics[hsmash=r, align=c, width=4.5cm, clip=True, trim=0.25cm 0.55cm 0.15cm 0.30cm]{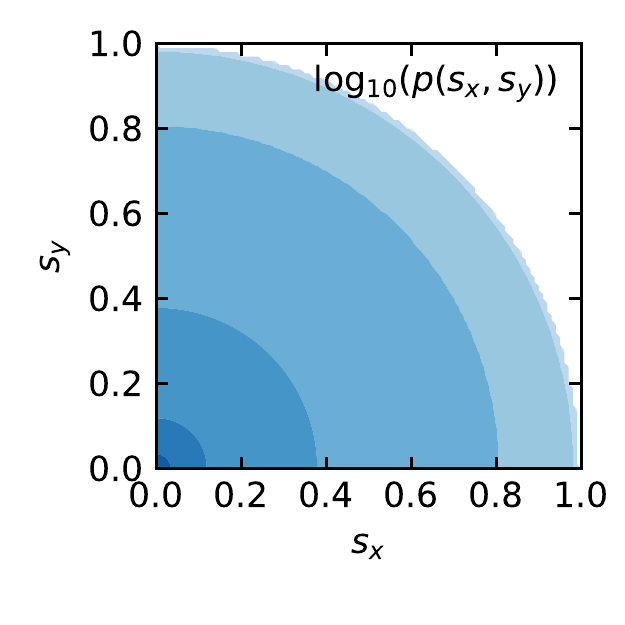} \\
        \hline
    \end{tabular}
    }
\end{table*}

In what follows, we assume fixed distributions for the source-frame component masses, redshift, spins, inclination, orientation, and arrival time of GWs from binary systems.
These are described in Table~\ref{tab:fixed population models}.
In order to focus on isotropy, we only infer the parameters of the distribution over right ascension ($\alpha$) and declination ($\delta$).
Studies of possible correlations between the direction to the source and other source properties are left to future work.

We only consider events from O3 (GWTC-2~\cite{GWTC-2}, GWTC-2.1~\cite{GWTC-2d1}, and GWTC-3~\cite{GWTC-3}), as the publicly available set of simulated signals processed with real searches used to estimate the catalog's sensitivity only covers O3~\cite{GWTC-3-injections}.
We approximate the catalog selection by requiring the false alarm rate (FAR) from at least one pipeline within GWTC-3 to be \result{$\leq 1/\mathrm{year}$}.\footnote{We include all searches present in GWTC-3: both modeled (GstLAL, MBTA, PyCBC broad, and PyCBC BBH) and unmodeled (cWB) searches. See Ref.~\cite{GWTC-3} for more details about individual searches.}
With this selection threshold, we retain \result{63} events from O3.
See Appendix~\ref{sec:events and injections} for more details.


\subsection{Formalism}
\label{sec:formalism}

\begin{table*}
    \caption{
        Population models for the distribution over right ascension ($\alpha$) and declination ($\delta$).
        See text for more detailed definitions of each model's parameters.
        We denote the uniform distribution between $X$ and $Y$ as $U(X, Y)$, the exponential distribution with scale parameter $Z$ as $\mathrm{Exp}[Z]$ ($p(x) = Z^{-1}e^{-x/Z}$), and the multivariate Normal distribution with mean vector $\mu$ and covariance matrix $\Xi$ as $\mathcal{N}(\mu, \Xi)$.
        Where relevant, we denote the area of pixel $i$ with $A_i$.
    }
    \label{tab:fitted population models}
    {\renewcommand{\arraystretch}{1.5}
    \begin{tabular}{@{\extracolsep{0.2cm}} cccc p{2.5cm}}
        \hline
        \hline
            Variates & Name & Parameters & Functional Form & \multicolumn{1}{c}{Example} \\
        \hline
            \multirow{19}{*}{$\Omega \equiv \alpha$, $\delta$}
              & \multirow{4}{*}{\thead{Rotated Hemisphere \\ (RH)}}
                & $f \sim U(0, 1)$
                  & \multirow{4}{*}{\thead{Rotate by Euler angles ($\phi$, $\theta$, $\psi$). \\ $\alpha, \delta \rightarrow \tilde{\alpha}, \tilde{\delta}$ \\ In the rotated frame \\ $p = (f\Theta(\tilde\delta > 0) + (1-f)\Theta(\tilde\delta < 0))/2\pi$}}
                    & \multirow{4}{*}{\includegraphics[hsmash=r, align=c, width=2.5cm, clip=True, trim=0.0cm 0.0cm 1.4cm 0.0cm]{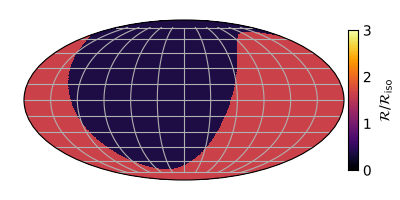}} \\
              & & $\phi \sim U(0, 2\pi)$ \\
              & & $\theta \sim U(0, 2\pi)$ \\
              & & $\psi = 0$ \\
              \cline{2-5}
              & \multirow{3}{*}{\thead{Simple Dipole \\ (SD)}}
                & $|\vec{b}| \sim U(0, 1)$
                  & \multirow{3}{*}{\thead{$p = (1+\vec{b}\cdot\hat{\Omega})/4\pi$ \\ $|\vec{b}| \leq 1$}}
                    & \multirow{3}{*}{\thead{\includegraphics[hsmash=r, align=c, width=2.5cm, clip=True, trim=0.0cm 0.0cm 1.4cm 0.0cm]{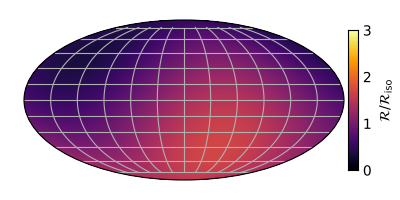}}} \\
              & & $\arctan( b_y / b_x) = \phi \sim U(0, 2\pi)$ \\
              & & $b_z / |b| = \cos\theta \sim U(-1, +1)$ \\
              \cline{2-5}
              & \thead{Healpix pixelization \\ (HP: $N_\mathrm{pix}=12, 48, 192$)}
                & \multirow{4}{*}{\thead{$f_i \sim \mathrm{Exp}(A_i^{-1})$ \\ $\forall \ i \in [1, \cdots, N_\mathrm{pix}]$}}
                  & \multirow{10}{*}{\thead{$p = \sum\limits_i^{N_\mathrm{pix}} f_i A_i^{-1}\Theta((\alpha, \delta) \in A_i)$ \\ $\sum\limits_i^{N_\mathrm{pix}} f_i = 1$}}
                    & \includegraphics[hsmash=r, align=c, width=2.5cm, clip=True, trim=0.0cm 0.0cm 1.4cm 0.0cm]{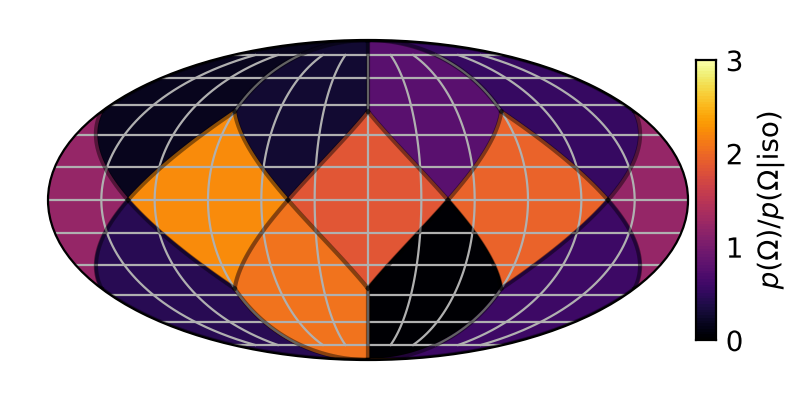} \\
              \cline{2-2}
              & \thead{IAU Constellations \\ (IAU: $N_\mathrm{pix}=89$)}
                &
                  &
                    & \includegraphics[hsmash=r, align=c, width=2.5cm, clip=True, trim=0.0cm 0.0cm 1.4cm 0.0cm]{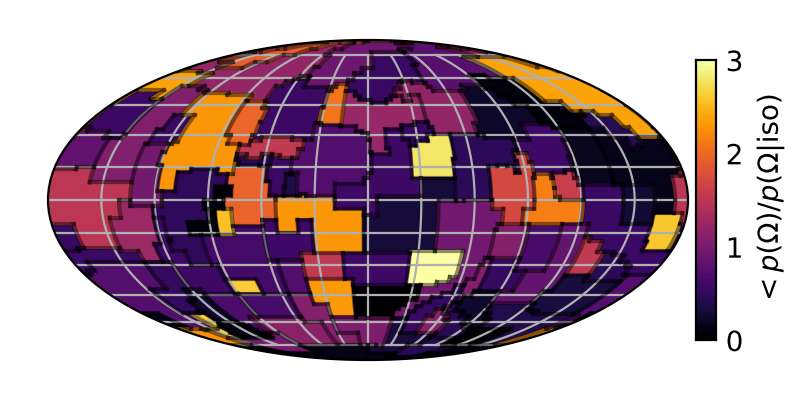} \\
              \cline{2-3}
              & \multirow{5}{*}{\thead{Gaussian Random Field \\ (GRF)}} 
                & $\log(f_i/A_i) \sim \mathcal{N}(0, \Xi_{ij})$
                  &
                    & \multirow{5}{*}{\includegraphics[hsmash=r, align=c, width=2.5cm, clip=True, trim=0.0cm 0.0cm 1.4cm 0.0cm]{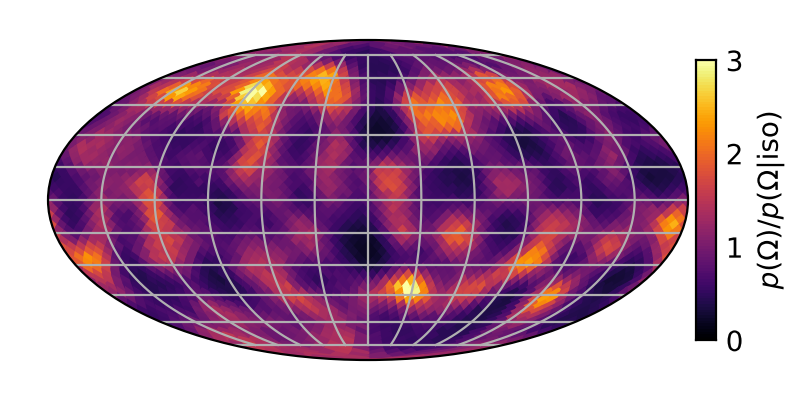}} \\
              & & $\Xi_{ij} = \sigma_\mathrm{wn}^2 \delta_{ij} + \sigma^2 e^{-(\Delta\theta_{ij})^2/\vartheta^2}$ \\
              & & $\vartheta \sim U(\vartheta_\mathrm{min}, \pi/3)$ \\
              & & $\sigma \sim U(0, 3)$ \\
              & & $\sigma_\mathrm{wn} = \sigma/10$ \\
              \cline{2-5}
              & \multirow{4}{*}{\thead{Exponentiated \\ Spherical Harmonics \\ (ESH: $l_\mathrm{max}=1,2,3,4$)}}
                & $\mathbb{R}\{b_{l\leq l_\mathrm{max}}^{m=0}\} \sim U(-10, +10)$
                  & \multirow{4}{*}{\thead{
$p \propto \exp\left(\sum\limits_{lm} b_l^m Y_l^m(\alpha, \delta)\right)$ \\
$b_l^{-m} = (b_l^{+m})^\ast$
}}
                    & \multirow{4}{*}{\includegraphics[hsmash=r, align=c, width=2.5cm, clip=True, trim=0.0cm 0.0cm 1.4cm 0.0cm]{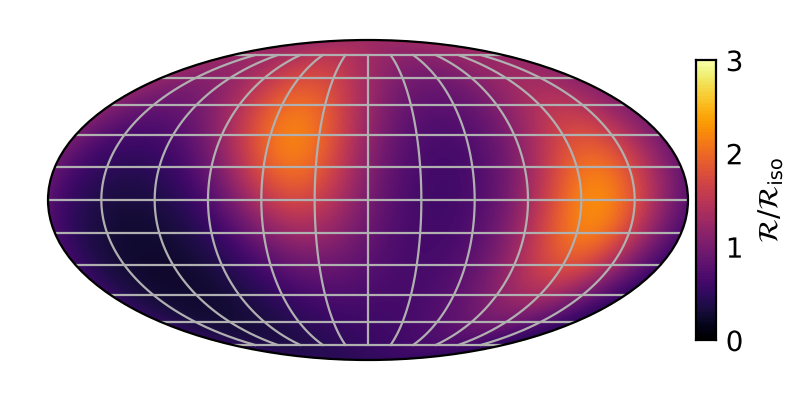}} \\

              & & $\mathbb{I}\{b_{l\leq l_\mathrm{max}}^{m=0}\} = 0$ \\
              & & $\mathbb{R}\{b_{l\leq l_\mathrm{max}}^{m>0}\} \sim U(-10, +10)$ \\
              & & $\mathbb{I}\{b_{l\leq l_\mathrm{max}}^{m>0}\} \sim U(-10, +10)$ \\
        \hline
    \end{tabular}
    }
\end{table*}

We employ hierarchical Bayesian inference to infer the rate density of merging compact binaries:
\begin{equation}
    \frac{dN}{d\theta} = \mathcal{R} p(\theta|\Lambda)
\end{equation}
where each event is described by parameters $\theta$ (masses, redshift, right ascension, declination, etc), the population distribution $p(\theta|\Lambda)$ is described by some set of parameters $\Lambda$ (minimum and maximum masses, anisotropy parameters, etc), and $\mathcal{R}$ acts as an overall normalization constant.

Specifically, we sample from the rate-marginalized inhomogeneous Poisson likelihood for the observed data $\{D_i\}$ from $N$ events
\begin{equation}\label{eq:likelihood}
    p(\{D_i\}|\Lambda) = \prod\limits_{i}^{N} \frac{\int d\theta\, p(D_i|\theta) p(\theta|\Lambda)}{\int d\theta\, P(\mathrm{det}|\theta) p(\theta|\Lambda)}
\end{equation}
with a corresponding prior for $\Lambda$.
Here, $P(\mathrm{det}|\theta)$ is the (time-averaged) probability of detecting a signal with parameters $\theta$.
Eq.~\ref{eq:likelihood} implicitly assumes $p(\mathcal{R}) \sim 1/\mathcal{R}$ within the marginalization over $\mathcal{R}$.
We estimate the numerators in Eq.~\ref{eq:likelihood} via Monte Carlo importance sampling of single-event posterior samples for each event, and the denominator with a set of detected simulated signals (Appendix~\ref{sec:events and injections}).
See, e.g., Refs.~\cite{Loredo:2004, Mandel:2010, Mandel:2019, Essick:2022} and references therein for more details.


\subsection{Cartography}
\label{sec:results}

Within the hierarchical framework, we consider several different representations of the distributions over the sky.
Broadly, these can be classified as those based on pixelizations (like the Rotated Hemisphere model in Sec.~\ref{sec:counting experiments}) and those based on spherical harmonic decompositions.
Table~\ref{tab:fitted population models} summarizes our models, their parameters, and the priors chosen for those parameters.
While there is no fundamental difference between the two approaches, each introduces different priors on the types of variation over the sky.
Nonetheless, as we will see, we obtain comparable results regardless of the precise model choices.


\subsubsection{Pixelized Representations}
\label{sec:pixelized representations}

\begin{figure*}
    \begin{tabular}{c p{0.47\textwidth} p{0.47\textwidth}}
        {\large $N_\mathrm{pix}$} & {\hspace{3.2cm} \Large mean} & {\hspace{2.7cm} \Large significance} \\
        {\large 12}
          & \includegraphics[hsmash=r, align=c, width=0.47\textwidth]{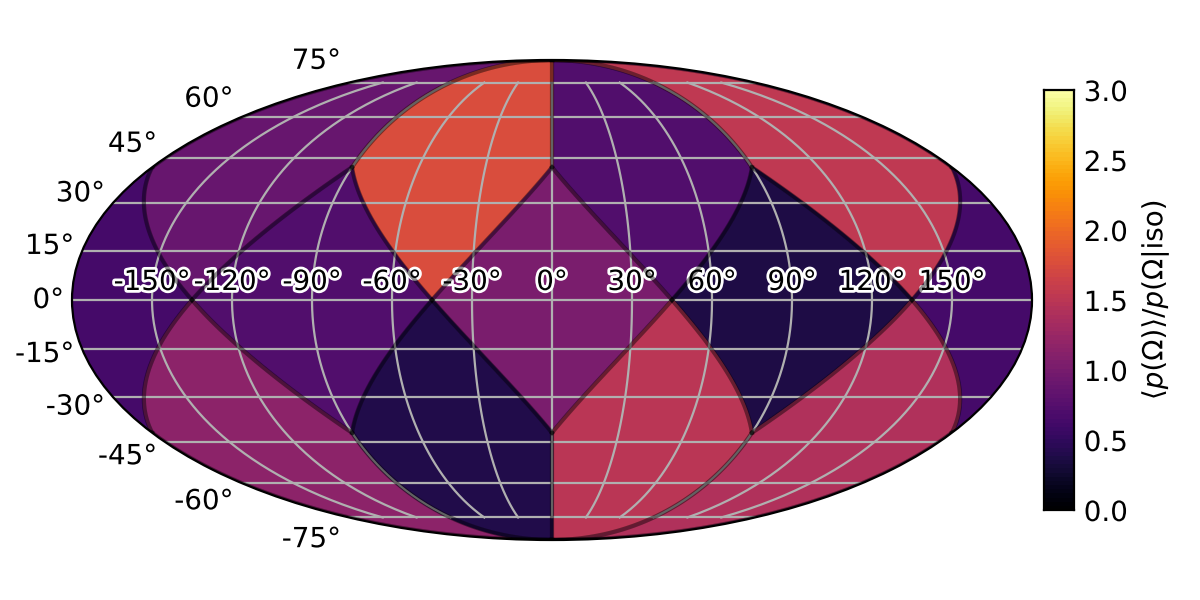}
          & \includegraphics[hsmash=r, align=c, width=0.47\textwidth]{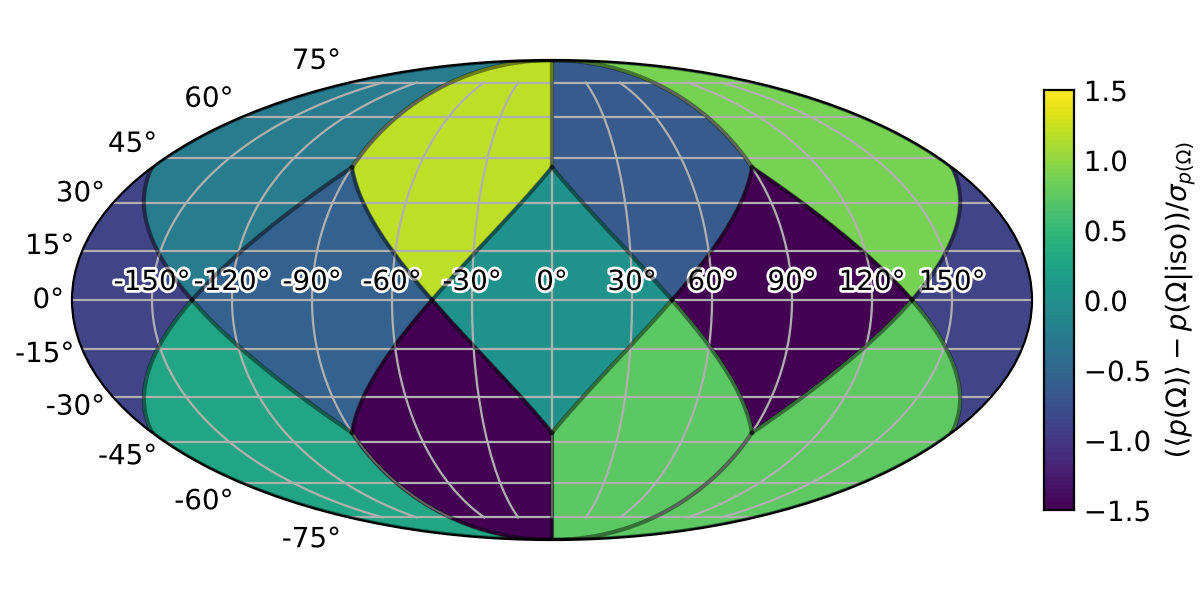} \\
        {\large 48}
          & \includegraphics[hsmash=r, align=c, width=0.47\textwidth]{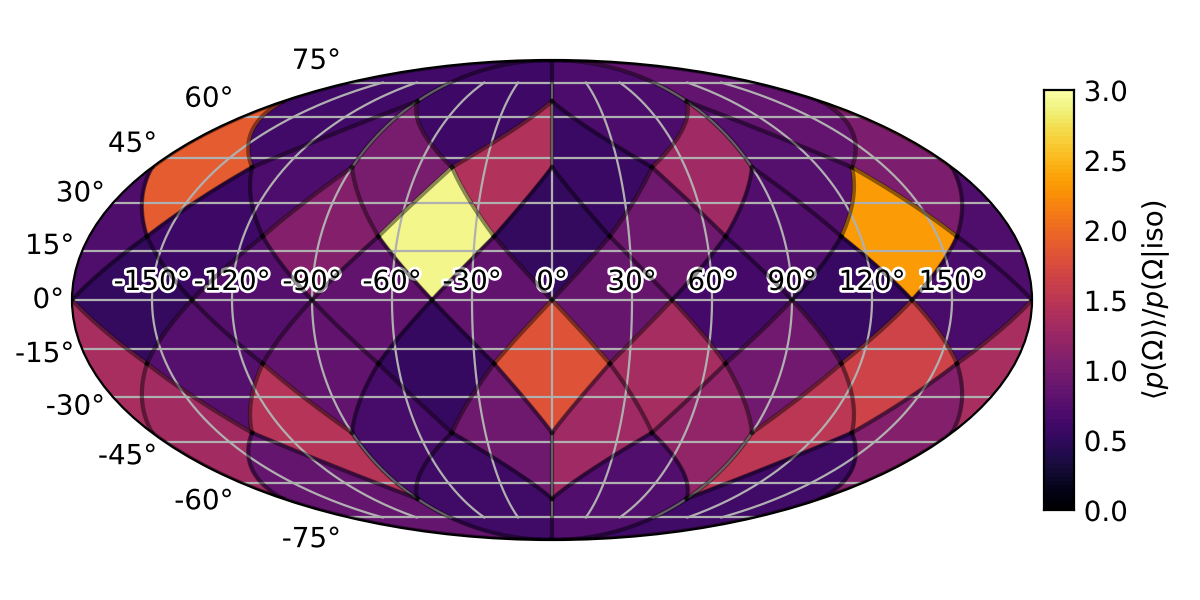}
          & \includegraphics[hsmash=r, align=c, width=0.47\textwidth]{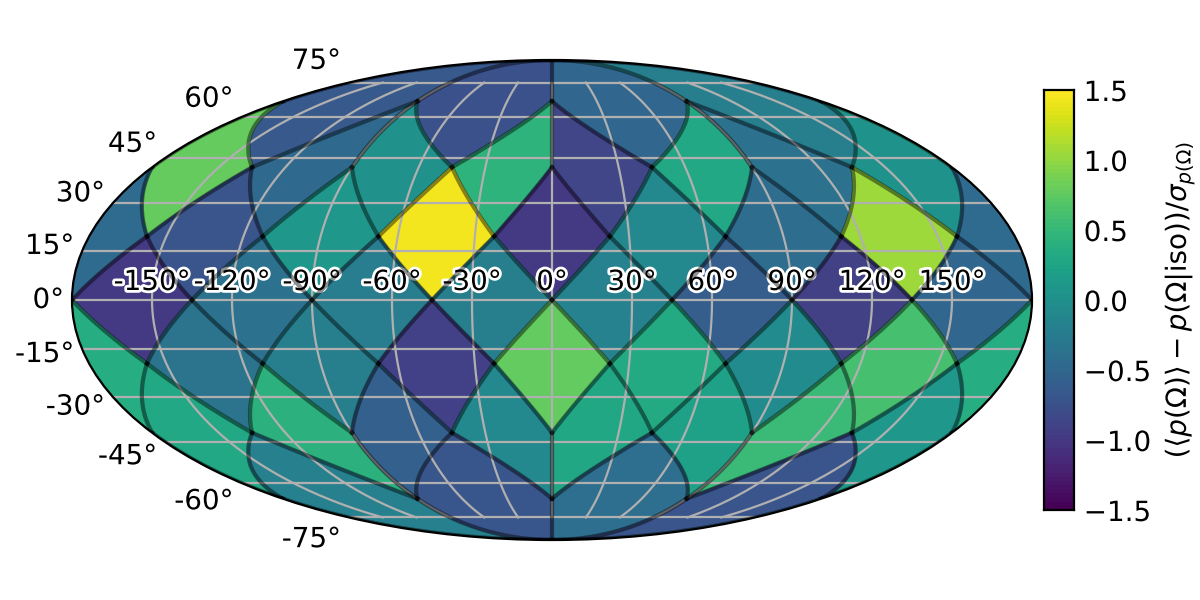} \\
        {\large 196}
          & \includegraphics[hsmash=r, align=c, width=0.47\textwidth]{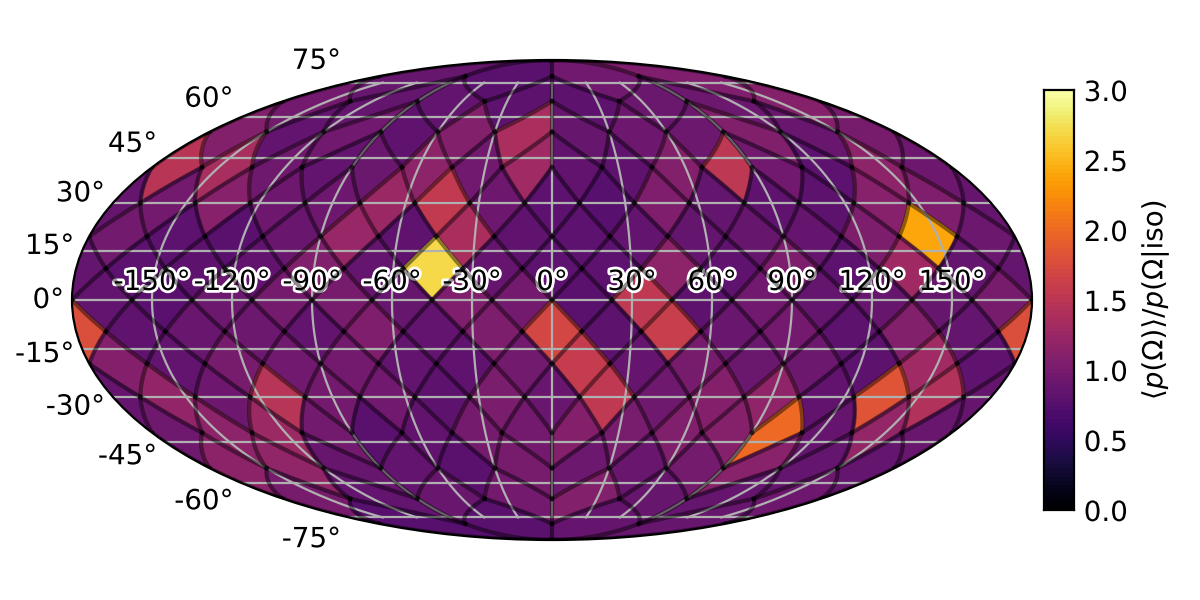}
          & \includegraphics[hsmash=r, align=c, width=0.47\textwidth]{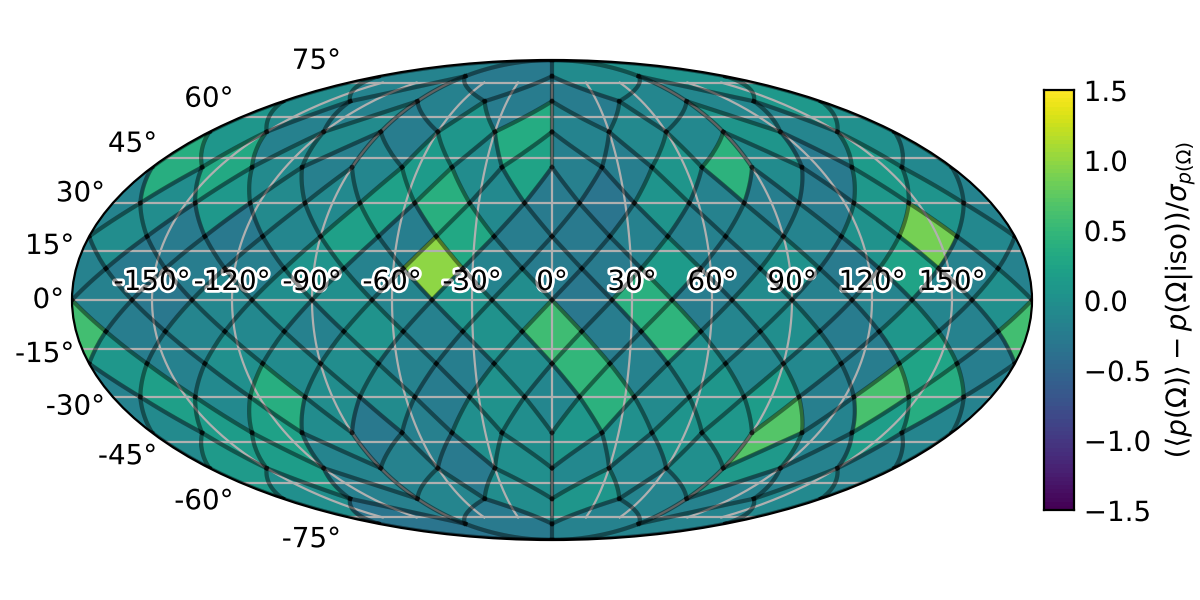} \\
        {\large 89}
          & \includegraphics[hsmash=r, align=c, width=0.47\textwidth]{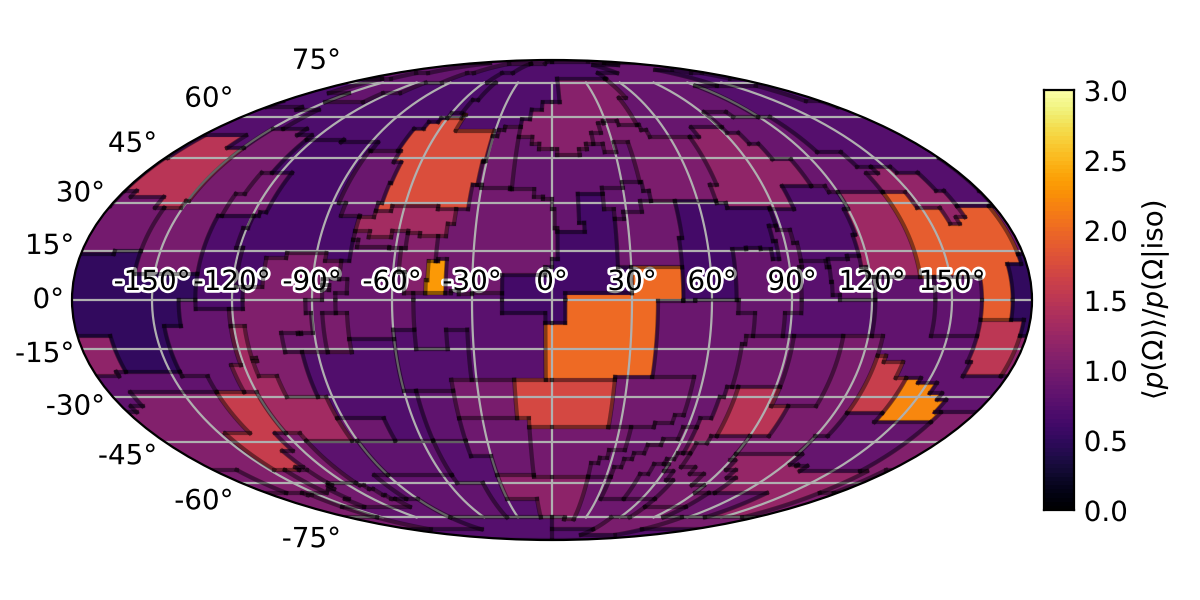}
          & \includegraphics[hsmash=r, align=c, width=0.47\textwidth]{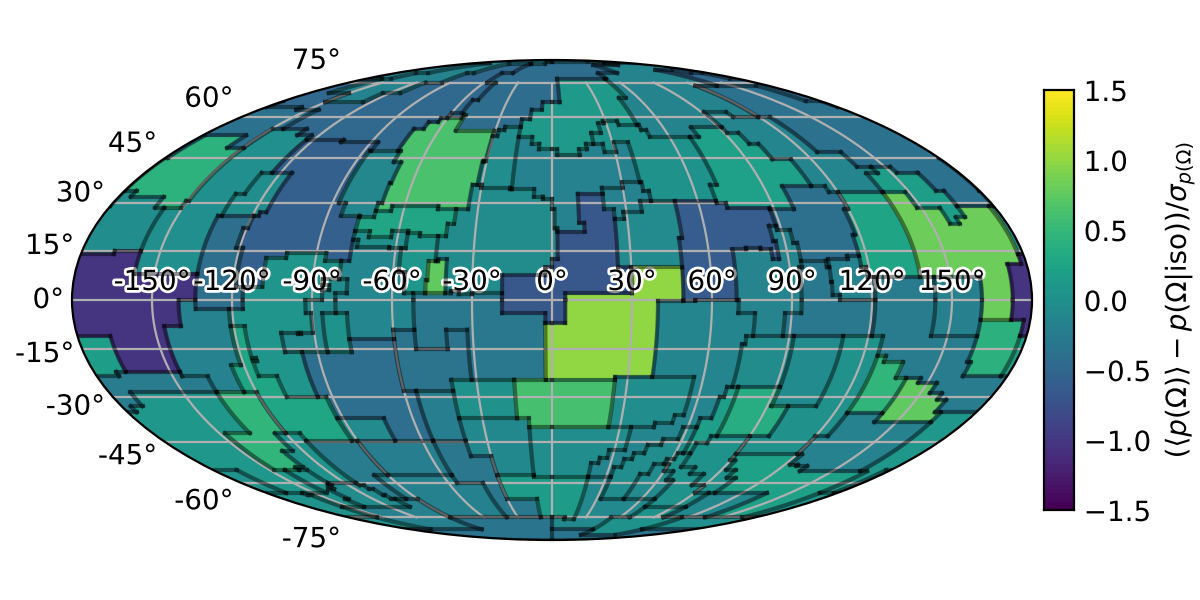}
    \end{tabular}
    \caption{
        Mollweide projections of the posterior for the rate density across the sky with pixelized representations.
        (\emph{top to bottom}) Healpix pixelizations with 12, 48, and 192 pixels as well as a pixelization based on the 88 IAU Constellations (89 pixels; Serpens is divided into two disjoint regions).
        (\emph{left}) The average rate density \textit{a posteriori} scaled by the rate for an isotropic distribution.
        (\emph{right}) A measure of statistical significance: the difference between the average rate density and the rate for an isotropic distribution scaled by the standard deviation of the rate density in each pixel \textit{a posteriori}.
    }
    \label{fig:hierarchical pixel maps}
\end{figure*}

\begin{figure*}
    \begin{tabular}{c p{0.47\textwidth} p{0.47\textwidth}}
        {\Large $l_\mathrm{max}$} & {\hspace{3.2cm} \Large mean} & {\hspace{2.7cm} \Large significance} \\
        {\large 1}
          & \includegraphics[hsmash=r, align=c, width=0.47\textwidth]{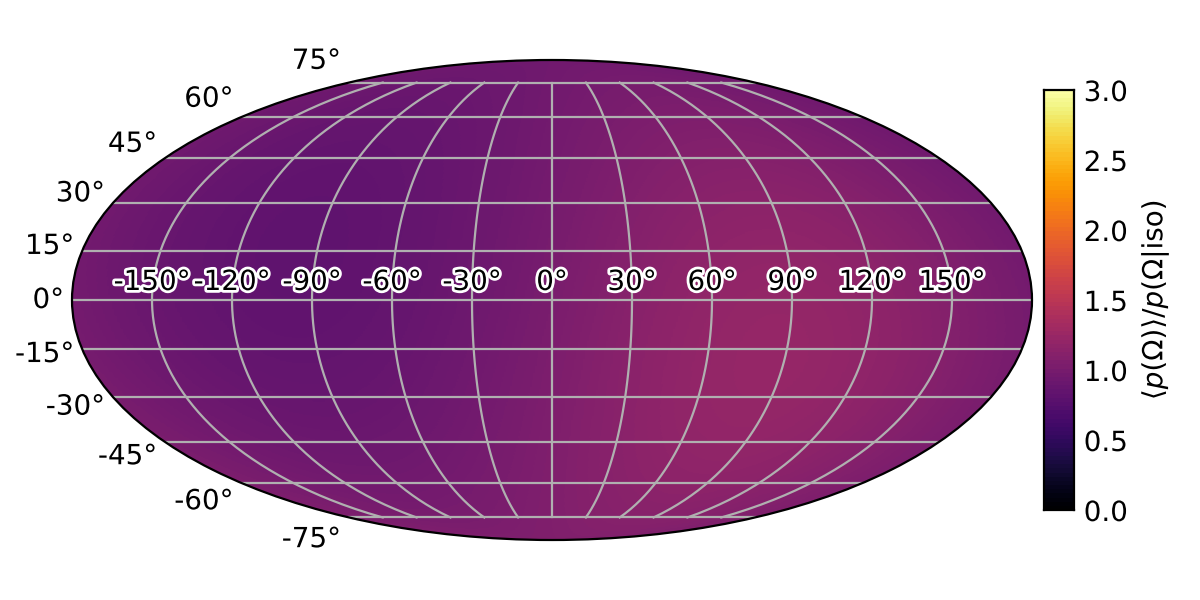}
          & \includegraphics[hsmash=r, align=c, width=0.47\textwidth]{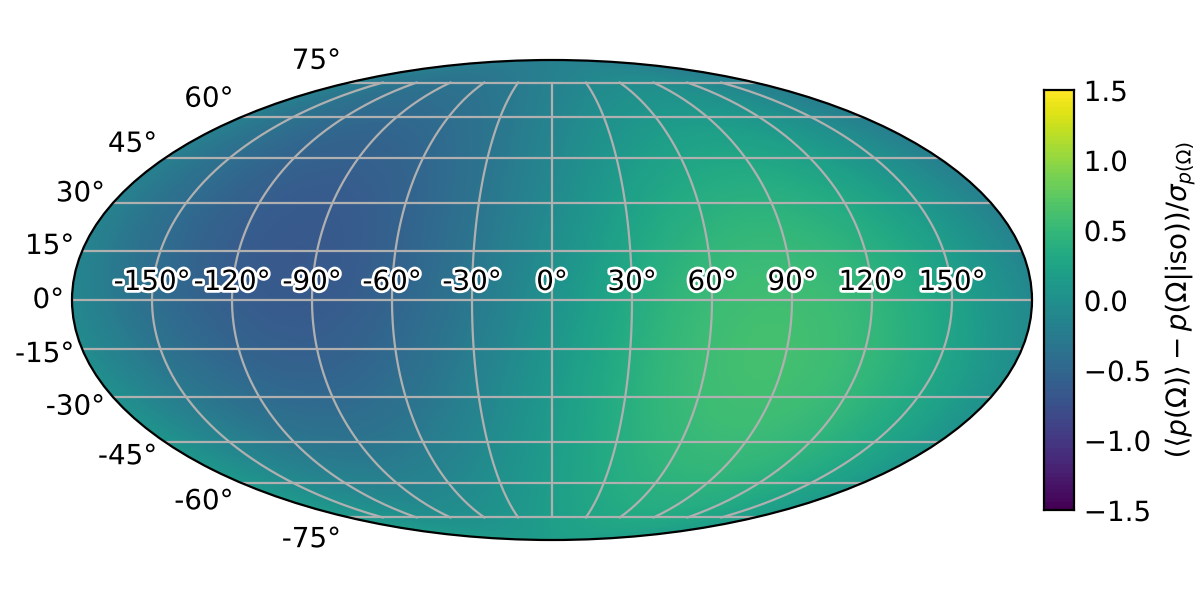} \\
        {\large 2}
          & \includegraphics[hsmash=r, align=c, width=0.47\textwidth]{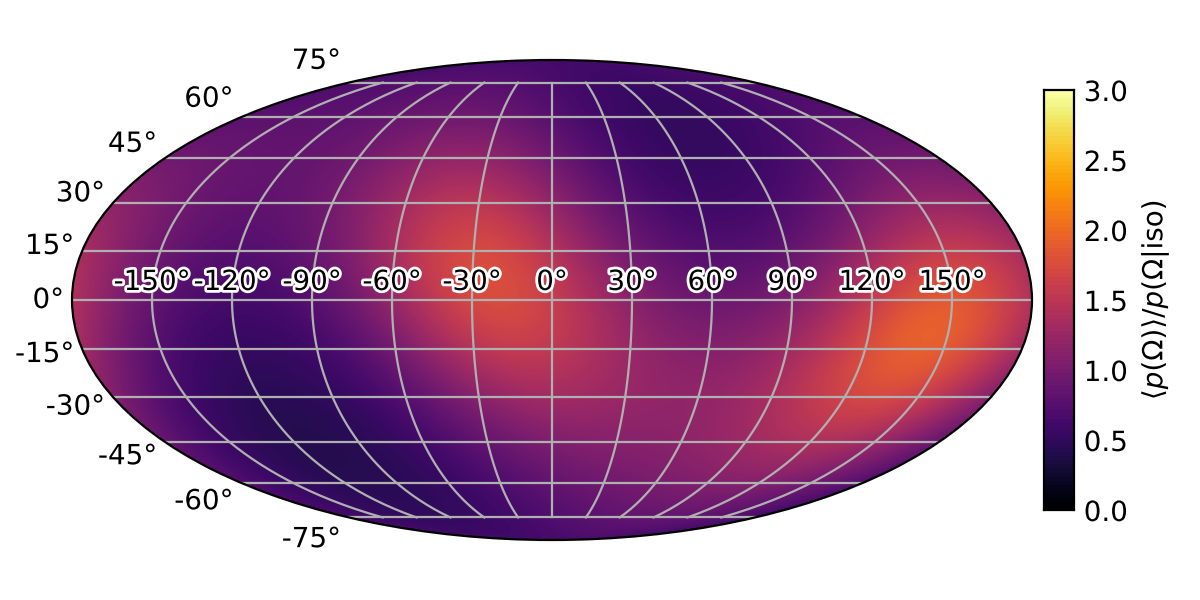}
          & \includegraphics[hsmash=r, align=c, width=0.47\textwidth]{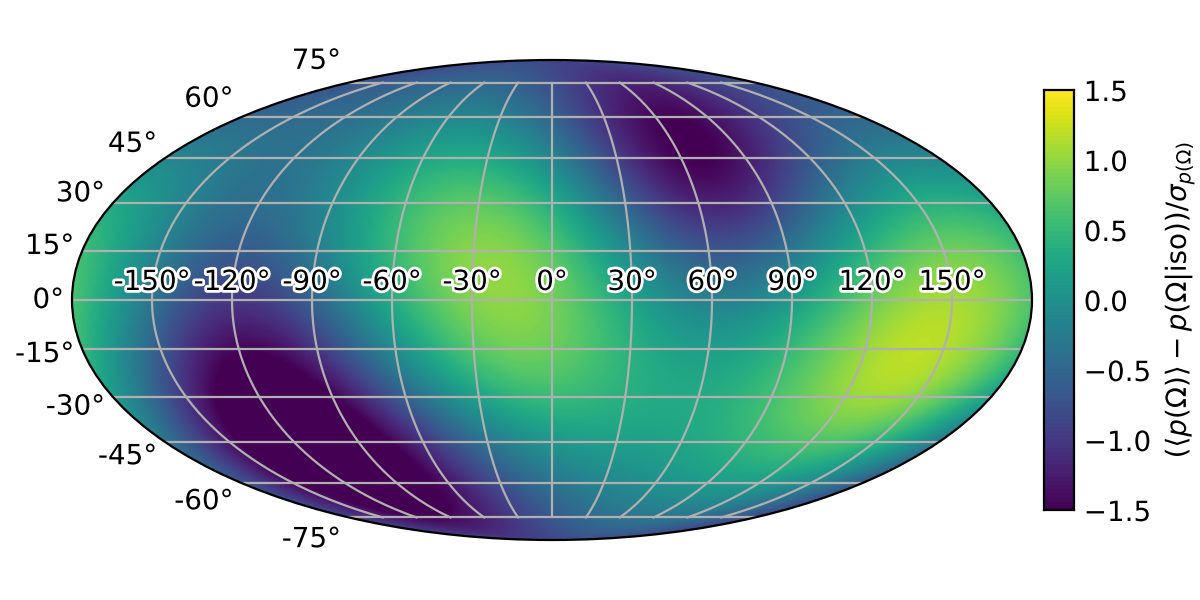} \\
        {\large 3}
          & \includegraphics[hsmash=r, align=c, width=0.47\textwidth]{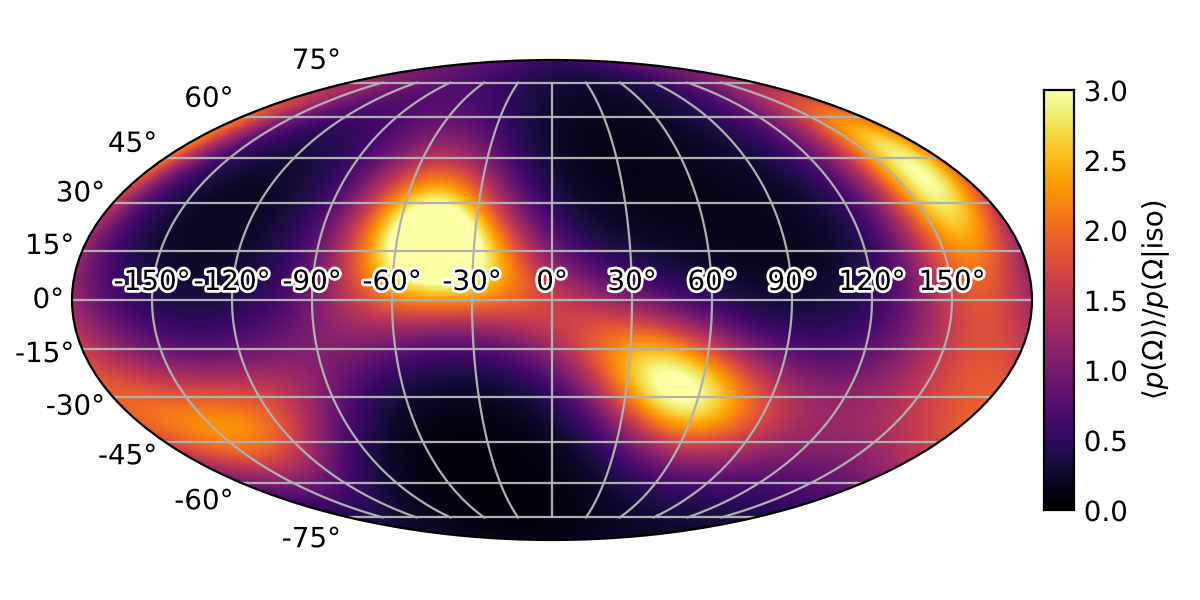}
          & \includegraphics[hsmash=r, align=c, width=0.47\textwidth]{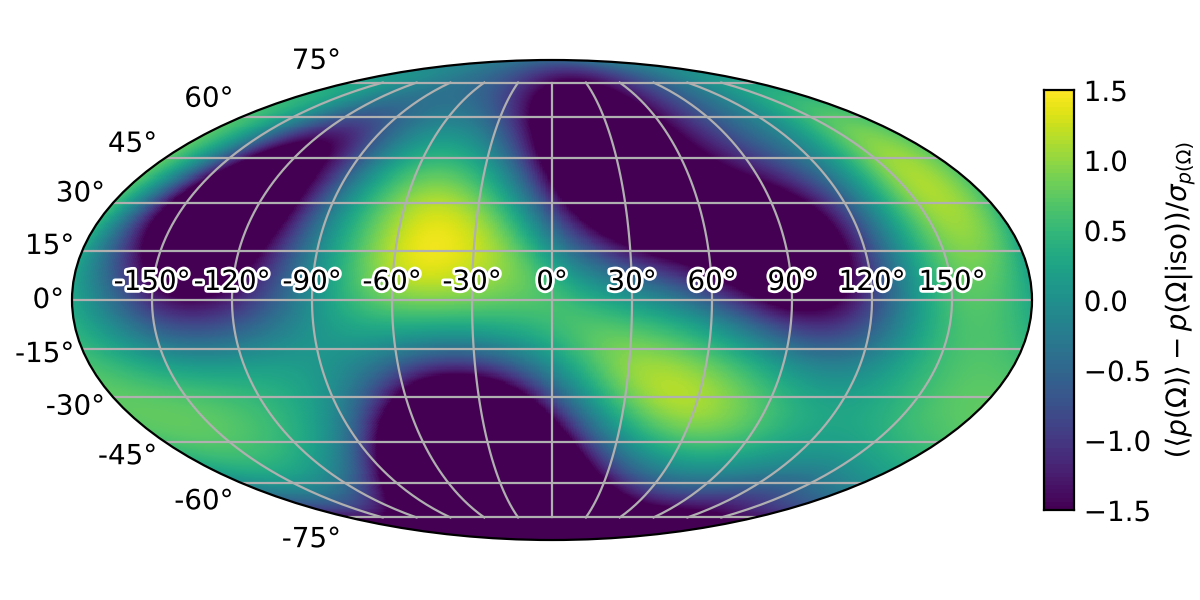} \\
    \end{tabular}
    \caption{
        Mollweide projections of the posterior rate density across the sky when it is represented by a spherical harmonic decomposition.
        Analogous to Fig.~\ref{fig:hierarchical pixel maps}.
        (\emph{top to bottom}) Spherical harmonics are included up to $l_\mathrm{max}=1$, 2, and 3.
        As more harmonics are included, we observe structure across the sky similar to what is found with the pixelized representations.
        However, as in Fig.~\ref{fig:hierarchical pixel maps}, the fluctuations in the posterior are larger than the difference between the mean \textit{a posteriori} and an isotropic distribution.
    }
    \label{fig:hierarchical spherical harmonic maps}
\end{figure*}

The simple Rotated Hemisphere model (Section~\ref{sec:counting experiments}) found weak evidence for isotropy, but this could be due to the assumptions implicit in the model.
We now focus on pixelization schemes that allow for more complex anisotropies.
Fig.~\ref{fig:hierarchical pixel maps} shows Mollweide projections of $p(\Omega)$ derived from different modeling assumptions.
Specifically, we employ the Healpix pixelization scheme~\cite{Gorski:2005} and consider models with 12, 48, and 192 pixels, respectively.
In each model, the rate density in each pixel is independently, exponentially distributed \textit{a priori}.
The exponential distribution is convenient because it only has support for non-negative rate densities.
Furthermore, the independent priors for each pixel give models with more pixels more freedom.
That is, as the number of pixels increases, the prior not only allows for, but actually prefers increased variation across the sky.

We additionally consider a radically different pixelization scheme: the 88 constellations\footnote{Although there are only 88 constellations, we fit the rate in 89 pixels, dividing Serpens (Ser) into its two disjoint regions.} defined by the International Astronomical Union (IAU)~\cite{Delporte:1930}.
Although we expect the IAU constellations to be completely unrelated to GW events, which come from much greater distances than the stars that make up the constellations, they provide a convenient and memorable alternative pixelization.
As with the Healpix models, the rate in each constellation is independently exponentially distributed \textit{a priori}.
Because we parametrize the model in terms of the rate (count per steradian) within each constellation, this implies that the expected number of events from a constellation scales with the constellation's area.

Unlike our Rotated Hemisphere model and Ref.~\cite{Stiskalek:2020}, we do not consider rotations of these pixelizations.
Ref.~\cite{Stiskalek:2020} only used 12 pixels and introduced 3 Euler angles to attempt to control for model systematics associated with the placement of the 12 pixels.
We instead use models with more pixels and different methods of partitioning the sky to test for model systematics.

We find similar features with all models.
Although there are ``hot pixels'' throughout the sky for each, on average (left column of Fig.~\ref{fig:hierarchical pixel maps}) there is a consistently hot pixel near $(\alpha, \delta)=(-45^\circ, +10^\circ)$.
This hot spot lies within Equuleus (Equ: the ``little horse''), which has the second smallest area of any of the IAU constellations and is associated with a handful of relatively well localized events (see Appendix~\ref{sec:events and injections}).
Even though the horse is little, at face value it may play a big role in GW anisotropy measurements.
However, while hot pixels can at times correspond to rates several times larger than the rate for an isotropic distribution, there is still significant uncertainty in the posterior.
In fact, the expected value of the rate in any pixel is always less than $\pm1.5$ standard deviations away from isotropy \textit{a posteriori} (right column of Fig.~\ref{fig:hierarchical pixel maps}).

We note that the size of the deviations from isotropy are less significant within models with more pixels.
This is because there are fewer expected events per pixel and therefore greater relative uncertainty in each pixel's rate density.
Indeed, while there are always some pixels for which the rate is not confidently bounded away from zero, there are (many) more poorly constrained pixels for models with larger $N_\mathrm{pix}$.
See Appendix~\ref{sec:perturbation} for more discussion.

While comparisons based on only one-dimensional marginal posteriors do not actually represent a full test of isotropy (the rate density must be consistent with isotropy in all pixels simultaneously, not just separately for each pixel), this is nevertheless suggestive.
We also eschew \BayesIsoAni~for these models due to the possibly strong dependence on our prior choices (see discussions in, e.g., Refs.~\cite{Essick:2020, Isi:2022}).
We quantify constraints on anisotropies in more detail in Section~\ref{sec:quantification}.


\subsubsection{Spherical Harmonic Representations}
\label{sec:spherical harmonic representations}

\begin{figure*}
    \centering
    \includegraphics[hsmash=c, width=1.0\textwidth, clip=True, trim=1.00cm 1.25cm 0.5cm 2.00cm]{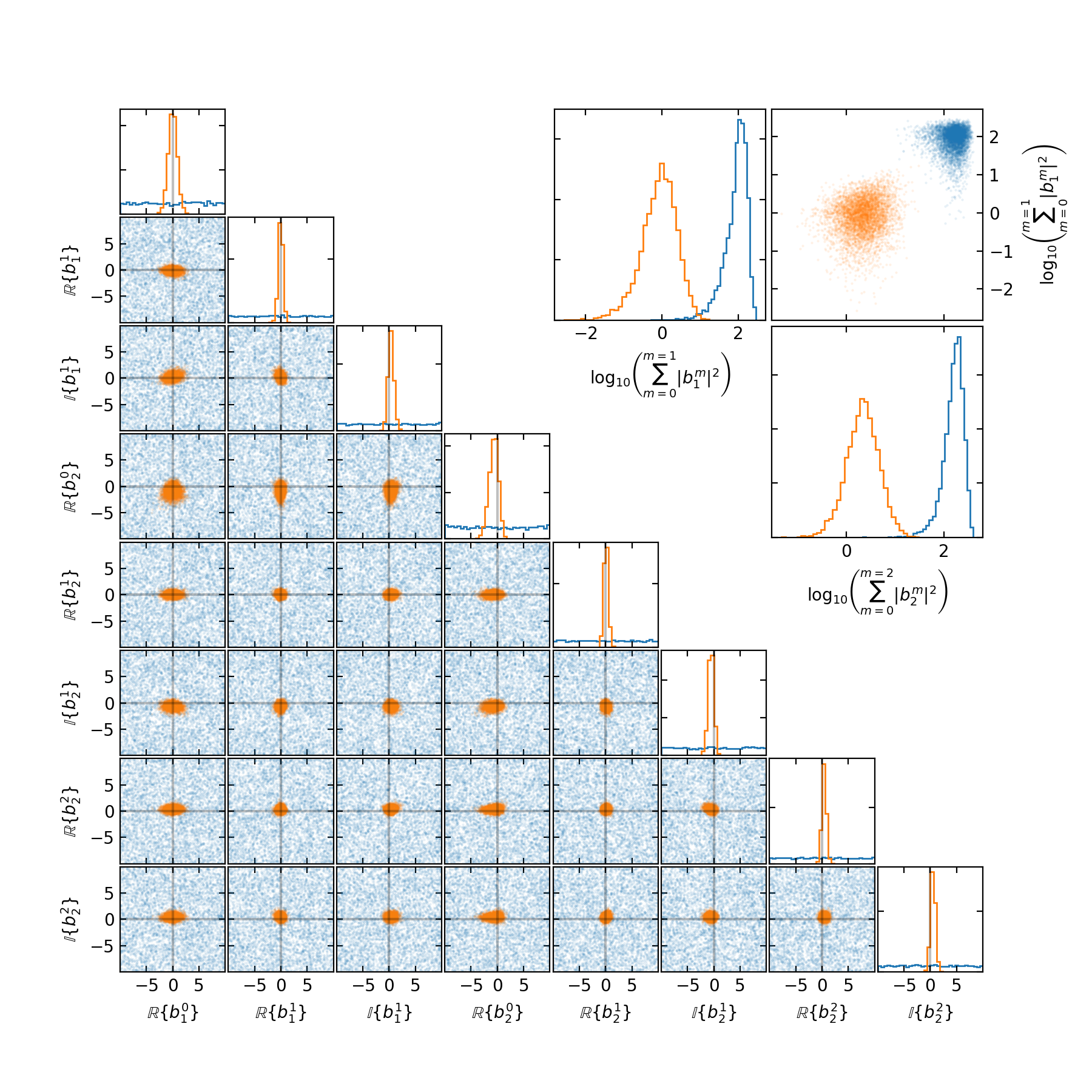}
    \caption{
        Prior (\emph{blue}) and posterior (\emph{orange}) for the spherical harmonic coefficients ($b_l^m$) with $l_\mathrm{max}=2$.
        Other $l_\mathrm{max}$ produce similar behavior.
        (\emph{lower left}) Distributions over $b_l^m$.
        Although the marginal posteriors for some $b_{l>0}^m$ peak at nonzero values, they are all consistent with isotropy.
        Our priors are, perhaps, unrealistically broad, but were intentionally chosen to be much broader than the posterior.
        (\emph{upper right}) Distributions over the power in each angular harmonic.
        Even though isotropy is strongly disfavored \textit{a priori}, the data constrain the power in higher harmonics to be small compared to the prior.
    }
    \label{fig:ESH corner plots}
\end{figure*}

We now turn our attention to representations of $p(\Omega)$ based on spherical harmonic decompositions.
There are many ways to construct a representation of a positive semi-definite function defined on $\mathbb{S}_2$ in terms of spherical harmonics.
We explored several models of the form
\begin{equation}\label{eq:spherical harmonic sum}
    p(\Omega) = F\left(\sum\limits_{l=0}^{l_\mathrm{max}} \sum\limits_{m=-l}^{m=+l} b_l^m Y_l^m(\Omega) \right)
\end{equation}
with the additional constraint $b_l^{-m} = (b_l^{+m})^\ast$ to ensure the sum is real, where $(\cdot)^\ast$ denotes complex conjugation.

To begin, we consider a Simple Dipole model described by a single vector $\vec{b}$ so that
\begin{equation}
    p(\Omega) = \frac{1}{4\pi} \left( 1 + \vec{b} \cdot \hat{\Omega} \right),
\end{equation}
with $|\vec{b}| \leq 1$.
This model is similar to the Rotated Hemisphere model from Sec.~\ref{sec:counting experiments}, but avoids sharp features in the rate density.
It corresponds to $F(x) = x$ and $l_\mathrm{max}=1$ in Eq.~\ref{eq:spherical harmonic sum}.
With a uniform prior over $|\vec{b}|$ and isotropic priors on its orientation, we constrain \result{$|\vec{b}| \leq 0.5$ at 90\% credibility}, in agreement with the Rotated Hemisphere model.
We also find \result{$\BayesIsoAni = 2.5$}, slightly smaller than the Rotated Hemisphere model.
This is because the simple dipole lacks sharp features in $p(\Omega)$ and therefore larger anisotropies are harder to constrain.

We now additionally consider $l_\mathrm{max} \geq 1$.
Although we find consistent results with several choices of $F(x)$,\footnote{
Ref.~\cite{Payne:2020} chose $F(x) = x^2$, and we also explored $F(x) = |x|$.
However, both of these approaches complicate the interpretation of the model as they introduce strong degeneracies.
That is, multiple distinct sets of $b_l^m$ can produce similar $p(\Omega)$.
For example, a distribution with only $b_1^0 \neq 0$ produces similar $p(\Omega)$ to a distribution with only $b_2^0 \neq 0$.
It is this mixing between different $l$ can be difficult to interpret.
These degeneracies render the posterior for $\{b_l^m\}$ multimodal, which in part motivated Ref.~\cite{Payne:2020}'s choice to limit the magnitude of $b_l^m$ to small values \textit{a priori}.
} we focus on $F(x) = e^x$, which we refer to as the Exponentiated Spherical Harmonic model.
That is, we model the logarithm of the probability density with a spherical harmonic decomposition.
This preserves the parity of all $Y_l^m$, thereby removing many of the degeneracies introduced by other choices and simplifying the interpretation of posterior constraints for $b_l^m$.  

We consider independent, uniform priors for the real and imaginary parts of $b_l^m$ (subject to the reality constraint) up to several maximum harmonic numbers ($l_\mathrm{max}$).
Just as larger $N_\mathrm{pix}$ allow for more model freedom, larger $l_\mathrm{max}$ allow the spherical harmonic decomposition to represent more complex distributions over the sky.
Fig.~\ref{fig:hierarchical spherical harmonic maps} shows maps constructed with this Exponentiated Spherical Harmonic model for $l_\mathrm{max} = 1$, $2$, and $3$.

As a rule of thumb, constraints on low-$l$ coefficients weaken as $l_\mathrm{max}$ increases.
However, we consistently find that the $l=1$ (dipole) coefficients are constrained to be rather small, consistent with the Rotated Hemisphere and Simple Dipole models.
Constraints on higher harmonics are weaker, but they are also all constrained to be relatively small.
Fig.~\ref{fig:ESH corner plots} shows the prior and posterior for individual $b_l^m$ when $l_\mathrm{max}=2$.
We again eschew \BayesIsoAni~for this model because of ambiguity in the choices for the prior bounds on the $\{b_l^m\}$.
Indeed, because the posterior is consistent with isotropy, we can make \BayesIsoAni~as large as we like by simply increasing the extent of the prior.

When $l_\mathrm{max} \geq 2$, we begin to see structure appear across the sky on average \textit{a posteriori} (Fig.~\ref{fig:hierarchical spherical harmonic maps}).
This resembles the structure observed with pixelized representation, and, like the pixelized representations, there are large fluctuations in the posterior that render the difference between the marginal means and isotropy statistically insignificant.
Another way to view this is to examine the power in each harmonic.
Fig.~\ref{fig:ESH corner plots} shows these distributions as well.
Indeed, the power allowed in each harmonic \textit{a posteriori} is larger for higher harmonics, but it is always much smaller than the prior, showing that the data favor isotropy.


\subsection{Gaussian Random Fields}
\label{sec:gaussian process}

\begin{figure}
    \centering
    \includegraphics[hsmash=c, width=1.0\columnwidth]{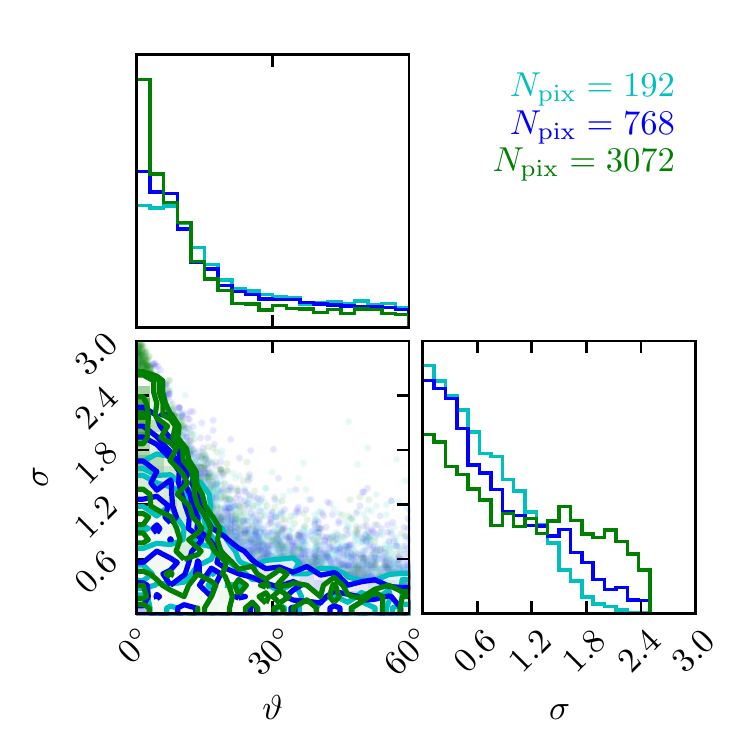}
    \caption{
        Posterior probability for GRF parameters ($\sigma$: marginal uncertainty; $\vartheta$: correlation length) assuming independent, uniform priors for each (see Table~\ref{tab:fitted population models}) when we use Healpix decompositions with (\emph{light blue}) 192, (\emph{dark blue}) 768, and (\emph{green}) 3072 pixels.
        The distributions do not depend strongly on the number of pixels used.
    }
    \label{fig:GP corner plots}
\end{figure}

\begin{figure*}
    \begin{tabular}{c p{0.47\textwidth} p{0.47\textwidth}}
        {\Large $\vartheta_\mathrm{min}$} & {\hspace{3.2cm} \Large mean} & {\hspace{2.7cm} \Large significance} \\
        {\large $20^\circ$}
          & \includegraphics[hsmash=r, align=c, width=0.47\textwidth]{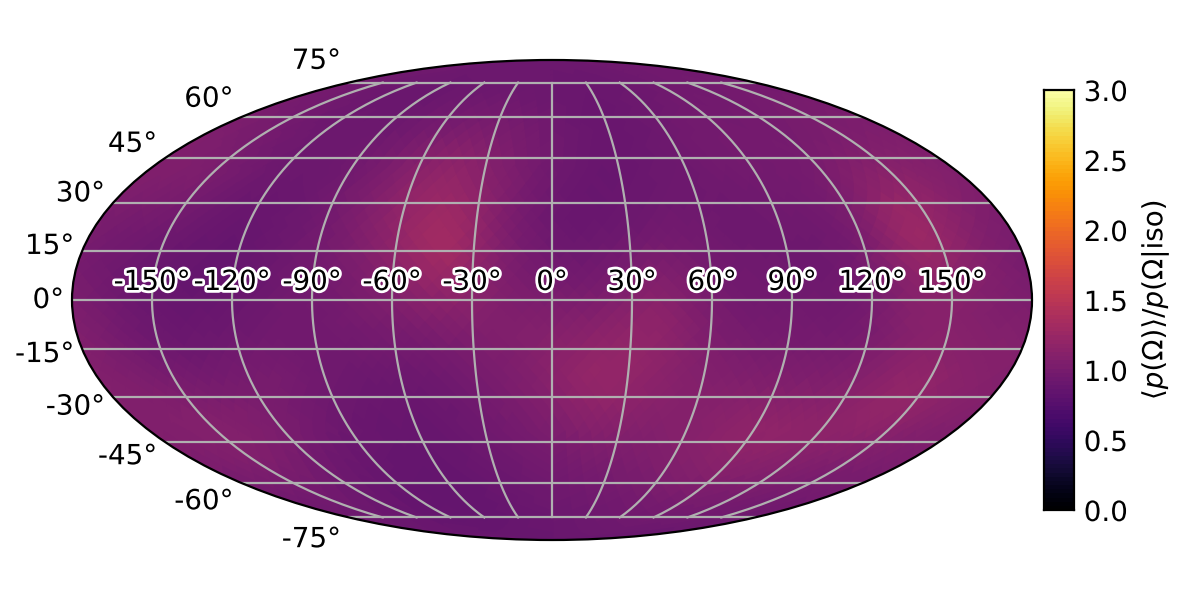}
          & \includegraphics[hsmash=r, align=c, width=0.47\textwidth]{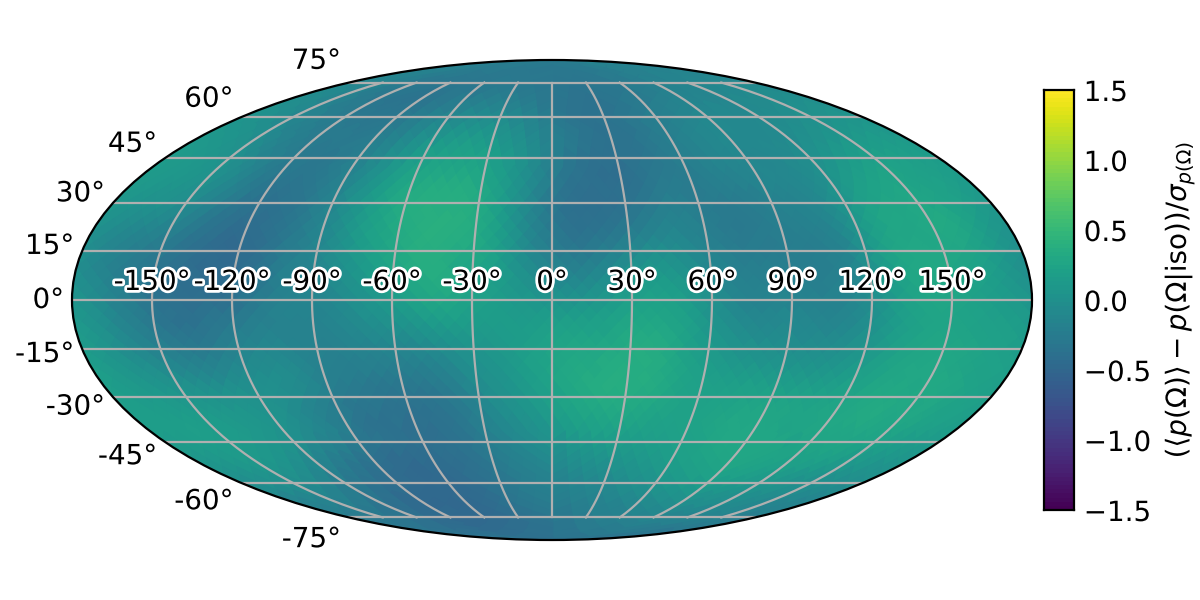} \\
        {\large $10^\circ$}
          & \includegraphics[hsmash=r, align=c, width=0.47\textwidth]{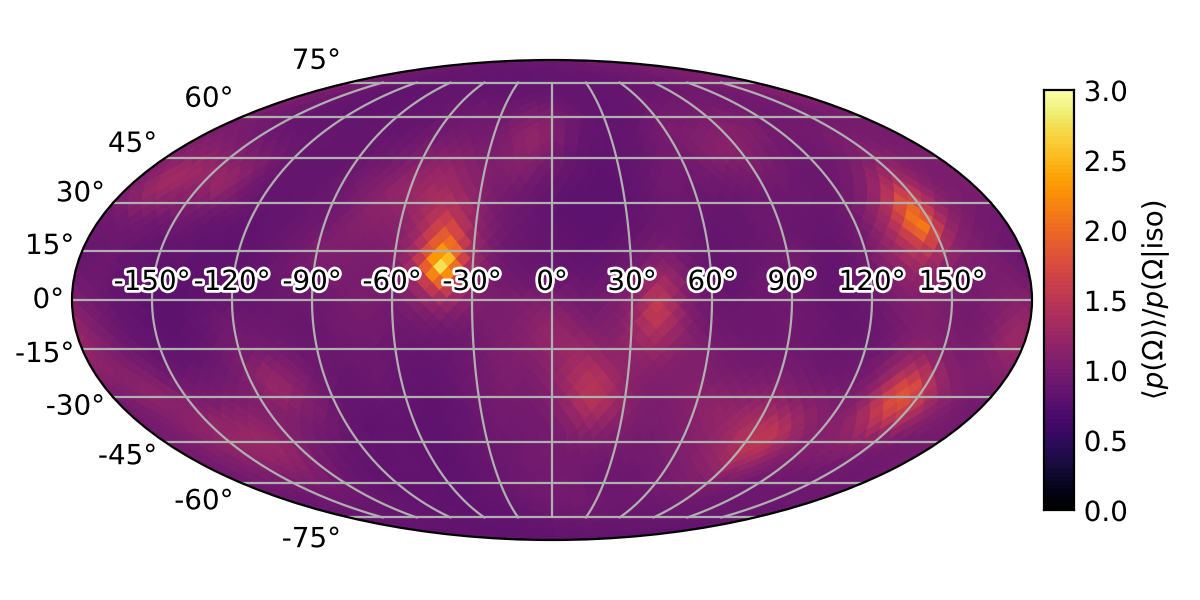}
          & \includegraphics[hsmash=r, align=c, width=0.47\textwidth]{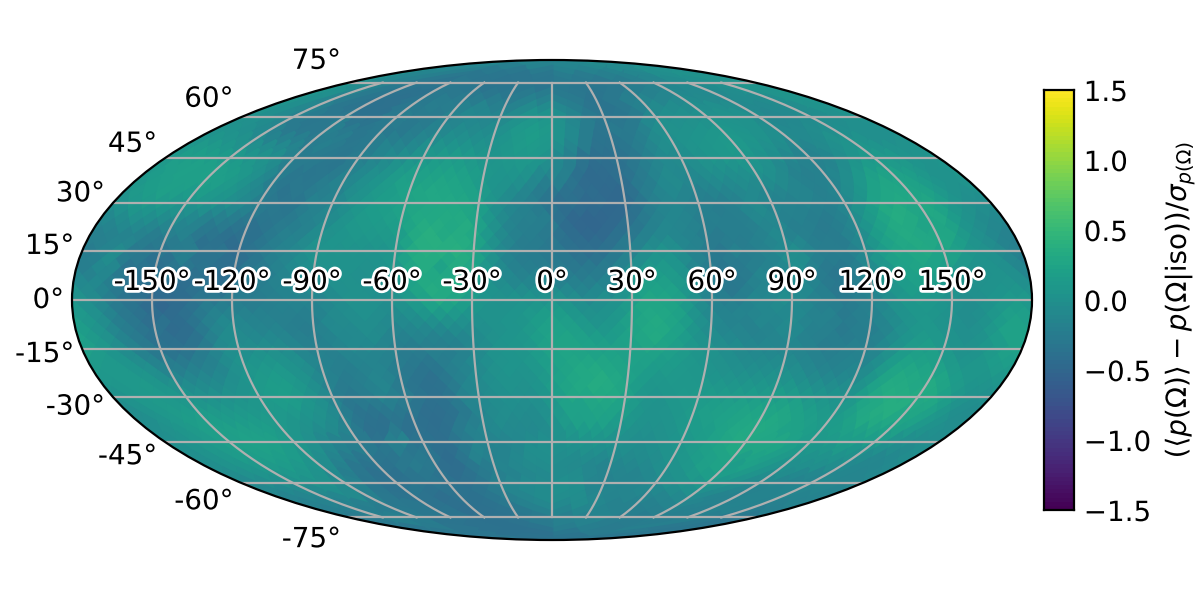} \\
        {\large $5^\circ$}
          & \includegraphics[hsmash=r, align=c, width=0.47\textwidth]{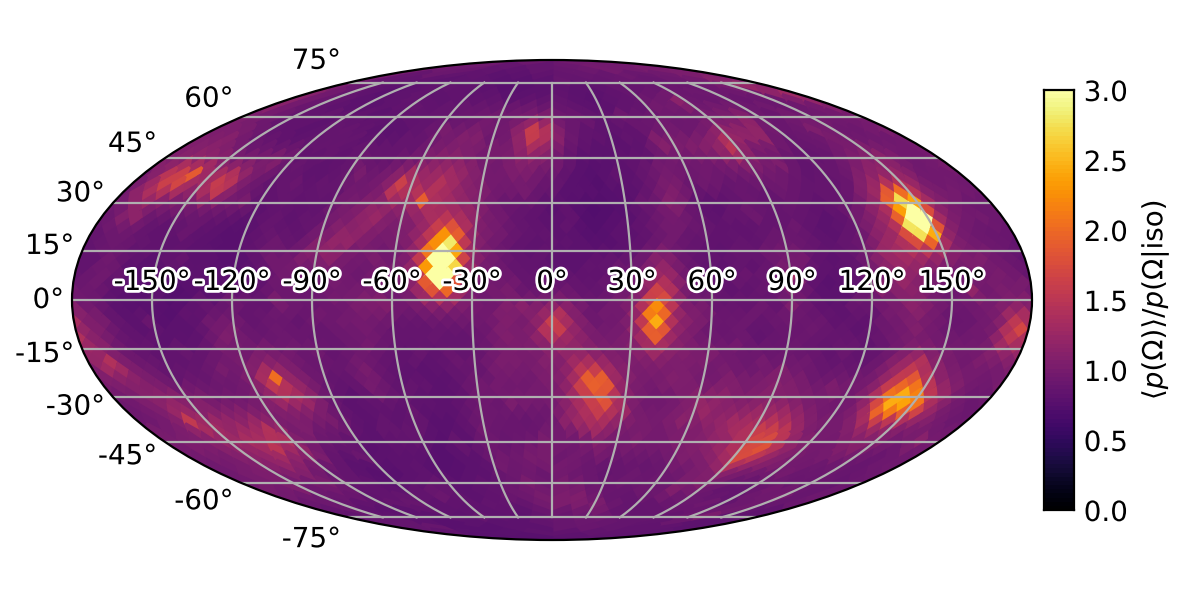}
          & \includegraphics[hsmash=r, align=c, width=0.47\textwidth]{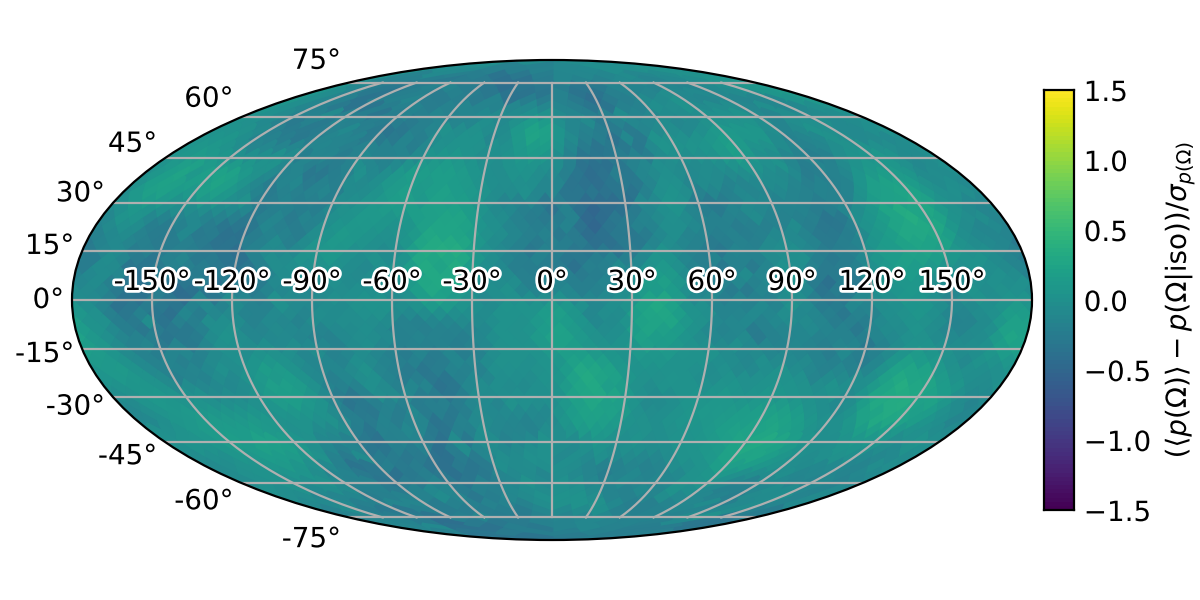} \\
    \end{tabular}
    \caption{
        Mollweide projections of the mean and significance of the rate density marginalized over correlation parameters within the GRF model with $N_\mathrm{pix}=3072$.
        We observe generally consistent results with the rest of our models, with a decrease in the difference between the posterior mean and isotropy as the minimum allowed correlation length ($\vartheta_\mathrm{min}$) increases.
    }
    \label{fig:GP maps}
\end{figure*}

We complete our survey of the impact of model choices by modeling the (logarithm of the) rate density as a Gaussian random field (GRF, also known as a Gaussian process~\cite{Rasmussen:2005}).

Specifically, we assume a Healpix pixelization scheme with many pixels but, importantly, do not assign independent priors to each pixel.
Instead, we assume the rate density in each pixel is correlated with neighboring pixels according to a covariance kernel
\begin{align}
    \mathrm{Cov}\left[p(\Omega_i), p(\Omega_j)\right]
        & \equiv \Xi_{ij} \nonumber \\
        & = \sigma_\mathrm{wn}^2 \delta_{ij} + \sigma^2 \exp\left( -\frac{(\Delta \theta_{ij})^2}{\vartheta^2} \right)
\end{align}
composed of a white noise component (uncorrelated variance within each pixel scaled by $\sigma_\mathrm{wn}^2$) and a squared exponential component (described by a marginal variance $\sigma^2$ and correlation length scale $\vartheta$) that correlates neighboring pixels based on the angular separation between their centers ($\Delta \theta_{ij}$).
We fix $\sigma_\mathrm{wn} = \sigma/10$, as we wish pixels to be significantly correlated and only include the white-noise variance for numerical stability.
While this choice was made primarily to guarantee numerical stability within Cholesky decompositions of (at times) ill-conditioned covariance matrices with large $\vartheta$, it also introduces a natural resolution scale at which $\sigma_\mathrm{wn}$ from many pixels tends to dominate the variance in $p(\Omega)$ over the sky.
For large $\vartheta$ (strong squared-exponential correlations), we expect $\sigma_\mathrm{wn}$ to contribute a significant fraction of the overall variability when $N_\mathrm{pix} \gtrsim 100$ if $\sigma_\mathrm{wn} = \sigma/10$.
However, for $\vartheta \sim 60^\circ$, this increases to $N_\mathrm{pix} \sim 1100$, and for $\vartheta \sim 10^\circ$ it increases to $N_\mathrm{pix} \sim 41,000$.
We therefore expect our results to not depend strongly on the choice $\sigma_\mathrm{wn} = \sigma/10$ given the range of $\vartheta$ included and the number of pixels used.
We confirmed this by also investigating $\sigma_\mathrm{wn} = \sigma/100$, $\sigma_\mathrm{wn} = \sigma/1000$, finding consistent results.

Just as our GRF model is related to our pixelized models with a different prior, it can also be expressed in terms of a spherical harmonic representation.
Specifically, the $b_l^m$ are independently, Normally distributed within a GRF prior, and their individual variances depend on the form of the covariance kernel (see, e.g., Ref.~\cite{Land:2015, Farr:2018}).
The GRF prior controls how the prior uncertainty in $b_l^m$ decreases as $l$ increases; the contribution of high-$l$ modes are limited and the resulting rate density is smooth.
Similarly, the same prior controls how quickly the rate density is allowed to vary from pixel to pixel.

The key advantages of the GRF approach are that it is straightforward to learn the correlation parameters at the same time we fit the data and that it does not depend strongly on how many pixels or harmonics are included.
That is, we need not make strong (and poorly understood) prior choices about how many pixels or $b_l^m$ to include.
The data itself will determine which correlations are preferred.
Fig.~\ref{fig:GP corner plots} shows the resulting posteriors for the GRF parameters.
The data prefer small $\sigma$ when \result{$\vartheta \gtrsim 15^\circ$}, and are consistent with the isotropic limit ($\sigma \rightarrow 0$) for all $\vartheta$.

The data do not strongly constrain the correlation length, although the constraints on $\sigma$ are more stringent for longer $\vartheta$.
In other words, if neighboring pixels are significantly correlated, then the data are less consistent with large fluctuations in the rate density across the sky.
This is similar to the fact that we are able to more tightly constrain the low-$l$ coefficients in the spherical harmonic model compared to high-$l$ coefficients.

Finally, Fig.~\ref{fig:GP maps} shows Mollweide projections analogous to Figs.~\ref{fig:hierarchical pixel maps} and~\ref{fig:hierarchical spherical harmonic maps} when we impose several lower limits on the correlation length ($\vartheta \geq \vartheta_\mathrm{min}$).
The key differences between Figs.~\ref{fig:GP maps} and~\ref{fig:hierarchical pixel maps} are that the most extreme excursions of the posterior's mean are smaller for the GRF models due to the correlations between neighboring pixels from the prior.
We also note that, correspondingly, the fluctuations that do occur in the GRF model appear even less significant.

Nevertheless, we see features reminiscent of individual events within the posterior process's mean when $\vartheta_\mathrm{min}$ is small.
This is not unexpected, as the posterior mean is related to the sum of individual events' posteriors when the anisotropies are small.
Appendix~\ref{sec:perturbation} describes this in more detail.


\subsection{Quantifying Constraints on Anisotropy}
\label{sec:quantification}

As we have discussed, it can be difficult to interpret Bayes factors within our models because of ambiguity in the choices of prior ranges.
For example, \BayesIsoAni~can be made as large or as small as one would like within the Exponentiated Spherical Harmonic model by changing the prior ranges allowed for each $b_l^m$.
Therefore, we propose a more direct measure of the extent of anisotropies: the variance of the rate density across the sky,
\begin{equation}
    \sigma^2_{p(\Omega)} \equiv \frac{1}{4\pi}\int d\Omega \left(p(\Omega) - \frac{1}{4\pi}\right)^2.
\end{equation}
This is closely related to the GRF model's $\sigma$ parameter.
Isotropy corresponds to the limit $\sigma_{p(\Omega)}=0$.

In both the Healpix and Exponentiated Spherical Harmonic models, we find that the posterior supports larger amounts of variability as we increase $N_\mathrm{pix}$ or $l_\mathrm{max}$.
That is, the data does not strongly constrain rapid oscillations within the distribution over the sky, and the variability in the inferred distribution is dominated by the prior induced over these high-$l$ modes.
In particular, both the Healpix and Exponentiated Spherical Harmonic \textit{a priori} have vanishingly small support for small variance over the sky.
This carries over to the posterior, and any upper limit on the variance will strongly depend on the prior.

The GRF model, on the other hand, naturally avoids this issue by simultaneously sampling over both the correlation parameters and the distribution over the sky.
Because the GRF model contains support for all correlation lengths ($\vartheta$), as opposed to a fixed choice of $N_\mathrm{pix}$ or $l_\mathrm{max}$, we do not observe vanishing support for small variances.
Indeed, we obtain a consistent upper limit for all the variability measures within the GRF model when $\vartheta_\mathrm{min} \gtrsim 10^\circ$ regardless of the number of pixels used.
The distribution of merging binaries produces \result{$\sigma_{p(\Omega)} \lesssim 16\%$} of the isotropic rate at \result{90\%} credibility when \result{$\vartheta \geq 10^\circ$}.
That is, the rate density fluctuates by \result{$\lesssim 16\%$} across the sky.
When \result{$\vartheta \geq 20^\circ$}, this is reduced to \result{$\lesssim 3.5\%$}.


\section{Discussion}
\label{sec:discussion}

Using \result{63} confidently detected GW sources from the LVK's third observing run, we constrained the distribution of merging binaries across the celestial sphere.
Our constraints improve upon previous work that used the 11 events from GWTC-1, finding constraints on anisotropies ($\BayesIsoAni$) that are a factor of a few stronger.
However, because of ambiguity in the interpretation of $\BayesIsoAni$ due to arbitrary prior choices, we instead quantify constraints on anisotropies with a direct measure of how much $p(\Omega)$ varies over the sky.
Modeling anisotropies as a Gaussian random field, we constrain the fluctuations to be \result{$\leq 16\%$} if the field is correlated with a length scale \result{$\geq 10^\circ$}.
That is, the distribution of merging binaries varies by \result{$\lesssim 16\%$} of the isotropic rate at \result{90\% credibility}.

We also observe consistently hot pixels within all of our models of $p(\Omega)$.
While none of these are statistically significant, the brightest pixel is in the direction of the constellation Equuleus.
Our hot pixels tend to fall near the equator on average, and they do not match the hot spots found in previous work with GWTC-1~\cite{Stiskalek:2020, Payne:2020}.
This is consistent with the expectation that the distribution is isotropic, and we are in effect ``fitting noise'' when we construct maps of the mean $p(\Omega)$.

Nonetheless, it may be interesting to extend this work in the future.
In particular, we have only studied isotropy, and the cosmological principle also predicts homogeneity.
It may be of interest to not only consider clustering in three spatial dimensions,\footnote{The rate of GW sources almost certainly evolves over cosmic time~\cite{Fishbach:2021a, GWTC-3-RnP}. This means we will need to consider the effect of lookback time when considering inhomogeneities in the spatial distribution.} but also correlations between intrinsic source properties (masses, spins, etc) and extrinsic properties (location, orientation, etc).
Furthermore, correlating anisotropies and/or inhomogeneities in GW catalogs with other catalogs will be of increasing importance.
Under the assumption that GW events only come from galaxies, current galaxy catalogs have been used to provide a weak constraint on the Hubble parameter~\cite{GWTC-3-Cosmo}.
With larger GW catalogs, it may be possible to directly test the assumption that GWs only come from galaxies, or to determine which types of galaxies are more likely to host GW sources~\cite{Chen:2016, Singer:2016}.
More generally, this may constrain cosmic structure, and clustering scales in GW catalogs could connect to the mass scales of typical host galaxies~\cite{Fishbach:2021}.
Of course, there may also be synergies from connecting the distribution of nearby, well-resolved systems with the stochastic GW background from the myriad more distance sources~\cite{Callister:2020}.

The LVK also searches for unmodeled ``burst'' events.
If such events are detected and no obvious source presents itself, determining whether the sources are isotropically distributed or correlated with local structure can inform the distance to the events and therefore their energy scale.
Demonstrating the ability to perform such a measurement and determining the size of the catalog needed to rule out isotropy\footnote{Note that even a single event may rule out the correlation with local structure, and therefore ruling out isotropy when the events do correlate with local structure is likely to be of greater interest.} may be worth establishing before such events are detected.

For all these reasons and more, it is worth studying in greater detail which properties of individual events make them informative and over what angular scales.
Indeed, as the size of the catalog grows, we may wish to know whether the isotropy constraints will always be dominated by the best-localized events or whether the legion of poorly localized events will eventually dominate through sheer force of numbers.
We have also shown that most of the information about (an)isotropy is carried by the best-localized events.
As searches become more sensitive to quieter events and/or events detected in a single interferometer, we may expect the rate at which isotropy constraints improve to slow as a larger fraction of GW catalogs will have large, uninformative localizations.\footnote{Appendix~\ref{sec:perturbation} introduces an eigenvalue analysis of which anisotropies can be best constrained with current data. The magnitude of the eigenvalues rapidly decreases, suggesting that it may take many more events to precisely constrain high-$l$ modes compared to low-$l$ modes.} 

Here we used \result{63} confident BNS, NSBH, and BBH detections from O3 to place limits on the anisotropy of gravitational wave events on the sky.
We do not find any evidence for anisotropy.
On the contrary, using flexible and data-driven models we bound the variability of the gravitational wave merger rate over the sky to \result{$\lesssim 16\%$} on scales larger than \result{$10^\circ$}, or \result{$\lesssim 3.5\%$} on scales larger than \result{$20^\circ$}.
As the GW catalog continues to grow, implementation of our methodology will lead to more definitive measurements. Understanding the homogeneity and isotropy of GW sources is an important astrophysical and cosmological probe of this newly discovered population.
There are still many unknowns about the distribution of merging binaries, and future catalogs will continue to provide surprises if we continue to look for them.


\acknowledgments

R.E. thanks the Canadian Institute for Advanced Research (CIFAR) for support.
Research at Perimeter Institute is supported in part by the Government of Canada through the Department of Innovation, Science and Economic Development Canada and by the Province of Ontario through the Ministry of Colleges and Universities.
M.F. is supported by NASA through NASA Hubble Fellowship grant HST-HF2-51455.001-A awarded by the Space Telescope Science Institute, which is operated by the Association of Universities for Research in Astronomy, Incorporated, under NASA contract NAS5-26555.
W.M.F. is partially supported by the Simons Foundation.
D.E.H. is supported by NSF grants PHY-2006645 and PHY-2110507, as well as the Kavli Institute for Cosmological Physics through an endowment from the Kavli Foundation and its founder Fred Kavli. D.E.H. also gratefully acknowledges support from the Marion and Stuart Rice Award.
E. K. is supported by the National Science Foundation (NSF) through award PHY-1764464 to the LIGO Laboratory. 

This material is based upon work supported by NSF's LIGO Laboratory which is a major facility fully funded by the National Science Foundation.
This work would not have been possible without the following software: \texttt{numpy}~\cite{numpy}, \texttt{scipy}~\cite{scipy}, \texttt{matplotlib}~\cite{matplotlib}, \texttt{jax}~\cite{jax}, and \texttt{numpyro}~\cite{pyro, numpyro}.


\bibliography{refs}

\begin{thebibliography}{62}%
\makeatletter
\providecommand \@ifxundefined [1]{%
 \@ifx{#1\undefined}
}%
\providecommand \@ifnum [1]{%
 \ifnum #1\expandafter \@firstoftwo
 \else \expandafter \@secondoftwo
 \fi
}%
\providecommand \@ifx [1]{%
 \ifx #1\expandafter \@firstoftwo
 \else \expandafter \@secondoftwo
 \fi
}%
\providecommand \natexlab [1]{#1}%
\providecommand \enquote  [1]{``#1''}%
\providecommand \bibnamefont  [1]{#1}%
\providecommand \bibfnamefont [1]{#1}%
\providecommand \citenamefont [1]{#1}%
\providecommand \href@noop [0]{\@secondoftwo}%
\providecommand \href [0]{\begingroup \@sanitize@url \@href}%
\providecommand \@href[1]{\@@startlink{#1}\@@href}%
\providecommand \@@href[1]{\endgroup#1\@@endlink}%
\providecommand \@sanitize@url [0]{\catcode `\\12\catcode `\$12\catcode
  `\&12\catcode `\#12\catcode `\^12\catcode `\_12\catcode `\%12\relax}%
\providecommand \@@startlink[1]{}%
\providecommand \@@endlink[0]{}%
\providecommand \url  [0]{\begingroup\@sanitize@url \@url }%
\providecommand \@url [1]{\endgroup\@href {#1}{\urlprefix }}%
\providecommand \urlprefix  [0]{URL }%
\providecommand \Eprint [0]{\href }%
\providecommand \doibase [0]{https://doi.org/}%
\providecommand \selectlanguage [0]{\@gobble}%
\providecommand \bibinfo  [0]{\@secondoftwo}%
\providecommand \bibfield  [0]{\@secondoftwo}%
\providecommand \translation [1]{[#1]}%
\providecommand \BibitemOpen [0]{}%
\providecommand \bibitemStop [0]{}%
\providecommand \bibitemNoStop [0]{.\EOS\space}%
\providecommand \EOS [0]{\spacefactor3000\relax}%
\providecommand \BibitemShut  [1]{\csname bibitem#1\endcsname}%
\let\auto@bib@innerbib\@empty
\bibitem [{\citenamefont {{Abbott}}\ \emph
  {et~al.}(2021{\natexlab{a}})\citenamefont {{Abbott}} \emph
  {et~al.}}]{GWTC-3-RnP}%
  \BibitemOpen
  \bibfield  {author} {\bibinfo {author} {\bibfnamefont {R.}~\bibnamefont
  {{Abbott}}} \emph {et~al.},\ }\bibfield  {title} {\bibinfo {title} {{The
  population of merging compact binaries inferred using gravitational waves
  through GWTC-3}},\ }\href@noop {} {\bibfield  {journal} {\bibinfo  {journal}
  {arXiv e-prints}\ ,\ \bibinfo {eid} {arXiv:2111.03634}} (\bibinfo {year}
  {2021}{\natexlab{a}})},\ \Eprint {https://arxiv.org/abs/2111.03634}
  {arXiv:2111.03634 [astro-ph.HE]} \BibitemShut {NoStop}%
\bibitem [{\citenamefont {{Abbott}}\ \emph
  {et~al.}(2021{\natexlab{b}})\citenamefont {{Abbott}} \emph
  {et~al.}}]{GWTC-3-Cosmo}%
  \BibitemOpen
  \bibfield  {author} {\bibinfo {author} {\bibfnamefont {R.}~\bibnamefont
  {{Abbott}}} \emph {et~al.},\ }\bibfield  {title} {\bibinfo {title}
  {{Constraints on the cosmic expansion history from GWTC-3}},\ }\href@noop {}
  {\bibfield  {journal} {\bibinfo  {journal} {arXiv e-prints}\ ,\ \bibinfo
  {eid} {arXiv:2111.03604}} (\bibinfo {year} {2021}{\natexlab{b}})},\ \Eprint
  {https://arxiv.org/abs/2111.03604} {arXiv:2111.03604 [astro-ph.CO]}
  \BibitemShut {NoStop}%
\bibitem [{\citenamefont {Mukherjee}\ \emph {et~al.}(2021)\citenamefont
  {Mukherjee}, \citenamefont {Wandelt}, \citenamefont {Nissanke},\ and\
  \citenamefont {Silvestri}}]{Mukherjee:2021}%
  \BibitemOpen
  \bibfield  {author} {\bibinfo {author} {\bibfnamefont {S.}~\bibnamefont
  {Mukherjee}}, \bibinfo {author} {\bibfnamefont {B.~D.}\ \bibnamefont
  {Wandelt}}, \bibinfo {author} {\bibfnamefont {S.~M.}\ \bibnamefont
  {Nissanke}},\ and\ \bibinfo {author} {\bibfnamefont {A.}~\bibnamefont
  {Silvestri}},\ }\bibfield  {title} {\bibinfo {title} {Accurate precision
  cosmology with redshift unknown gravitational wave sources},\ }\href
  {https://doi.org/10.1103/PhysRevD.103.043520} {\bibfield  {journal} {\bibinfo
   {journal} {Phys. Rev. D}\ }\textbf {\bibinfo {volume} {103}},\ \bibinfo
  {pages} {043520} (\bibinfo {year} {2021})}\BibitemShut {NoStop}%
\bibitem [{\citenamefont {{Vitale}}\ \emph {et~al.}(2022)\citenamefont
  {{Vitale}}, \citenamefont {{Biscoveanu}},\ and\ \citenamefont
  {{Talbot}}}]{Vitale:2022}%
  \BibitemOpen
  \bibfield  {author} {\bibinfo {author} {\bibfnamefont {S.}~\bibnamefont
  {{Vitale}}}, \bibinfo {author} {\bibfnamefont {S.}~\bibnamefont
  {{Biscoveanu}}},\ and\ \bibinfo {author} {\bibfnamefont {C.}~\bibnamefont
  {{Talbot}}},\ }\bibfield  {title} {\bibinfo {title} {{The orientations of the
  binary black holes in GWTC-3}},\ }\href@noop {} {\bibfield  {journal}
  {\bibinfo  {journal} {arXiv e-prints}\ ,\ \bibinfo {eid} {arXiv:2204.00968}}
  (\bibinfo {year} {2022})},\ \Eprint {https://arxiv.org/abs/2204.00968}
  {arXiv:2204.00968 [gr-qc]} \BibitemShut {NoStop}%
\bibitem [{\citenamefont {Libanore}\ \emph {et~al.}(2021)\citenamefont
  {Libanore}, \citenamefont {Artale}, \citenamefont {Karagiannis},
  \citenamefont {Liguori}, \citenamefont {Bartolo}, \citenamefont {Bouffanais},
  \citenamefont {Giacobbo}, \citenamefont {Mapelli},\ and\ \citenamefont
  {Matarrese}}]{Libanore:2021}%
  \BibitemOpen
  \bibfield  {author} {\bibinfo {author} {\bibfnamefont {S.}~\bibnamefont
  {Libanore}}, \bibinfo {author} {\bibfnamefont {M.~C.}\ \bibnamefont
  {Artale}}, \bibinfo {author} {\bibfnamefont {D.}~\bibnamefont {Karagiannis}},
  \bibinfo {author} {\bibfnamefont {M.}~\bibnamefont {Liguori}}, \bibinfo
  {author} {\bibfnamefont {N.}~\bibnamefont {Bartolo}}, \bibinfo {author}
  {\bibfnamefont {Y.}~\bibnamefont {Bouffanais}}, \bibinfo {author}
  {\bibfnamefont {N.}~\bibnamefont {Giacobbo}}, \bibinfo {author}
  {\bibfnamefont {M.}~\bibnamefont {Mapelli}},\ and\ \bibinfo {author}
  {\bibfnamefont {S.}~\bibnamefont {Matarrese}},\ }\bibfield  {title} {\bibinfo
  {title} {Gravitational wave mergers as tracers of large scale structures},\
  }\href {https://doi.org/10.1088/1475-7516/2021/02/035} {\bibfield  {journal}
  {\bibinfo  {journal} {Journal of Cosmology and Astroparticle Physics}\
  }\textbf {\bibinfo {volume} {2021}}\bibinfo  {number} { (02)},\ \bibinfo
  {pages} {035}}\BibitemShut {NoStop}%
\bibitem [{\citenamefont {Shao}\ \emph {et~al.}(2022)\citenamefont {Shao},
  \citenamefont {Cao}, \citenamefont {Fan},\ and\ \citenamefont
  {Wu}}]{Shao:2022}%
  \BibitemOpen
\bibfield  {number} {  }\bibfield  {author} {\bibinfo {author} {\bibfnamefont
  {X.}~\bibnamefont {Shao}}, \bibinfo {author} {\bibfnamefont {Z.}~\bibnamefont
  {Cao}}, \bibinfo {author} {\bibfnamefont {X.}~\bibnamefont {Fan}},\ and\
  \bibinfo {author} {\bibfnamefont {S.}~\bibnamefont {Wu}},\ }\bibfield
  {title} {\bibinfo {title} {Probing the large-scale structure of the universe
  through gravitational wave observations},\ }\href
  {https://doi.org/10.1088/1674-4527/ac32b4} {\bibfield  {journal} {\bibinfo
  {journal} {Research in Astronomy and Astrophysics}\ }\textbf {\bibinfo
  {volume} {22}},\ \bibinfo {pages} {015006} (\bibinfo {year}
  {2022})}\BibitemShut {NoStop}%
\bibitem [{\citenamefont {{Chen}}\ and\ \citenamefont
  {{Holz}}(2016)}]{Chen:2016}%
  \BibitemOpen
  \bibfield  {author} {\bibinfo {author} {\bibfnamefont {H.-Y.}\ \bibnamefont
  {{Chen}}}\ and\ \bibinfo {author} {\bibfnamefont {D.~E.}\ \bibnamefont
  {{Holz}}},\ }\bibfield  {title} {\bibinfo {title} {{Finding the One:
  Identifying the Host Galaxies of Gravitational-Wave Sources}},\ }\href@noop
  {} {\bibfield  {journal} {\bibinfo  {journal} {arXiv e-prints}\ ,\ \bibinfo
  {eid} {arXiv:1612.01471}} (\bibinfo {year} {2016})},\ \Eprint
  {https://arxiv.org/abs/1612.01471} {arXiv:1612.01471 [astro-ph.HE]}
  \BibitemShut {NoStop}%
\bibitem [{\citenamefont {Singer}\ \emph {et~al.}(2016)\citenamefont {Singer},
  \citenamefont {Chen}, \citenamefont {Holz}, \citenamefont {Farr},
  \citenamefont {Price}, \citenamefont {Raymond}, \citenamefont {Cenko},
  \citenamefont {Gehrels}, \citenamefont {Cannizzo}, \citenamefont {Kasliwal},
  \citenamefont {Nissanke}, \citenamefont {Coughlin}, \citenamefont {Farr},
  \citenamefont {Urban}, \citenamefont {Vitale}, \citenamefont {Veitch},
  \citenamefont {Graff}, \citenamefont {Berry}, \citenamefont {Mohapatra},\
  and\ \citenamefont {Mandel}}]{Singer:2016}%
  \BibitemOpen
  \bibfield  {author} {\bibinfo {author} {\bibfnamefont {L.~P.}\ \bibnamefont
  {Singer}}, \bibinfo {author} {\bibfnamefont {H.-Y.}\ \bibnamefont {Chen}},
  \bibinfo {author} {\bibfnamefont {D.~E.}\ \bibnamefont {Holz}}, \bibinfo
  {author} {\bibfnamefont {W.~M.}\ \bibnamefont {Farr}}, \bibinfo {author}
  {\bibfnamefont {L.~R.}\ \bibnamefont {Price}}, \bibinfo {author}
  {\bibfnamefont {V.}~\bibnamefont {Raymond}}, \bibinfo {author} {\bibfnamefont
  {S.~B.}\ \bibnamefont {Cenko}}, \bibinfo {author} {\bibfnamefont
  {N.}~\bibnamefont {Gehrels}}, \bibinfo {author} {\bibfnamefont
  {J.}~\bibnamefont {Cannizzo}}, \bibinfo {author} {\bibfnamefont {M.~M.}\
  \bibnamefont {Kasliwal}}, \bibinfo {author} {\bibfnamefont {S.}~\bibnamefont
  {Nissanke}}, \bibinfo {author} {\bibfnamefont {M.}~\bibnamefont {Coughlin}},
  \bibinfo {author} {\bibfnamefont {B.}~\bibnamefont {Farr}}, \bibinfo {author}
  {\bibfnamefont {A.~L.}\ \bibnamefont {Urban}}, \bibinfo {author}
  {\bibfnamefont {S.}~\bibnamefont {Vitale}}, \bibinfo {author} {\bibfnamefont
  {J.}~\bibnamefont {Veitch}}, \bibinfo {author} {\bibfnamefont
  {P.}~\bibnamefont {Graff}}, \bibinfo {author} {\bibfnamefont {C.~P.~L.}\
  \bibnamefont {Berry}}, \bibinfo {author} {\bibfnamefont {S.}~\bibnamefont
  {Mohapatra}},\ and\ \bibinfo {author} {\bibfnamefont {I.}~\bibnamefont
  {Mandel}},\ }\bibfield  {title} {\bibinfo {title} {{GOING} {THE} {DISTANCE}:
  {MAPPING} {HOST} {GALAXIES} {OF} {LIGO} {AND} {VIRGO} {SOURCES} {IN} {THREE}
  {DIMENSIONS} {USING} {LOCAL} {COSMOGRAPHY} {AND} {TARGETED} {FOLLOW}-{UP}},\
  }\href {https://doi.org/10.3847/2041-8205/829/1/l15} {\bibfield  {journal}
  {\bibinfo  {journal} {The Astrophysical Journal}\ }\textbf {\bibinfo {volume}
  {829}},\ \bibinfo {pages} {L15} (\bibinfo {year} {2016})}\BibitemShut
  {NoStop}%
\bibitem [{\citenamefont {{Fishbach}}\ and\ \citenamefont
  {{Kalogera}}(2021)}]{Fishbach:2021}%
  \BibitemOpen
  \bibfield  {author} {\bibinfo {author} {\bibfnamefont {M.}~\bibnamefont
  {{Fishbach}}}\ and\ \bibinfo {author} {\bibfnamefont {V.}~\bibnamefont
  {{Kalogera}}},\ }\bibfield  {title} {\bibinfo {title} {{The Time Delay
  Distribution and Formation Metallicity of LIGO-Virgos Binary Black Holes}},\
  }\href {https://doi.org/10.3847/2041-8213/ac05c4} {\bibfield  {journal}
  {\bibinfo  {journal} {\apj Lett.}\ }\textbf {\bibinfo {volume} {914}},\
  \bibinfo {eid} {L30} (\bibinfo {year} {2021})},\ \Eprint
  {https://arxiv.org/abs/2105.06491} {arXiv:2105.06491 [astro-ph.HE]}
  \BibitemShut {NoStop}%
\bibitem [{\citenamefont {{Zevin}}\ \emph {et~al.}(2022)\citenamefont
  {{Zevin}}, \citenamefont {{Nugent}}, \citenamefont {{Adhikari}},
  \citenamefont {{Fong}}, \citenamefont {{Holz}},\ and\ \citenamefont
  {{Kelley}}}]{Zevin:2022}%
  \BibitemOpen
  \bibfield  {author} {\bibinfo {author} {\bibfnamefont {M.}~\bibnamefont
  {{Zevin}}}, \bibinfo {author} {\bibfnamefont {A.~E.}\ \bibnamefont
  {{Nugent}}}, \bibinfo {author} {\bibfnamefont {S.}~\bibnamefont
  {{Adhikari}}}, \bibinfo {author} {\bibfnamefont {W.-f.}\ \bibnamefont
  {{Fong}}}, \bibinfo {author} {\bibfnamefont {D.~E.}\ \bibnamefont {{Holz}}},\
  and\ \bibinfo {author} {\bibfnamefont {L.~Z.}\ \bibnamefont {{Kelley}}},\
  }\bibfield  {title} {\bibinfo {title} {{Observational Inference on the Delay
  Time Distribution of Short Gamma-ray Bursts}},\ }\href@noop {} {\bibfield
  {journal} {\bibinfo  {journal} {arXiv e-prints}\ ,\ \bibinfo {eid}
  {arXiv:2206.02814}} (\bibinfo {year} {2022})},\ \Eprint
  {https://arxiv.org/abs/2206.02814} {arXiv:2206.02814 [astro-ph.HE]}
  \BibitemShut {NoStop}%
\bibitem [{\citenamefont {Aasi}\ \emph {et~al.}(2015)\citenamefont {Aasi} \emph
  {et~al.}}]{LIGO}%
  \BibitemOpen
  \bibfield  {author} {\bibinfo {author} {\bibfnamefont {J.}~\bibnamefont
  {Aasi}} \emph {et~al.} (\bibinfo {collaboration} {LIGO Scientific}),\
  }\bibfield  {title} {\bibinfo {title} {{Advanced LIGO}},\ }\href
  {https://doi.org/10.1088/0264-9381/32/7/074001} {\bibfield  {journal}
  {\bibinfo  {journal} {Class. Quant. Grav.}\ }\textbf {\bibinfo {volume}
  {32}},\ \bibinfo {pages} {074001} (\bibinfo {year} {2015})},\ \Eprint
  {https://arxiv.org/abs/1411.4547} {arXiv:1411.4547 [gr-qc]} \BibitemShut
  {NoStop}%
\bibitem [{\citenamefont {Acernese}\ \emph {et~al.}(2015)\citenamefont
  {Acernese} \emph {et~al.}}]{Virgo}%
  \BibitemOpen
  \bibfield  {author} {\bibinfo {author} {\bibfnamefont {F.}~\bibnamefont
  {Acernese}} \emph {et~al.} (\bibinfo {collaboration} {VIRGO}),\ }\bibfield
  {title} {\bibinfo {title} {{Advanced Virgo: a second-generation
  interferometric gravitational wave detector}},\ }\href
  {https://doi.org/10.1088/0264-9381/32/2/024001} {\bibfield  {journal}
  {\bibinfo  {journal} {Class. Quant. Grav.}\ }\textbf {\bibinfo {volume}
  {32}},\ \bibinfo {pages} {024001} (\bibinfo {year} {2015})},\ \Eprint
  {https://arxiv.org/abs/1408.3978} {arXiv:1408.3978 [gr-qc]} \BibitemShut
  {NoStop}%
\bibitem [{\citenamefont {Abbott}\ \emph
  {et~al.}(2021{\natexlab{a}})\citenamefont {Abbott} \emph
  {et~al.}}]{O3-burst}%
  \BibitemOpen
  \bibfield  {author} {\bibinfo {author} {\bibfnamefont {R.}~\bibnamefont
  {Abbott}} \emph {et~al.} (\bibinfo {collaboration} {The LIGO Scientific
  Collaboration, the Virgo Collaboration, and the KAGRA Collaboration}),\
  }\bibfield  {title} {\bibinfo {title} {All-sky search for short
  gravitational-wave bursts in the third advanced ligo and advanced virgo
  run},\ }\href {https://doi.org/10.1103/PhysRevD.104.122004} {\bibfield
  {journal} {\bibinfo  {journal} {Phys. Rev. D}\ }\textbf {\bibinfo {volume}
  {104}},\ \bibinfo {pages} {122004} (\bibinfo {year}
  {2021}{\natexlab{a}})}\BibitemShut {NoStop}%
\bibitem [{\citenamefont {Andrade}\ \emph {et~al.}(2019)\citenamefont
  {Andrade}, \citenamefont {Bengaly}, \citenamefont {Alcaniz},\ and\
  \citenamefont {Capozziello}}]{Andrade:2019}%
  \BibitemOpen
  \bibfield  {author} {\bibinfo {author} {\bibfnamefont {U.}~\bibnamefont
  {Andrade}}, \bibinfo {author} {\bibfnamefont {C.~A.~P.}\ \bibnamefont
  {Bengaly}}, \bibinfo {author} {\bibfnamefont {J.~S.}\ \bibnamefont
  {Alcaniz}},\ and\ \bibinfo {author} {\bibfnamefont {S.}~\bibnamefont
  {Capozziello}},\ }\bibfield  {title} {\bibinfo {title} {{Revisiting the
  statistical isotropy of GRB sky distribution}},\ }\href
  {https://doi.org/10.1093/mnras/stz2754} {\bibfield  {journal} {\bibinfo
  {journal} {Monthly Notices of the Royal Astronomical Society}\ }\textbf
  {\bibinfo {volume} {490}},\ \bibinfo {pages} {4481} (\bibinfo {year}
  {2019})},\ \Eprint
  {https://arxiv.org/abs/https://academic.oup.com/mnras/article-pdf/490/4/4481/30459240/stz2754.pdf}
  {https://academic.oup.com/mnras/article-pdf/490/4/4481/30459240/stz2754.pdf}
  \BibitemShut {NoStop}%
\bibitem [{\citenamefont {Dent}\ \emph {et~al.}(2021)\citenamefont {Dent},
  \citenamefont {Gabella}, \citenamefont {Holley-Bockelmann},\ and\
  \citenamefont {Kephart}}]{PhysRevD.104.044030}%
  \BibitemOpen
  \bibfield  {author} {\bibinfo {author} {\bibfnamefont {J.~B.}\ \bibnamefont
  {Dent}}, \bibinfo {author} {\bibfnamefont {W.~E.}\ \bibnamefont {Gabella}},
  \bibinfo {author} {\bibfnamefont {K.}~\bibnamefont {Holley-Bockelmann}},\
  and\ \bibinfo {author} {\bibfnamefont {T.~W.}\ \bibnamefont {Kephart}},\
  }\bibfield  {title} {\bibinfo {title} {Gravitational waves from a black hole
  orbiting in a wormhole geometry},\ }\href
  {https://doi.org/10.1103/PhysRevD.104.044030} {\bibfield  {journal} {\bibinfo
   {journal} {Phys. Rev. D}\ }\textbf {\bibinfo {volume} {104}},\ \bibinfo
  {pages} {044030} (\bibinfo {year} {2021})}\BibitemShut {NoStop}%
\bibitem [{\citenamefont {Xu}\ \emph {et~al.}(2022)\citenamefont {Xu},
  \citenamefont {Ezquiaga},\ and\ \citenamefont {Holz}}]{Xu:2021}%
  \BibitemOpen
  \bibfield  {author} {\bibinfo {author} {\bibfnamefont {F.}~\bibnamefont
  {Xu}}, \bibinfo {author} {\bibfnamefont {J.~M.}\ \bibnamefont {Ezquiaga}},\
  and\ \bibinfo {author} {\bibfnamefont {D.~E.}\ \bibnamefont {Holz}},\
  }\bibfield  {title} {\bibinfo {title} {Please repeat: Strong lensing of
  gravitational waves as a probe of compact binary and galaxy populations},\
  }\href {https://doi.org/10.3847/1538-4357/ac58f8} {\bibfield  {journal}
  {\bibinfo  {journal} {The Astrophysical Journal}\ }\textbf {\bibinfo {volume}
  {929}},\ \bibinfo {pages} {9} (\bibinfo {year} {2022})}\BibitemShut {NoStop}%
\bibitem [{\citenamefont {{{\c{C}}al{\i}{\c{s}}kan}}\ \emph
  {et~al.}(2022)\citenamefont {{{\c{C}}al{\i}{\c{s}}kan}}, \citenamefont
  {{Mar{\'\i}a Ezquiaga}}, \citenamefont {{Hannuksela}},\ and\ \citenamefont
  {{Holz}}}]{Caliskan:2022}%
  \BibitemOpen
  \bibfield  {author} {\bibinfo {author} {\bibfnamefont {M.}~\bibnamefont
  {{{\c{C}}al{\i}{\c{s}}kan}}}, \bibinfo {author} {\bibfnamefont
  {J.}~\bibnamefont {{Mar{\'\i}a Ezquiaga}}}, \bibinfo {author} {\bibfnamefont
  {O.~A.}\ \bibnamefont {{Hannuksela}}},\ and\ \bibinfo {author} {\bibfnamefont
  {D.~E.}\ \bibnamefont {{Holz}}},\ }\bibfield  {title} {\bibinfo {title}
  {{Lensing or luck? False alarm probabilities for gravitational lensing of
  gravitational waves}},\ }\href@noop {} {\bibfield  {journal} {\bibinfo
  {journal} {arXiv e-prints}\ ,\ \bibinfo {eid} {arXiv:2201.04619}} (\bibinfo
  {year} {2022})},\ \Eprint {https://arxiv.org/abs/2201.04619}
  {arXiv:2201.04619 [astro-ph.CO]} \BibitemShut {NoStop}%
\bibitem [{\citenamefont {Ezquiaga}\ \emph {et~al.}(2021)\citenamefont
  {Ezquiaga}, \citenamefont {Hu}, \citenamefont {Lagos},\ and\ \citenamefont
  {Lin}}]{Ezquiaga:2021}%
  \BibitemOpen
  \bibfield  {author} {\bibinfo {author} {\bibfnamefont {J.~M.}\ \bibnamefont
  {Ezquiaga}}, \bibinfo {author} {\bibfnamefont {W.}~\bibnamefont {Hu}},
  \bibinfo {author} {\bibfnamefont {M.}~\bibnamefont {Lagos}},\ and\ \bibinfo
  {author} {\bibfnamefont {M.-X.}\ \bibnamefont {Lin}},\ }\bibfield  {title}
  {\bibinfo {title} {Gravitational wave propagation beyond general relativity:
  waveform distortions and echoes},\ }\href
  {https://doi.org/10.1088/1475-7516/2021/11/048} {\bibfield  {journal}
  {\bibinfo  {journal} {Journal of Cosmology and Astroparticle Physics}\
  }\textbf {\bibinfo {volume} {2021}}\bibinfo  {number} { (11)},\ \bibinfo
  {pages} {048}}\BibitemShut {NoStop}%
\bibitem [{\citenamefont {{Ezquiaga}}\ \emph {et~al.}(2022)\citenamefont
  {{Ezquiaga}}, \citenamefont {{Hu}}, \citenamefont {{Lagos}}, \citenamefont
  {{Lin}},\ and\ \citenamefont {{Xu}}}]{Ezquiaga:2022}%
  \BibitemOpen
\bibfield  {number} {  }\bibfield  {author} {\bibinfo {author} {\bibfnamefont
  {J.~M.}\ \bibnamefont {{Ezquiaga}}}, \bibinfo {author} {\bibfnamefont
  {W.}~\bibnamefont {{Hu}}}, \bibinfo {author} {\bibfnamefont {M.}~\bibnamefont
  {{Lagos}}}, \bibinfo {author} {\bibfnamefont {M.-X.}\ \bibnamefont {{Lin}}},\
  and\ \bibinfo {author} {\bibfnamefont {F.}~\bibnamefont {{Xu}}},\ }\bibfield
  {title} {\bibinfo {title} {{Modified gravitational wave propagation with
  higher modes and its degeneracies with lensing}},\ }\href@noop {} {\bibfield
  {journal} {\bibinfo  {journal} {arXiv e-prints}\ ,\ \bibinfo {eid}
  {arXiv:2203.13252}} (\bibinfo {year} {2022})},\ \Eprint
  {https://arxiv.org/abs/2203.13252} {arXiv:2203.13252 [gr-qc]} \BibitemShut
  {NoStop}%
\bibitem [{\citenamefont {Abbott}\ \emph {et~al.}(2019)\citenamefont {Abbott}
  \emph {et~al.}}]{GWTC-1}%
  \BibitemOpen
  \bibfield  {author} {\bibinfo {author} {\bibfnamefont {B.~P.}\ \bibnamefont
  {Abbott}} \emph {et~al.} (\bibinfo {collaboration} {LIGO Scientific
  Collaboration and Virgo Collaboration}),\ }\bibfield  {title} {\bibinfo
  {title} {Gwtc-1: A gravitational-wave transient catalog of compact binary
  mergers observed by ligo and virgo during the first and second observing
  runs},\ }\href {https://doi.org/10.1103/PhysRevX.9.031040} {\bibfield
  {journal} {\bibinfo  {journal} {Phys. Rev. X}\ }\textbf {\bibinfo {volume}
  {9}},\ \bibinfo {pages} {031040} (\bibinfo {year} {2019})}\BibitemShut
  {NoStop}%
\bibitem [{\citenamefont {Stiskalek}\ \emph {et~al.}(2020)\citenamefont
  {Stiskalek}, \citenamefont {Veitch},\ and\ \citenamefont
  {Messenger}}]{Stiskalek:2020}%
  \BibitemOpen
  \bibfield  {author} {\bibinfo {author} {\bibfnamefont {R.}~\bibnamefont
  {Stiskalek}}, \bibinfo {author} {\bibfnamefont {J.}~\bibnamefont {Veitch}},\
  and\ \bibinfo {author} {\bibfnamefont {C.}~\bibnamefont {Messenger}},\
  }\bibfield  {title} {\bibinfo {title} {{Are stellar-mass binary black hole
  mergers isotropically distributed?}},\ }\href
  {https://doi.org/10.1093/mnras/staa3613} {\bibfield  {journal} {\bibinfo
  {journal} {Monthly Notices of the Royal Astronomical Society}\ }\textbf
  {\bibinfo {volume} {501}},\ \bibinfo {pages} {970} (\bibinfo {year}
  {2020})},\ \Eprint
  {https://arxiv.org/abs/https://academic.oup.com/mnras/article-pdf/501/1/970/35102504/staa3613.pdf}
  {https://academic.oup.com/mnras/article-pdf/501/1/970/35102504/staa3613.pdf}
  \BibitemShut {NoStop}%
\bibitem [{\citenamefont {Chen}\ \emph {et~al.}(2017)\citenamefont {Chen},
  \citenamefont {Essick}, \citenamefont {Vitale}, \citenamefont {Holz},\ and\
  \citenamefont {Katsavounidis}}]{Chen:2017}%
  \BibitemOpen
  \bibfield  {author} {\bibinfo {author} {\bibfnamefont {H.-Y.}\ \bibnamefont
  {Chen}}, \bibinfo {author} {\bibfnamefont {R.}~\bibnamefont {Essick}},
  \bibinfo {author} {\bibfnamefont {S.}~\bibnamefont {Vitale}}, \bibinfo
  {author} {\bibfnamefont {D.~E.}\ \bibnamefont {Holz}},\ and\ \bibinfo
  {author} {\bibfnamefont {E.}~\bibnamefont {Katsavounidis}},\ }\bibfield
  {title} {\bibinfo {title} {{OBSERVATIONAL} {SELECTION} {EFFECTS} {WITH}
  {GROUND}-{BASED} {GRAVITATIONAL} {WAVE} {DETECTORS}},\ }\href
  {https://doi.org/10.3847/1538-4357/835/1/31} {\bibfield  {journal} {\bibinfo
  {journal} {The Astrophysical Journal}\ }\textbf {\bibinfo {volume} {835}},\
  \bibinfo {pages} {31} (\bibinfo {year} {2017})}\BibitemShut {NoStop}%
\bibitem [{\citenamefont {Payne}\ \emph {et~al.}(2020)\citenamefont {Payne},
  \citenamefont {Banagiri}, \citenamefont {Lasky},\ and\ \citenamefont
  {Thrane}}]{Payne:2020}%
  \BibitemOpen
  \bibfield  {author} {\bibinfo {author} {\bibfnamefont {E.}~\bibnamefont
  {Payne}}, \bibinfo {author} {\bibfnamefont {S.}~\bibnamefont {Banagiri}},
  \bibinfo {author} {\bibfnamefont {P.~D.}\ \bibnamefont {Lasky}},\ and\
  \bibinfo {author} {\bibfnamefont {E.}~\bibnamefont {Thrane}},\ }\bibfield
  {title} {\bibinfo {title} {Searching for anisotropy in the distribution of
  binary black hole mergers},\ }\href
  {https://doi.org/10.1103/PhysRevD.102.102004} {\bibfield  {journal} {\bibinfo
   {journal} {Phys. Rev. D}\ }\textbf {\bibinfo {volume} {102}},\ \bibinfo
  {pages} {102004} (\bibinfo {year} {2020})}\BibitemShut {NoStop}%
\bibitem [{\citenamefont {Cavaglia}\ and\ \citenamefont
  {Modi}(2020)}]{Cavaglia:2020}%
  \BibitemOpen
  \bibfield  {author} {\bibinfo {author} {\bibfnamefont {M.}~\bibnamefont
  {Cavaglia}}\ and\ \bibinfo {author} {\bibfnamefont {A.}~\bibnamefont
  {Modi}},\ }\bibfield  {title} {\bibinfo {title} {{Two-dimensional correlation
  function of binary black hole coalescences}},\ }\href
  {https://doi.org/10.3390/universe6070093} {\bibfield  {journal} {\bibinfo
  {journal} {Universe}\ }\textbf {\bibinfo {volume} {6}},\ \bibinfo {pages}
  {93} (\bibinfo {year} {2020})},\ \Eprint {https://arxiv.org/abs/2005.06004}
  {arXiv:2005.06004 [astro-ph.HE]} \BibitemShut {NoStop}%
\bibitem [{\citenamefont {{Renzini}}\ \emph {et~al.}(2022)\citenamefont
  {{Renzini}}, \citenamefont {{Goncharov}}, \citenamefont {{Jenkins}},\ and\
  \citenamefont {{Meyers}}}]{Renzini:2022}%
  \BibitemOpen
  \bibfield  {author} {\bibinfo {author} {\bibfnamefont {A.~I.}\ \bibnamefont
  {{Renzini}}}, \bibinfo {author} {\bibfnamefont {B.}~\bibnamefont
  {{Goncharov}}}, \bibinfo {author} {\bibfnamefont {A.~C.}\ \bibnamefont
  {{Jenkins}}},\ and\ \bibinfo {author} {\bibfnamefont {P.~M.}\ \bibnamefont
  {{Meyers}}},\ }\bibfield  {title} {\bibinfo {title} {{Stochastic
  Gravitational-Wave Backgrounds: Current Detection Efforts and Future
  Prospects}},\ }\href {https://doi.org/10.3390/galaxies10010034} {\bibfield
  {journal} {\bibinfo  {journal} {Galaxies}\ }\textbf {\bibinfo {volume}
  {10}},\ \bibinfo {pages} {34} (\bibinfo {year} {2022})},\ \Eprint
  {https://arxiv.org/abs/2202.00178} {arXiv:2202.00178 [gr-qc]} \BibitemShut
  {NoStop}%
\bibitem [{\citenamefont {Callister}\ \emph {et~al.}(2020)\citenamefont
  {Callister}, \citenamefont {Fishbach}, \citenamefont {Holz},\ and\
  \citenamefont {Farr}}]{Callister:2020}%
  \BibitemOpen
  \bibfield  {author} {\bibinfo {author} {\bibfnamefont {T.}~\bibnamefont
  {Callister}}, \bibinfo {author} {\bibfnamefont {M.}~\bibnamefont {Fishbach}},
  \bibinfo {author} {\bibfnamefont {D.~E.}\ \bibnamefont {Holz}},\ and\
  \bibinfo {author} {\bibfnamefont {W.~M.}\ \bibnamefont {Farr}},\ }\bibfield
  {title} {\bibinfo {title} {Shouts and murmurs: Combining individual
  gravitational-wave sources with the stochastic background to measure the
  history of binary black hole mergers},\ }\href
  {https://doi.org/10.3847/2041-8213/ab9743} {\bibfield  {journal} {\bibinfo
  {journal} {The Astrophysical Journal}\ }\textbf {\bibinfo {volume} {896}},\
  \bibinfo {pages} {L32} (\bibinfo {year} {2020})}\BibitemShut {NoStop}%
\bibitem [{\citenamefont {Cordes}\ and\ \citenamefont
  {Chatterjee}(2019)}]{Cordes:2019}%
  \BibitemOpen
  \bibfield  {author} {\bibinfo {author} {\bibfnamefont {J.~M.}\ \bibnamefont
  {Cordes}}\ and\ \bibinfo {author} {\bibfnamefont {S.}~\bibnamefont
  {Chatterjee}},\ }\bibfield  {title} {\bibinfo {title} {Fast radio bursts: An
  extragalactic enigma},\ }\href
  {https://doi.org/10.1146/annurev-astro-091918-104501} {\bibfield  {journal}
  {\bibinfo  {journal} {Annual Review of Astronomy and Astrophysics}\ }\textbf
  {\bibinfo {volume} {57}},\ \bibinfo {pages} {417} (\bibinfo {year} {2019})},\
  \Eprint
  {https://arxiv.org/abs/https://doi.org/10.1146/annurev-astro-091918-104501}
  {https://doi.org/10.1146/annurev-astro-091918-104501} \BibitemShut {NoStop}%
\bibitem [{\citenamefont {Abeysekara}\ \emph {et~al.}(2014)\citenamefont
  {Abeysekara} \emph {et~al.}}]{Abeysekara:2014}%
  \BibitemOpen
  \bibfield  {author} {\bibinfo {author} {\bibfnamefont {A.~U.}\ \bibnamefont
  {Abeysekara}} \emph {et~al.},\ }\bibfield  {title} {\bibinfo {title}
  {{OBSERVATION} {OF} {SMALL}-{SCALE} {ANISOTROPY} {IN} {THE} {ARRIVAL}
  {DIRECTION} {DISTRIBUTION} {OF} {TeV} {COSMIC} {RAYS} {WITH} {HAWC}},\ }\href
  {https://doi.org/10.1088/0004-637x/796/2/108} {\bibfield  {journal} {\bibinfo
   {journal} {The Astrophysical Journal}\ }\textbf {\bibinfo {volume} {796}},\
  \bibinfo {pages} {108} (\bibinfo {year} {2014})}\BibitemShut {NoStop}%
\bibitem [{\citenamefont {Illuminati}(2016)}]{Illuminati:2016}%
  \BibitemOpen
  \bibfield  {author} {\bibinfo {author} {\bibfnamefont {G.}~\bibnamefont
  {Illuminati}},\ }\bibfield  {title} {\bibinfo {title} {Study of the high
  energy cosmic rays large scale anisotropies with the {ANTARES} neutrino
  telescope},\ }\href {https://doi.org/10.1088/1742-6596/689/1/012012}
  {\bibfield  {journal} {\bibinfo  {journal} {Journal of Physics: Conference
  Series}\ }\textbf {\bibinfo {volume} {689}},\ \bibinfo {pages} {012012}
  (\bibinfo {year} {2016})}\BibitemShut {NoStop}%
\bibitem [{\citenamefont {Aab}\ \emph {et~al.}(2018)\citenamefont {Aab} \emph
  {et~al.}}]{Aab:2018}%
  \BibitemOpen
  \bibfield  {author} {\bibinfo {author} {\bibfnamefont {A.}~\bibnamefont
  {Aab}} \emph {et~al.},\ }\bibfield  {title} {\bibinfo {title} {Large-scale
  cosmic-ray anisotropies above 4 {EeV} measured by the pierre auger
  observatory},\ }\href {https://doi.org/10.3847/1538-4357/aae689} {\bibfield
  {journal} {\bibinfo  {journal} {The Astrophysical Journal}\ }\textbf
  {\bibinfo {volume} {868}},\ \bibinfo {pages} {4} (\bibinfo {year}
  {2018})}\BibitemShut {NoStop}%
\bibitem [{\citenamefont {{McNally}}\ \emph {et~al.}(2021)\citenamefont
  {{McNally}}, \citenamefont {{Abbasi}}, \citenamefont {{Desiati}},
  \citenamefont {{D{\'\i}az V{\'e}lez}}, \citenamefont {{Aguado}},
  \citenamefont {{Gruchot}}, \citenamefont {{Moy}}, \citenamefont {{Simmons}},
  \citenamefont {{Thorpe}},\ and\ \citenamefont {{Woodward}}}]{McNally:2021}%
  \BibitemOpen
  \bibfield  {author} {\bibinfo {author} {\bibfnamefont {F.}~\bibnamefont
  {{McNally}}}, \bibinfo {author} {\bibfnamefont {R.}~\bibnamefont {{Abbasi}}},
  \bibinfo {author} {\bibfnamefont {P.}~\bibnamefont {{Desiati}}}, \bibinfo
  {author} {\bibfnamefont {J.~C.}\ \bibnamefont {{D{\'\i}az V{\'e}lez}}},
  \bibinfo {author} {\bibfnamefont {T.}~\bibnamefont {{Aguado}}}, \bibinfo
  {author} {\bibfnamefont {K.}~\bibnamefont {{Gruchot}}}, \bibinfo {author}
  {\bibfnamefont {A.}~\bibnamefont {{Moy}}}, \bibinfo {author} {\bibfnamefont
  {A.}~\bibnamefont {{Simmons}}}, \bibinfo {author} {\bibfnamefont
  {A.}~\bibnamefont {{Thorpe}}},\ and\ \bibinfo {author} {\bibfnamefont
  {H.}~\bibnamefont {{Woodward}}},\ }\bibfield  {title} {\bibinfo {title}
  {{Observation of Cosmic Ray Anisotropy with Nine Years of IceCube Data}},\
  }\href@noop {} {\bibfield  {journal} {\bibinfo  {journal} {arXiv e-prints}\
  ,\ \bibinfo {eid} {arXiv:2107.11454}} (\bibinfo {year} {2021})},\ \Eprint
  {https://arxiv.org/abs/2107.11454} {arXiv:2107.11454 [astro-ph.HE]}
  \BibitemShut {NoStop}%
\bibitem [{\citenamefont {Abbott}\ \emph
  {et~al.}(2021{\natexlab{b}})\citenamefont {Abbott} \emph {et~al.}}]{GWTC-2}%
  \BibitemOpen
  \bibfield  {author} {\bibinfo {author} {\bibfnamefont {R.}~\bibnamefont
  {Abbott}} \emph {et~al.} (\bibinfo {collaboration} {LIGO Scientific
  Collaboration and Virgo Collaboration}),\ }\bibfield  {title} {\bibinfo
  {title} {Gwtc-2: Compact binary coalescences observed by ligo and virgo
  during the first half of the third observing run},\ }\href
  {https://doi.org/10.1103/PhysRevX.11.021053} {\bibfield  {journal} {\bibinfo
  {journal} {Phys. Rev. X}\ }\textbf {\bibinfo {volume} {11}},\ \bibinfo
  {pages} {021053} (\bibinfo {year} {2021}{\natexlab{b}})}\BibitemShut
  {NoStop}%
\bibitem [{\citenamefont {{Abbott}}\ \emph
  {et~al.}(2021{\natexlab{a}})\citenamefont {{Abbott}} \emph
  {et~al.}}]{GWTC-2d1}%
  \BibitemOpen
  \bibfield  {author} {\bibinfo {author} {\bibfnamefont {R.}~\bibnamefont
  {{Abbott}}} \emph {et~al.},\ }\bibfield  {title} {\bibinfo {title}
  {{GWTC-2.1: Deep Extended Catalog of Compact Binary Coalescences Observed by
  LIGO and Virgo During the First Half of the Third Observing Run}},\
  }\href@noop {} {\bibfield  {journal} {\bibinfo  {journal} {arXiv e-prints}\
  ,\ \bibinfo {eid} {arXiv:2108.01045}} (\bibinfo {year}
  {2021}{\natexlab{a}})},\ \Eprint {https://arxiv.org/abs/2108.01045}
  {arXiv:2108.01045 [gr-qc]} \BibitemShut {NoStop}%
\bibitem [{\citenamefont {{Abbott}}\ \emph
  {et~al.}(2021{\natexlab{b}})\citenamefont {{Abbott}} \emph
  {et~al.}}]{GWTC-3}%
  \BibitemOpen
  \bibfield  {author} {\bibinfo {author} {\bibfnamefont {R.}~\bibnamefont
  {{Abbott}}} \emph {et~al.},\ }\bibfield  {title} {\bibinfo {title} {{GWTC-3:
  Compact Binary Coalescences Observed by LIGO and Virgo During the Second Part
  of the Third Observing Run}},\ }\href@noop {} {\bibfield  {journal} {\bibinfo
   {journal} {arXiv e-prints}\ ,\ \bibinfo {eid} {arXiv:2111.03606}} (\bibinfo
  {year} {2021}{\natexlab{b}})},\ \Eprint {https://arxiv.org/abs/2111.03606}
  {arXiv:2111.03606 [gr-qc]} \BibitemShut {NoStop}%
\bibitem [{\citenamefont {{Abbott}}\ \emph
  {et~al.}(2021{\natexlab{c}})\citenamefont {{Abbott}} \emph
  {et~al.}}]{GWTC-3-injections}%
  \BibitemOpen
  \bibfield  {author} {\bibinfo {author} {\bibfnamefont {R.}~\bibnamefont
  {{Abbott}}} \emph {et~al.},\ }\bibfield  {title} {\bibinfo {title} {{GWTC-3:
  Compact Binary Coalescences Observed by LIGO and Virgo During the Second Part
  of the Third Observing Run — O3 search sensitivity estimates}},\ }\href
  {https://doi.org/10.5281/zenodo.5546676} {10.5281/zenodo.5546676} (\bibinfo
  {year} {2021}{\natexlab{c}})\BibitemShut {NoStop}%
\bibitem [{\citenamefont {{Farah}}\ \emph {et~al.}(2022)\citenamefont
  {{Farah}}, \citenamefont {{Fishbach}}, \citenamefont {{Essick}},
  \citenamefont {{Holz}},\ and\ \citenamefont {{Galaudage}}}]{Farah:2021}%
  \BibitemOpen
  \bibfield  {author} {\bibinfo {author} {\bibfnamefont {A.~M.}\ \bibnamefont
  {{Farah}}}, \bibinfo {author} {\bibfnamefont {M.}~\bibnamefont {{Fishbach}}},
  \bibinfo {author} {\bibfnamefont {R.}~\bibnamefont {{Essick}}}, \bibinfo
  {author} {\bibfnamefont {D.~E.}\ \bibnamefont {{Holz}}},\ and\ \bibinfo
  {author} {\bibfnamefont {S.}~\bibnamefont {{Galaudage}}},\ }\bibfield
  {title} {\bibinfo {title} {{Bridging the Gap: Categorizing Gravitational-Wave
  Events at the Transition Between Neutron Stars and Black Holes}},\ }\href
  {https://doi.org/10.3847/1538-4357/ac5f03} {\bibfield  {journal} {\bibinfo
  {journal} {The Astrophysical Journal}\ }\textbf {\bibinfo {volume} {931}},\
  \bibinfo {pages} {108} (\bibinfo {year} {2022})}\BibitemShut {NoStop}%
\bibitem [{\citenamefont {{Aghanim, N.}}\ \emph {et~al.}(2020)\citenamefont
  {{Aghanim, N.}} \emph {et~al.}}]{Planck:2020}%
  \BibitemOpen
  \bibfield  {author} {\bibinfo {author} {\bibnamefont {{Aghanim, N.}}} \emph
  {et~al.},\ }\bibfield  {title} {\bibinfo {title} {Planck 2018 results - vi.
  cosmological parameters},\ }\href
  {https://doi.org/10.1051/0004-6361/201833910} {\bibfield  {journal} {\bibinfo
   {journal} {A\&A}\ }\textbf {\bibinfo {volume} {641}},\ \bibinfo {pages} {A6}
  (\bibinfo {year} {2020})}\BibitemShut {NoStop}%
\bibitem [{\citenamefont {{Loredo}}(2004)}]{Loredo:2004}%
  \BibitemOpen
  \bibfield  {author} {\bibinfo {author} {\bibfnamefont {T.~J.}\ \bibnamefont
  {{Loredo}}},\ }\bibfield  {title} {\bibinfo {title} {{Accounting for Source
  Uncertainties in Analyses of Astronomical Survey Data}},\ }in\ \href
  {https://doi.org/10.1063/1.1835214} {\emph {\bibinfo {booktitle} {Bayesian
  Inference and Maximum Entropy Methods in Science and Engineering: 24th
  International Workshop on Bayesian Inference and Maximum Entropy Methods in
  Science and Engineering}}},\ \bibinfo {series} {American Institute of Physics
  Conference Series}, Vol.\ \bibinfo {volume} {735},\ \bibinfo {editor} {edited
  by\ \bibinfo {editor} {\bibfnamefont {R.}~\bibnamefont {{Fischer}}}, \bibinfo
  {editor} {\bibfnamefont {R.}~\bibnamefont {{Preuss}}},\ and\ \bibinfo
  {editor} {\bibfnamefont {U.~V.}\ \bibnamefont {{Toussaint}}}}\ (\bibinfo
  {year} {2004})\ pp.\ \bibinfo {pages} {195--206},\ \Eprint
  {https://arxiv.org/abs/astro-ph/0409387} {arXiv:astro-ph/0409387 [astro-ph]}
  \BibitemShut {NoStop}%
\bibitem [{\citenamefont {Mandel}(2010)}]{Mandel:2010}%
  \BibitemOpen
  \bibfield  {author} {\bibinfo {author} {\bibfnamefont {I.}~\bibnamefont
  {Mandel}},\ }\bibfield  {title} {\bibinfo {title} {Parameter estimation on
  gravitational waves from multiple coalescing binaries},\ }\href
  {https://doi.org/10.1103/PhysRevD.81.084029} {\bibfield  {journal} {\bibinfo
  {journal} {Phys. Rev. D}\ }\textbf {\bibinfo {volume} {81}},\ \bibinfo
  {pages} {084029} (\bibinfo {year} {2010})}\BibitemShut {NoStop}%
\bibitem [{\citenamefont {Mandel}\ \emph {et~al.}(2019)\citenamefont {Mandel},
  \citenamefont {Farr},\ and\ \citenamefont {Gair}}]{Mandel:2019}%
  \BibitemOpen
  \bibfield  {author} {\bibinfo {author} {\bibfnamefont {I.}~\bibnamefont
  {Mandel}}, \bibinfo {author} {\bibfnamefont {W.~M.}\ \bibnamefont {Farr}},\
  and\ \bibinfo {author} {\bibfnamefont {J.~R.}\ \bibnamefont {Gair}},\
  }\bibfield  {title} {\bibinfo {title} {{Extracting distribution parameters
  from multiple uncertain observations with selection biases}},\ }\href
  {https://doi.org/10.1093/mnras/stz896} {\bibfield  {journal} {\bibinfo
  {journal} {Monthly Notices of the Royal Astronomical Society}\ }\textbf
  {\bibinfo {volume} {486}},\ \bibinfo {pages} {1086} (\bibinfo {year}
  {2019})},\ \Eprint
  {https://arxiv.org/abs/https://academic.oup.com/mnras/article-pdf/486/1/1086/28390969/stz896.pdf}
  {https://academic.oup.com/mnras/article-pdf/486/1/1086/28390969/stz896.pdf}
  \BibitemShut {NoStop}%
\bibitem [{\citenamefont {{Essick}}\ and\ \citenamefont
  {{Farr}}(2022)}]{Essick:2022}%
  \BibitemOpen
  \bibfield  {author} {\bibinfo {author} {\bibfnamefont {R.}~\bibnamefont
  {{Essick}}}\ and\ \bibinfo {author} {\bibfnamefont {W.}~\bibnamefont
  {{Farr}}},\ }\bibfield  {title} {\bibinfo {title} {{Precision Requirements
  for Monte Carlo Sums within Hierarchical Bayesian Inference}},\ }\href@noop
  {} {\bibfield  {journal} {\bibinfo  {journal} {arXiv e-prints}\ ,\ \bibinfo
  {eid} {arXiv:2204.00461}} (\bibinfo {year} {2022})},\ \Eprint
  {https://arxiv.org/abs/2204.00461} {arXiv:2204.00461 [astro-ph.IM]}
  \BibitemShut {NoStop}%
\bibitem [{\citenamefont {{G{\'o}rski}}\ \emph {et~al.}(2005)\citenamefont
  {{G{\'o}rski}}, \citenamefont {{Hivon}}, \citenamefont {{Banday}},
  \citenamefont {{Wandelt}}, \citenamefont {{Hansen}}, \citenamefont
  {{Reinecke}},\ and\ \citenamefont {{Bartelmann}}}]{Gorski:2005}%
  \BibitemOpen
  \bibfield  {author} {\bibinfo {author} {\bibfnamefont {K.~M.}\ \bibnamefont
  {{G{\'o}rski}}}, \bibinfo {author} {\bibfnamefont {E.}~\bibnamefont
  {{Hivon}}}, \bibinfo {author} {\bibfnamefont {A.~J.}\ \bibnamefont
  {{Banday}}}, \bibinfo {author} {\bibfnamefont {B.~D.}\ \bibnamefont
  {{Wandelt}}}, \bibinfo {author} {\bibfnamefont {F.~K.}\ \bibnamefont
  {{Hansen}}}, \bibinfo {author} {\bibfnamefont {M.}~\bibnamefont
  {{Reinecke}}},\ and\ \bibinfo {author} {\bibfnamefont {M.}~\bibnamefont
  {{Bartelmann}}},\ }\bibfield  {title} {\bibinfo {title} {{HEALPix: A
  Framework for High-Resolution Discretization and Fast Analysis of Data
  Distributed on the Sphere}},\ }\href {https://doi.org/10.1086/427976}
  {\bibfield  {journal} {\bibinfo  {journal} {\apj}\ }\textbf {\bibinfo
  {volume} {622}},\ \bibinfo {pages} {759} (\bibinfo {year} {2005})},\ \Eprint
  {https://arxiv.org/abs/astro-ph/0409513} {arXiv:astro-ph/0409513 [astro-ph]}
  \BibitemShut {NoStop}%
\bibitem [{\citenamefont {Delporte}\ and\ \citenamefont
  {Union}(1930)}]{Delporte:1930}%
  \BibitemOpen
  \bibfield  {author} {\bibinfo {author} {\bibfnamefont {E.}~\bibnamefont
  {Delporte}}\ and\ \bibinfo {author} {\bibfnamefont {I.~A.}\ \bibnamefont
  {Union}},\ }\href {https://books.google.com/books?id=v3XvAAAAMAAJ} {\emph
  {\bibinfo {title} {D{\'e}limitation scientifique des constellations: (tables
  et cartes)}}},\ Report /Commission 3 of the International Astronomical Union\
  (\bibinfo  {publisher} {At the University Press},\ \bibinfo {year}
  {1930})\BibitemShut {NoStop}%
\bibitem [{\citenamefont {Essick}\ and\ \citenamefont
  {Landry}(2020)}]{Essick:2020}%
  \BibitemOpen
  \bibfield  {author} {\bibinfo {author} {\bibfnamefont {R.}~\bibnamefont
  {Essick}}\ and\ \bibinfo {author} {\bibfnamefont {P.}~\bibnamefont
  {Landry}},\ }\bibfield  {title} {\bibinfo {title} {Discriminating between
  neutron stars and black holes with imperfect knowledge of the maximum neutron
  star mass},\ }\href {https://doi.org/10.3847/1538-4357/abbd3b} {\bibfield
  {journal} {\bibinfo  {journal} {The Astrophysical Journal}\ }\textbf
  {\bibinfo {volume} {904}},\ \bibinfo {pages} {80} (\bibinfo {year}
  {2020})}\BibitemShut {NoStop}%
\bibitem [{\citenamefont {{Isi}}\ \emph {et~al.}(2022)\citenamefont {{Isi}},
  \citenamefont {{Farr}},\ and\ \citenamefont {{Chatziioannou}}}]{Isi:2022}%
  \BibitemOpen
  \bibfield  {author} {\bibinfo {author} {\bibfnamefont {M.}~\bibnamefont
  {{Isi}}}, \bibinfo {author} {\bibfnamefont {W.~M.}\ \bibnamefont {{Farr}}},\
  and\ \bibinfo {author} {\bibfnamefont {K.}~\bibnamefont {{Chatziioannou}}},\
  }\bibfield  {title} {\bibinfo {title} {{Comparing Bayes factors and
  hierarchical inference for testing general relativity with gravitational
  waves}},\ }\href@noop {} {\bibfield  {journal} {\bibinfo  {journal} {arXiv
  e-prints}\ ,\ \bibinfo {eid} {arXiv:2204.10742}} (\bibinfo {year} {2022})},\
  \Eprint {https://arxiv.org/abs/2204.10742} {arXiv:2204.10742 [gr-qc]}
  \BibitemShut {NoStop}%
\bibitem [{\citenamefont {Rasmussen}\ and\ \citenamefont
  {Williams}(2005)}]{Rasmussen:2005}%
  \BibitemOpen
  \bibfield  {author} {\bibinfo {author} {\bibfnamefont {C.}~\bibnamefont
  {Rasmussen}}\ and\ \bibinfo {author} {\bibfnamefont {C.}~\bibnamefont
  {Williams}},\ }\href {https://books.google.com/books?id=Tr34DwAAQBAJ} {\emph
  {\bibinfo {title} {Gaussian Processes for Machine Learning}}},\ Adaptive
  Computation and Machine Learning series\ (\bibinfo  {publisher} {MIT Press},\
  \bibinfo {year} {2005})\BibitemShut {NoStop}%
\bibitem [{\citenamefont {Lang}\ and\ \citenamefont
  {Schwab}(2015)}]{Land:2015}%
  \BibitemOpen
  \bibfield  {author} {\bibinfo {author} {\bibfnamefont {A.}~\bibnamefont
  {Lang}}\ and\ \bibinfo {author} {\bibfnamefont {C.}~\bibnamefont {Schwab}},\
  }\bibfield  {title} {\bibinfo {title} {{Isotropic Gaussian random fields on
  the sphere: Regularity, fast simulation and stochastic partial differential
  equations}},\ }\href {https://doi.org/10.1214/14-AAP1067} {\bibfield
  {journal} {\bibinfo  {journal} {The Annals of Applied Probability}\ }\textbf
  {\bibinfo {volume} {25}},\ \bibinfo {pages} {3047 } (\bibinfo {year}
  {2015})}\BibitemShut {NoStop}%
\bibitem [{\citenamefont {Farr}\ \emph {et~al.}(2018)\citenamefont {Farr},
  \citenamefont {Farr}, \citenamefont {Cowan}, \citenamefont {Haggard},\ and\
  \citenamefont {Robinson}}]{Farr:2018}%
  \BibitemOpen
  \bibfield  {author} {\bibinfo {author} {\bibfnamefont {B.}~\bibnamefont
  {Farr}}, \bibinfo {author} {\bibfnamefont {W.~M.}\ \bibnamefont {Farr}},
  \bibinfo {author} {\bibfnamefont {N.~B.}\ \bibnamefont {Cowan}}, \bibinfo
  {author} {\bibfnamefont {H.~M.}\ \bibnamefont {Haggard}},\ and\ \bibinfo
  {author} {\bibfnamefont {T.}~\bibnamefont {Robinson}},\ }\bibfield  {title}
  {\bibinfo {title} {exocartographer: A bayesian framework for mapping
  exoplanets in reflected light},\ }\href
  {https://doi.org/10.3847/1538-3881/aad775} {\bibfield  {journal} {\bibinfo
  {journal} {The Astronomical Journal}\ }\textbf {\bibinfo {volume} {156}},\
  \bibinfo {pages} {146} (\bibinfo {year} {2018})}\BibitemShut {NoStop}%
\bibitem [{\citenamefont {Fishbach}\ \emph {et~al.}(2021)\citenamefont
  {Fishbach}, \citenamefont {Doctor}, \citenamefont {Callister}, \citenamefont
  {Edelman}, \citenamefont {Ye}, \citenamefont {Essick}, \citenamefont {Farr},
  \citenamefont {Farr},\ and\ \citenamefont {Holz}}]{Fishbach:2021a}%
  \BibitemOpen
  \bibfield  {author} {\bibinfo {author} {\bibfnamefont {M.}~\bibnamefont
  {Fishbach}}, \bibinfo {author} {\bibfnamefont {Z.}~\bibnamefont {Doctor}},
  \bibinfo {author} {\bibfnamefont {T.}~\bibnamefont {Callister}}, \bibinfo
  {author} {\bibfnamefont {B.}~\bibnamefont {Edelman}}, \bibinfo {author}
  {\bibfnamefont {J.}~\bibnamefont {Ye}}, \bibinfo {author} {\bibfnamefont
  {R.}~\bibnamefont {Essick}}, \bibinfo {author} {\bibfnamefont {W.~M.}\
  \bibnamefont {Farr}}, \bibinfo {author} {\bibfnamefont {B.}~\bibnamefont
  {Farr}},\ and\ \bibinfo {author} {\bibfnamefont {D.~E.}\ \bibnamefont
  {Holz}},\ }\bibfield  {title} {\bibinfo {title} {When are {LIGO}/virgo's big
  black hole mergers?},\ }\href {https://doi.org/10.3847/1538-4357/abee11}
  {\bibfield  {journal} {\bibinfo  {journal} {The Astrophysical Journal}\
  }\textbf {\bibinfo {volume} {912}},\ \bibinfo {pages} {98} (\bibinfo {year}
  {2021})}\BibitemShut {NoStop}%
\bibitem [{\citenamefont {Harris}\ \emph {et~al.}(2020)\citenamefont {Harris},
  \citenamefont {Millman}, \citenamefont {van~der Walt}, \citenamefont
  {Gommers}, \citenamefont {Virtanen}, \citenamefont {Cournapeau},
  \citenamefont {Wieser}, \citenamefont {Taylor}, \citenamefont {Berg},
  \citenamefont {Smith}, \citenamefont {Kern}, \citenamefont {Picus},
  \citenamefont {Hoyer}, \citenamefont {van Kerkwijk}, \citenamefont {Brett},
  \citenamefont {Haldane}, \citenamefont {del R{\'{i}}o}, \citenamefont
  {Wiebe}, \citenamefont {Peterson}, \citenamefont {G{\'{e}}rard-Marchant},
  \citenamefont {Sheppard}, \citenamefont {Reddy}, \citenamefont {Weckesser},
  \citenamefont {Abbasi}, \citenamefont {Gohlke},\ and\ \citenamefont
  {Oliphant}}]{numpy}%
  \BibitemOpen
  \bibfield  {author} {\bibinfo {author} {\bibfnamefont {C.~R.}\ \bibnamefont
  {Harris}}, \bibinfo {author} {\bibfnamefont {K.~J.}\ \bibnamefont {Millman}},
  \bibinfo {author} {\bibfnamefont {S.~J.}\ \bibnamefont {van~der Walt}},
  \bibinfo {author} {\bibfnamefont {R.}~\bibnamefont {Gommers}}, \bibinfo
  {author} {\bibfnamefont {P.}~\bibnamefont {Virtanen}}, \bibinfo {author}
  {\bibfnamefont {D.}~\bibnamefont {Cournapeau}}, \bibinfo {author}
  {\bibfnamefont {E.}~\bibnamefont {Wieser}}, \bibinfo {author} {\bibfnamefont
  {J.}~\bibnamefont {Taylor}}, \bibinfo {author} {\bibfnamefont
  {S.}~\bibnamefont {Berg}}, \bibinfo {author} {\bibfnamefont {N.~J.}\
  \bibnamefont {Smith}}, \bibinfo {author} {\bibfnamefont {R.}~\bibnamefont
  {Kern}}, \bibinfo {author} {\bibfnamefont {M.}~\bibnamefont {Picus}},
  \bibinfo {author} {\bibfnamefont {S.}~\bibnamefont {Hoyer}}, \bibinfo
  {author} {\bibfnamefont {M.~H.}\ \bibnamefont {van Kerkwijk}}, \bibinfo
  {author} {\bibfnamefont {M.}~\bibnamefont {Brett}}, \bibinfo {author}
  {\bibfnamefont {A.}~\bibnamefont {Haldane}}, \bibinfo {author} {\bibfnamefont
  {J.~F.}\ \bibnamefont {del R{\'{i}}o}}, \bibinfo {author} {\bibfnamefont
  {M.}~\bibnamefont {Wiebe}}, \bibinfo {author} {\bibfnamefont
  {P.}~\bibnamefont {Peterson}}, \bibinfo {author} {\bibfnamefont
  {P.}~\bibnamefont {G{\'{e}}rard-Marchant}}, \bibinfo {author} {\bibfnamefont
  {K.}~\bibnamefont {Sheppard}}, \bibinfo {author} {\bibfnamefont
  {T.}~\bibnamefont {Reddy}}, \bibinfo {author} {\bibfnamefont
  {W.}~\bibnamefont {Weckesser}}, \bibinfo {author} {\bibfnamefont
  {H.}~\bibnamefont {Abbasi}}, \bibinfo {author} {\bibfnamefont
  {C.}~\bibnamefont {Gohlke}},\ and\ \bibinfo {author} {\bibfnamefont {T.~E.}\
  \bibnamefont {Oliphant}},\ }\bibfield  {title} {\bibinfo {title} {Array
  programming with {NumPy}},\ }\href
  {https://doi.org/10.1038/s41586-020-2649-2} {\bibfield  {journal} {\bibinfo
  {journal} {Nature}\ }\textbf {\bibinfo {volume} {585}},\ \bibinfo {pages}
  {357} (\bibinfo {year} {2020})}\BibitemShut {NoStop}%
\bibitem [{\citenamefont {Virtanen}\ \emph {et~al.}(2020)\citenamefont
  {Virtanen}, \citenamefont {Gommers}, \citenamefont {Oliphant}, \citenamefont
  {Haberland}, \citenamefont {Reddy}, \citenamefont {Cournapeau}, \citenamefont
  {Burovski}, \citenamefont {Peterson}, \citenamefont {Weckesser},
  \citenamefont {Bright}, \citenamefont {{van der Walt}}, \citenamefont
  {Brett}, \citenamefont {Wilson}, \citenamefont {Millman}, \citenamefont
  {Mayorov}, \citenamefont {Nelson}, \citenamefont {Jones}, \citenamefont
  {Kern}, \citenamefont {Larson}, \citenamefont {Carey}, \citenamefont {Polat},
  \citenamefont {Feng}, \citenamefont {Moore}, \citenamefont {{VanderPlas}},
  \citenamefont {Laxalde}, \citenamefont {Perktold}, \citenamefont {Cimrman},
  \citenamefont {Henriksen}, \citenamefont {Quintero}, \citenamefont {Harris},
  \citenamefont {Archibald}, \citenamefont {Ribeiro}, \citenamefont
  {Pedregosa}, \citenamefont {{van Mulbregt}},\ and\ \citenamefont {{SciPy 1.0
  Contributors}}}]{scipy}%
  \BibitemOpen
  \bibfield  {author} {\bibinfo {author} {\bibfnamefont {P.}~\bibnamefont
  {Virtanen}}, \bibinfo {author} {\bibfnamefont {R.}~\bibnamefont {Gommers}},
  \bibinfo {author} {\bibfnamefont {T.~E.}\ \bibnamefont {Oliphant}}, \bibinfo
  {author} {\bibfnamefont {M.}~\bibnamefont {Haberland}}, \bibinfo {author}
  {\bibfnamefont {T.}~\bibnamefont {Reddy}}, \bibinfo {author} {\bibfnamefont
  {D.}~\bibnamefont {Cournapeau}}, \bibinfo {author} {\bibfnamefont
  {E.}~\bibnamefont {Burovski}}, \bibinfo {author} {\bibfnamefont
  {P.}~\bibnamefont {Peterson}}, \bibinfo {author} {\bibfnamefont
  {W.}~\bibnamefont {Weckesser}}, \bibinfo {author} {\bibfnamefont
  {J.}~\bibnamefont {Bright}}, \bibinfo {author} {\bibfnamefont {S.~J.}\
  \bibnamefont {{van der Walt}}}, \bibinfo {author} {\bibfnamefont
  {M.}~\bibnamefont {Brett}}, \bibinfo {author} {\bibfnamefont
  {J.}~\bibnamefont {Wilson}}, \bibinfo {author} {\bibfnamefont {K.~J.}\
  \bibnamefont {Millman}}, \bibinfo {author} {\bibfnamefont {N.}~\bibnamefont
  {Mayorov}}, \bibinfo {author} {\bibfnamefont {A.~R.~J.}\ \bibnamefont
  {Nelson}}, \bibinfo {author} {\bibfnamefont {E.}~\bibnamefont {Jones}},
  \bibinfo {author} {\bibfnamefont {R.}~\bibnamefont {Kern}}, \bibinfo {author}
  {\bibfnamefont {E.}~\bibnamefont {Larson}}, \bibinfo {author} {\bibfnamefont
  {C.~J.}\ \bibnamefont {Carey}}, \bibinfo {author} {\bibfnamefont
  {{\.I}.}~\bibnamefont {Polat}}, \bibinfo {author} {\bibfnamefont
  {Y.}~\bibnamefont {Feng}}, \bibinfo {author} {\bibfnamefont {E.~W.}\
  \bibnamefont {Moore}}, \bibinfo {author} {\bibfnamefont {J.}~\bibnamefont
  {{VanderPlas}}}, \bibinfo {author} {\bibfnamefont {D.}~\bibnamefont
  {Laxalde}}, \bibinfo {author} {\bibfnamefont {J.}~\bibnamefont {Perktold}},
  \bibinfo {author} {\bibfnamefont {R.}~\bibnamefont {Cimrman}}, \bibinfo
  {author} {\bibfnamefont {I.}~\bibnamefont {Henriksen}}, \bibinfo {author}
  {\bibfnamefont {E.~A.}\ \bibnamefont {Quintero}}, \bibinfo {author}
  {\bibfnamefont {C.~R.}\ \bibnamefont {Harris}}, \bibinfo {author}
  {\bibfnamefont {A.~M.}\ \bibnamefont {Archibald}}, \bibinfo {author}
  {\bibfnamefont {A.~H.}\ \bibnamefont {Ribeiro}}, \bibinfo {author}
  {\bibfnamefont {F.}~\bibnamefont {Pedregosa}}, \bibinfo {author}
  {\bibfnamefont {P.}~\bibnamefont {{van Mulbregt}}},\ and\ \bibinfo {author}
  {\bibnamefont {{SciPy 1.0 Contributors}}},\ }\bibfield  {title} {\bibinfo
  {title} {{{SciPy} 1.0: Fundamental Algorithms for Scientific Computing in
  Python}},\ }\href {https://doi.org/10.1038/s41592-019-0686-2} {\bibfield
  {journal} {\bibinfo  {journal} {Nature Methods}\ }\textbf {\bibinfo {volume}
  {17}},\ \bibinfo {pages} {261} (\bibinfo {year} {2020})}\BibitemShut
  {NoStop}%
\bibitem [{\citenamefont {Hunter}(2007)}]{matplotlib}%
  \BibitemOpen
  \bibfield  {author} {\bibinfo {author} {\bibfnamefont {J.~D.}\ \bibnamefont
  {Hunter}},\ }\bibfield  {title} {\bibinfo {title} {Matplotlib: A 2d graphics
  environment},\ }\href {https://doi.org/10.1109/MCSE.2007.55} {\bibfield
  {journal} {\bibinfo  {journal} {Computing in Science \& Engineering}\
  }\textbf {\bibinfo {volume} {9}},\ \bibinfo {pages} {90} (\bibinfo {year}
  {2007})}\BibitemShut {NoStop}%
\bibitem [{\citenamefont {Bradbury}\ \emph {et~al.}(2018)\citenamefont
  {Bradbury}, \citenamefont {Frostig}, \citenamefont {Hawkins}, \citenamefont
  {Johnson}, \citenamefont {Leary}, \citenamefont {Maclaurin}, \citenamefont
  {Necula}, \citenamefont {Paszke}, \citenamefont {Vander{P}las}, \citenamefont
  {Wanderman-{M}ilne},\ and\ \citenamefont {Zhang}}]{jax}%
  \BibitemOpen
  \bibfield  {author} {\bibinfo {author} {\bibfnamefont {J.}~\bibnamefont
  {Bradbury}}, \bibinfo {author} {\bibfnamefont {R.}~\bibnamefont {Frostig}},
  \bibinfo {author} {\bibfnamefont {P.}~\bibnamefont {Hawkins}}, \bibinfo
  {author} {\bibfnamefont {M.~J.}\ \bibnamefont {Johnson}}, \bibinfo {author}
  {\bibfnamefont {C.}~\bibnamefont {Leary}}, \bibinfo {author} {\bibfnamefont
  {D.}~\bibnamefont {Maclaurin}}, \bibinfo {author} {\bibfnamefont
  {G.}~\bibnamefont {Necula}}, \bibinfo {author} {\bibfnamefont
  {A.}~\bibnamefont {Paszke}}, \bibinfo {author} {\bibfnamefont
  {J.}~\bibnamefont {Vander{P}las}}, \bibinfo {author} {\bibfnamefont
  {S.}~\bibnamefont {Wanderman-{M}ilne}},\ and\ \bibinfo {author}
  {\bibfnamefont {Q.}~\bibnamefont {Zhang}},\ }\href
  {http://github.com/google/jax} {\bibinfo {title} {{JAX}: composable
  transformations of {P}ython+{N}um{P}y programs}} (\bibinfo {year}
  {2018})\BibitemShut {NoStop}%
\bibitem [{\citenamefont {Bingham}\ \emph {et~al.}(2019)\citenamefont
  {Bingham}, \citenamefont {Chen}, \citenamefont {Jankowiak}, \citenamefont
  {Obermeyer}, \citenamefont {Pradhan}, \citenamefont {Karaletsos},
  \citenamefont {Singh}, \citenamefont {Szerlip}, \citenamefont {Horsfall},\
  and\ \citenamefont {Goodman}}]{pyro}%
  \BibitemOpen
  \bibfield  {author} {\bibinfo {author} {\bibfnamefont {E.}~\bibnamefont
  {Bingham}}, \bibinfo {author} {\bibfnamefont {J.~P.}\ \bibnamefont {Chen}},
  \bibinfo {author} {\bibfnamefont {M.}~\bibnamefont {Jankowiak}}, \bibinfo
  {author} {\bibfnamefont {F.}~\bibnamefont {Obermeyer}}, \bibinfo {author}
  {\bibfnamefont {N.}~\bibnamefont {Pradhan}}, \bibinfo {author} {\bibfnamefont
  {T.}~\bibnamefont {Karaletsos}}, \bibinfo {author} {\bibfnamefont
  {R.}~\bibnamefont {Singh}}, \bibinfo {author} {\bibfnamefont {P.~A.}\
  \bibnamefont {Szerlip}}, \bibinfo {author} {\bibfnamefont {P.}~\bibnamefont
  {Horsfall}},\ and\ \bibinfo {author} {\bibfnamefont {N.~D.}\ \bibnamefont
  {Goodman}},\ }\bibfield  {title} {\bibinfo {title} {Pyro: Deep universal
  probabilistic programming},\ }\href {http://jmlr.org/papers/v20/18-403.html}
  {\bibfield  {journal} {\bibinfo  {journal} {J. Mach. Learn. Res.}\ }\textbf
  {\bibinfo {volume} {20}},\ \bibinfo {pages} {28:1} (\bibinfo {year}
  {2019})}\BibitemShut {NoStop}%
\bibitem [{\citenamefont {Phan}\ \emph {et~al.}(2019)\citenamefont {Phan},
  \citenamefont {Pradhan},\ and\ \citenamefont {Jankowiak}}]{numpyro}%
  \BibitemOpen
  \bibfield  {author} {\bibinfo {author} {\bibfnamefont {D.}~\bibnamefont
  {Phan}}, \bibinfo {author} {\bibfnamefont {N.}~\bibnamefont {Pradhan}},\ and\
  \bibinfo {author} {\bibfnamefont {M.}~\bibnamefont {Jankowiak}},\ }\bibfield
  {title} {\bibinfo {title} {Composable effects for flexible and accelerated
  probabilistic programming in numpyro},\ }\href@noop {} {\bibfield  {journal}
  {\bibinfo  {journal} {arXiv preprint arXiv:1912.11554}\ } (\bibinfo {year}
  {2019})}\BibitemShut {NoStop}%
\bibitem [{\citenamefont {{Gair}}\ \emph {et~al.}(2022)\citenamefont {{Gair}},
  \citenamefont {{Antonelli}},\ and\ \citenamefont {{Barbieri}}}]{Gair:2022}%
  \BibitemOpen
  \bibfield  {author} {\bibinfo {author} {\bibfnamefont {J.~R.}\ \bibnamefont
  {{Gair}}}, \bibinfo {author} {\bibfnamefont {A.}~\bibnamefont
  {{Antonelli}}},\ and\ \bibinfo {author} {\bibfnamefont {R.}~\bibnamefont
  {{Barbieri}}},\ }\bibfield  {title} {\bibinfo {title} {{A Fisher matrix for
  gravitational-wave population inference}},\ }\href@noop {} {\bibfield
  {journal} {\bibinfo  {journal} {arXiv e-prints}\ ,\ \bibinfo {eid}
  {arXiv:2205.07893}} (\bibinfo {year} {2022})},\ \Eprint
  {https://arxiv.org/abs/2205.07893} {arXiv:2205.07893 [gr-qc]} \BibitemShut
  {NoStop}%
\bibitem [{\citenamefont {Edelman}\ \emph {et~al.}(2022)\citenamefont
  {Edelman}, \citenamefont {Doctor}, \citenamefont {Godfrey},\ and\
  \citenamefont {Farr}}]{Edelman:2022}%
  \BibitemOpen
  \bibfield  {author} {\bibinfo {author} {\bibfnamefont {B.}~\bibnamefont
  {Edelman}}, \bibinfo {author} {\bibfnamefont {Z.}~\bibnamefont {Doctor}},
  \bibinfo {author} {\bibfnamefont {J.}~\bibnamefont {Godfrey}},\ and\ \bibinfo
  {author} {\bibfnamefont {B.}~\bibnamefont {Farr}},\ }\bibfield  {title}
  {\bibinfo {title} {Ain't no mountain high enough: Semiparametric modeling of
  {LIGO}{\textendash}virgo's binary black hole mass distribution},\ }\href
  {https://doi.org/10.3847/1538-4357/ac3667} {\bibfield  {journal} {\bibinfo
  {journal} {The Astrophysical Journal}\ }\textbf {\bibinfo {volume} {924}},\
  \bibinfo {pages} {101} (\bibinfo {year} {2022})}\BibitemShut {NoStop}%
\bibitem [{\citenamefont {Buikema}\ \emph {et~al.}(2020)\citenamefont {Buikema}
  \emph {et~al.}}]{Buikema:2020}%
  \BibitemOpen
  \bibfield  {author} {\bibinfo {author} {\bibfnamefont {A.}~\bibnamefont
  {Buikema}} \emph {et~al.},\ }\bibfield  {title} {\bibinfo {title}
  {Sensitivity and performance of the advanced ligo detectors in the third
  observing run},\ }\href {https://doi.org/10.1103/PhysRevD.102.062003}
  {\bibfield  {journal} {\bibinfo  {journal} {Phys. Rev. D}\ }\textbf {\bibinfo
  {volume} {102}},\ \bibinfo {pages} {062003} (\bibinfo {year}
  {2020})}\BibitemShut {NoStop}%
\bibitem [{\citenamefont {{Abbott}}\ \emph
  {et~al.}(2021{\natexlab{d}})\citenamefont {{Abbott}} \emph
  {et~al.}}]{GWTC-2-pe-samples}%
  \BibitemOpen
  \bibfield  {author} {\bibinfo {author} {\bibfnamefont {R.}~\bibnamefont
  {{Abbott}}} \emph {et~al.},\ }\bibfield  {title} {\bibinfo {title} {Gwtc-2
  data release: Parameter estimation samples and skymaps},\ }\href
  {https://dcc.ligo.org/LIGO-P2000223-v7/public}
  {https://dcc.ligo.org/LIGO-P2000223-v7/public} (\bibinfo {year}
  {2021}{\natexlab{d}})\BibitemShut {NoStop}%
\bibitem [{\citenamefont {{Abbott}}\ \emph
  {et~al.}(2021{\natexlab{e}})\citenamefont {{Abbott}} \emph
  {et~al.}}]{GWTC-2d1-pe-samples}%
  \BibitemOpen
  \bibfield  {author} {\bibinfo {author} {\bibfnamefont {R.}~\bibnamefont
  {{Abbott}}} \emph {et~al.},\ }\bibfield  {title} {\bibinfo {title}
  {{GWTC-2.1: Deep Extended Catalog of Compact Binary Coalescences Observed by
  LIGO and Virgo During the First Half of the Third Observing Run - Parameter
  Estimation Data Release}},\ }\href {https://doi.org/10.5281/zenodo.5117703}
  {10.5281/zenodo.5117703} (\bibinfo {year} {2021}{\natexlab{e}})\BibitemShut
  {NoStop}%
\bibitem [{\citenamefont {Collaboration}\ \emph {et~al.}(2021)\citenamefont
  {Collaboration}, \citenamefont {Collaboration},\ and\ \citenamefont
  {Collaboration}}]{GWTC-3-pe-samples}%
  \BibitemOpen
  \bibfield  {author} {\bibinfo {author} {\bibfnamefont {L.~S.}\ \bibnamefont
  {Collaboration}}, \bibinfo {author} {\bibfnamefont {V.}~\bibnamefont
  {Collaboration}},\ and\ \bibinfo {author} {\bibfnamefont {K.}~\bibnamefont
  {Collaboration}},\ }\bibfield  {title} {\bibinfo {title} {{GWTC-3: Compact
  Binary Coalescences Observed by LIGO and Virgo During the Second Part of the
  Third Observing Run — Parameter estimation data release}},\ }\href
  {https://doi.org/10.5281/zenodo.5546663} {10.5281/zenodo.5546663} (\bibinfo
  {year} {2021})\BibitemShut {NoStop}%
\bibitem [{\citenamefont {{Abbott}}\ \emph {et~al.}(2020)\citenamefont
  {{Abbott}} \emph {et~al.}}]{GWTC-1-pe-samples}%
  \BibitemOpen
  \bibfield  {author} {\bibinfo {author} {\bibfnamefont {R.}~\bibnamefont
  {{Abbott}}} \emph {et~al.},\ }\bibfield  {title} {\bibinfo {title} {Parameter
  estimation sample release for gwtc-1},\ }\href
  {https://dcc.ligo.org/LIGO-P1800370/public}
  {https://dcc.ligo.org/LIGO-P1800370/public} (\bibinfo {year}
  {2020})\BibitemShut {NoStop}%
\end{thebibliography}%


\onecolumngrid

\newpage

\appendix


\section{Perturbative Analysis for Small Anisotropies}
\label{sec:perturbation}

We consider in detail how the data constrain different degrees of freedom in the distribution of merging binaries.
Ref.~\cite{Gair:2022} introduced expressions for the Fisher information matrix that describes how constraining the data is expected to be on average.
We instead consider the constraints from a particular realization of data by perturbing the likelihood directly.

\begin{figure}
    \includegraphics[hsmash=c, width=1.0\textwidth]{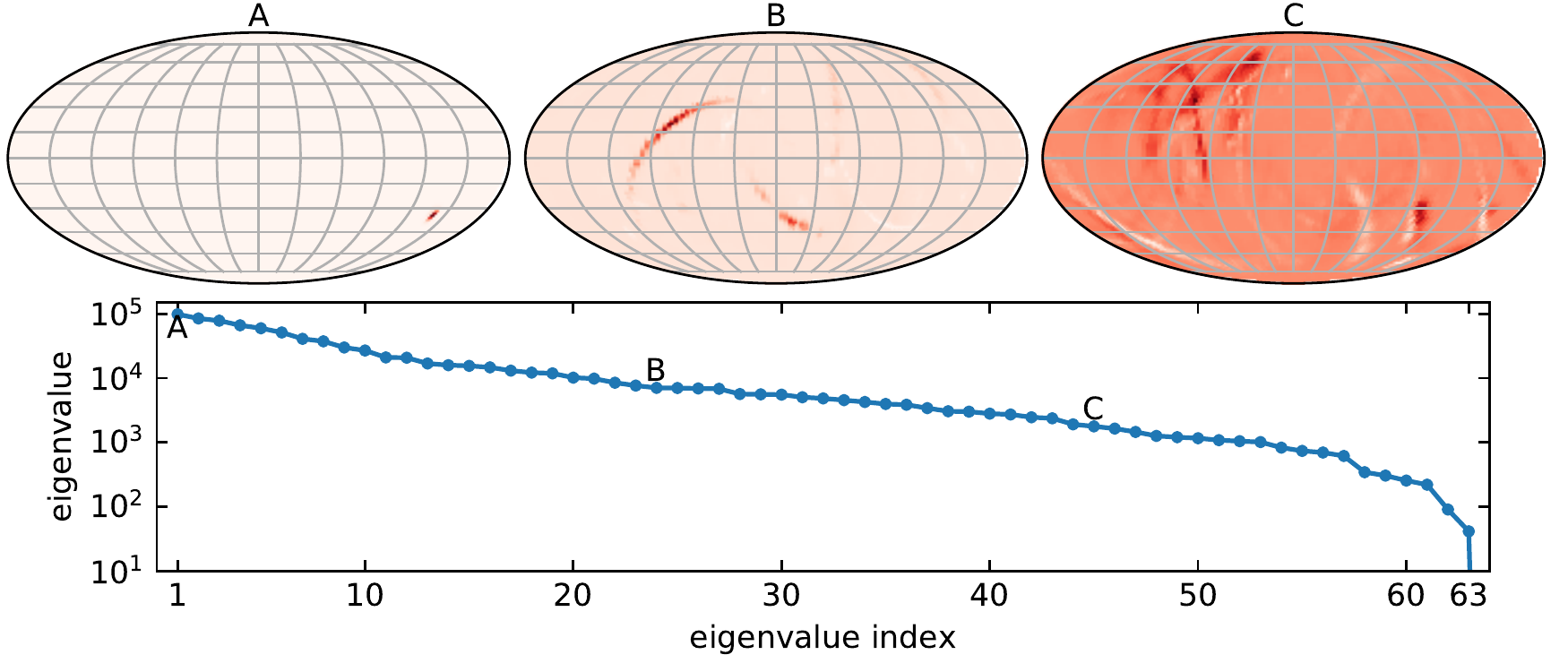}
    \caption{
        Non-vanishing eigenvalues and example eigenvectors from the perturbative analysis of \result{63} events from O3.
        (\emph{top, left to right}) The best-constrained eigenvector and two less constrained eigenvectors.
        (\emph{bottom}) The distribution of eigenvalues, which decays roughly exponentially up to the 63$^\mathrm{rd}$ eigenvalue.
        After that, the eigenvalues for the remaining eigenvectors are many orders of magnitude smaller.
    }
    \label{fig:eigenvalues}
\end{figure}

In particular, we construct a model that perturbs a ``base distribution'' over the single-event parameters $\theta$ by a small amount.
That is, we consider a rate density
\begin{equation}
    \frac{dN}{d\theta} = \mathcal{R} p(\theta|\Lambda) (1 + \eta(\theta))
\end{equation}
with $|\eta| \ll 1 \ \forall \ \theta$.
Inserting this into the inhomogeneous Poisson likelihood~\cite{Loredo:2004, Mandel:2010, Mandel:2019, Gair:2022}, we obtain
\begin{align}
    \ln p(\{D_i\}|\mathcal{R},\Lambda,\eta)
      & = N \ln \left[\mathcal{R}\right] - \mathcal{R} \int d\theta\, P(\mathrm{det}|\theta) p(\theta|\Lambda)(1+\eta(\theta)) +  \sum\limits_i^N \ln \left[\int d\theta\, p(D_i|\theta)p(\theta|\Lambda)(1+\eta(\theta))\right] \nonumber \\
      & = N \ln \left[\mathcal{R}\right] - \mathcal{R} \int d\theta\, P(\mathrm{det}|\theta) p(\theta|\Lambda) + \sum\limits_i^N \ln\left[\int d\theta\,p(D_i|\theta)p(\theta|\Lambda)\right] \nonumber \\
      & \quad \quad - \mathcal{R} \int d\theta\, P(\mathrm{det}|\theta) p(\theta|\Lambda) \eta(\theta) + \sum\limits_i^N \left[ \frac{\int d\theta\,p(D_i|\theta)p(\theta|\Lambda)\eta(\theta)}{\int d\theta\,p(D_i|\theta)p(\theta|\Lambda)} - \frac{1}{2}\left( \frac{\int d\theta\,p(D_i|\theta)p(\theta|\Lambda)\eta(\theta)}{\int d\theta\,p(D_i|\theta)p(\theta|\Lambda)} \right)^2 + \cdots \right]
\end{align}
We recognize
\begin{equation}
    \frac{p(D_i|\theta)p(\theta|\Lambda)}{\int d\theta\,p(D_i|\theta)p(\theta|\Lambda)} = p(\theta|D_i, \Lambda)
\end{equation}
and, retaining only terms up to second order in $\eta$, obtain
\begin{multline}
    \ln p(\{D_i\}|\mathcal{R},\Lambda,\eta) - p(\{D_i\}|\mathcal{R},\Lambda,\eta=0)
        = - \int d\theta \eta(\theta) \left[\mathcal{R} P(\mathrm{det}|\Lambda) p(\theta|\mathrm{det}, \Lambda) - \sum\limits_i^N p(\theta|D_i, \Lambda)\right] \\
        \quad \quad -\frac{1}{2} \int d\theta d\theta^\prime \eta(\theta) \left[ \sum\limits_i^N p(\theta|D_i, \Lambda)p(\theta^\prime|D_i,\Lambda) \right] \eta(\theta^\prime)
\end{multline}
We see, then, that the inhomogeneous Poisson likelihood naturally induces a Gaussian process over small perturbations away from a base distribution.
In particular, the Gaussian process has a positive semi-definite inverse covariance matrix
\begin{equation}
    \mathrm{Cov}^{-1}[\eta(\theta), \eta(\theta^\prime)] = \sum\limits_i^N p(\theta|D_i, \Lambda)p(\theta^\prime|D_i,\Lambda)
\end{equation}
Examining the mean vector in more detail, we see that it is proportional to the difference of two terms.
Taking the maximum likelihood estimate for $\mathcal{R}$ conditioned on $\Lambda$ and $N$, we expect $\mathcal{R} P(\mathrm{det}|\Lambda) = N$.
Therefore, the mean vector is proportional to the difference between the distribution over $\theta$ for detectable sources and the average of the single-event posteriors.
This is intuitively appealing and explains why stacking (adding) posteriors can often produce useful diagnostics even if it is not the correct way to perform a hierarchical inference~\cite{Loredo:2004}.
This is also why the mean of the GRF model in Sec.~\ref{sec:gaussian process} at times displays features reminiscent of individual events.
Some events are well localized relative to the $p(\theta|\mathrm{det},\Lambda)$ and therefore the mean looks as if we simply summed the posteriors of each event (compare Fig.~\ref{fig:GP maps} to Fig.~\ref{fig:overlaid event posteriors}).

We can also consider which types of features are constrained by the data by examining the eigenvectors and eigenvalues of the inverse covariance matrix.
While this analysis is completely general,\footnote{Similar ``semi-parametric'' analyses have been conducted for the mass distribution, although they implemented a spline model for the deviations from the base model~\cite{Edelman:2022}. However, considering the full Gaussian process induced by the likelihood and adopting a conjugate prior may allow for a clearer determination of exactly which features are driven by the data and which are driven by the prior assumptions, particularly when the correlations in the prior span high-dimensional spaces. Furthermore, this type of perturbative analysis can be conducted completely \textit{post hoc} given any base distribution, even semi-parametric or non-parametric representations of $p(\theta|\Lambda)$.} we specialize to the case at hand: an isotropic base distribution with masses, spins, and redshifts distributed as in Table~\ref{tab:fixed population models}.
We also only perturb the distribution over the sky.
Immediately, we see that the only pixels that are constrained by the data are those that have non-zero probability of containing at least one event under the base model ($p(\theta|D_i,\Lambda) \neq 0$ for at least one $D_i$).
It is natural to control these poorly constrained eigenvectors with a Gaussian process prior like the one introduced in Sec.~\ref{sec:gaussian process}.
Indeed, if we fix the base distribution (including the rate) then we can construct a posterior for $\eta$ analytically.

What's more, the magnitude of the inverse-covariance matrix's eigenvalues rapidly decays.
Fig.~\ref{fig:eigenvalues} demonstrates this with our selection of \result{63} events from O3.
As such, we can always expect there to be many eigenvectors that are dominated by the prior for any finite catalog.
Fig.~\ref{fig:eigenvalues} also shows a few eigenvectors.
Typically, the best-constrained eigenvectors resemble well-localized individual events, or just a few pixels on the sky, while less constrained eigenvectors resemble the overlap of multiple events.

\section{Selected Events and Catalog Sensitivity}
\label{sec:events and injections}

\begin{figure}
    \includegraphics[hsmash=c, width=1.0\textwidth]{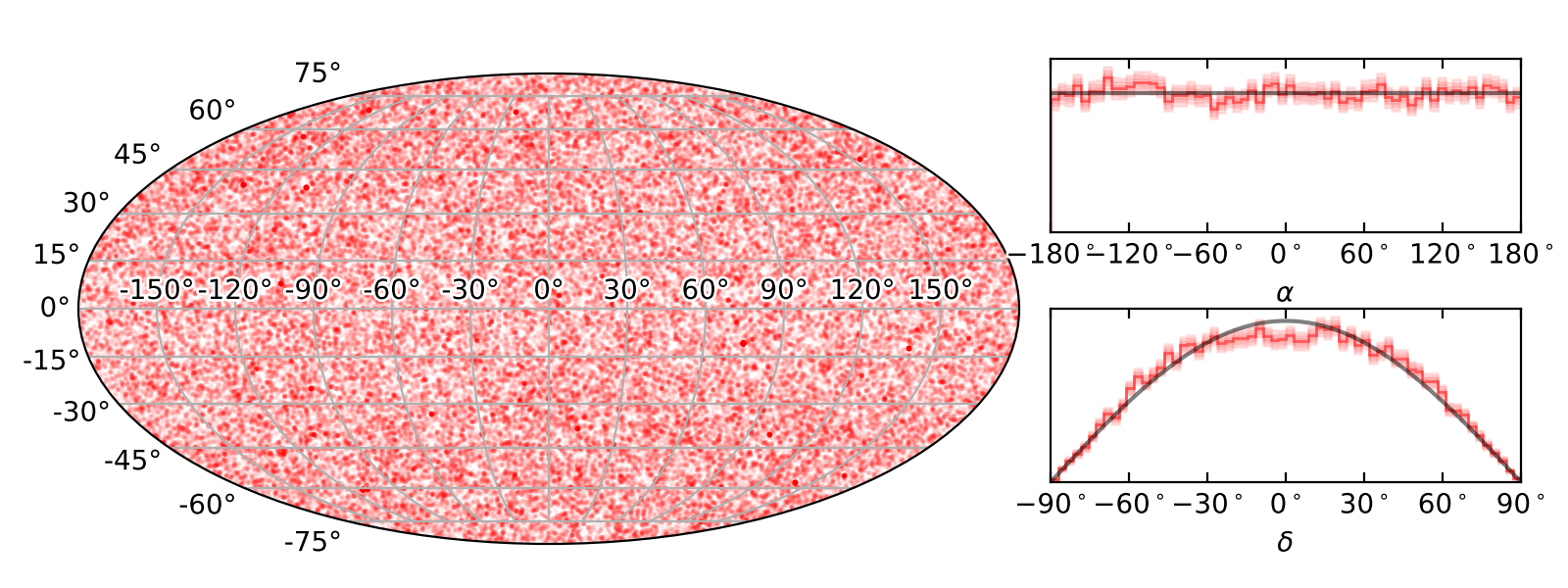}
    \caption{
        Distribution of detected injections from O3~\cite{GWTC-3-injections} assuming an isotropic population with mass, spin, and redshift distributions listed in Table~\ref{tab:fixed population models}.
        (\emph{left}) Scatter plot of detected events in a Mollweide projection.
        (\emph{right}) Marginal distributions of the detected population (\emph{red}) and the isotropic distribution (\emph{black}) for reference.
        Shaded regions correspond to 1-, 2-, and 3-$\sigma$ uncertainty on the detected distribution's marginals from the finite number of injections.
    }
    \label{fig:sensitivity}
\end{figure}

For completeness, we present the estimates of our survey's sensitivity (false alarm rate for any search \result{$\leq 1/\mathrm{year}$}) across the sky assuming the mass, spin, and redshift populations in Table~\ref{tab:fixed population models}.
Fig.~\ref{fig:sensitivity} shows the distribution of detected events from an isotropic source distribution.

While the search sensitivity is nearly uniform, we do observe slight excesses of detected injections from the mid-latitudes and a dearth of detections near the equator, in agreement with Fig. 1 of Ref.~\cite{Chen:2017}.
We also note that the diurnal cycle identified by Ref.~\cite{Chen:2017} during the first observing run (O1) is not apparent in O3.
This is likely due to a combination of factors: the detector duty cycles were higher in O3 than in O1~\cite{Buikema:2020}, and O3 lasted for nearly a full calendar year, thereby washing out the impact of a diurnal cycle (determined by the Earth's orientation to the Sun) when projected on the celestial sphere.

\begin{figure}
    \includegraphics[hsmash=c, width=0.66\textwidth]{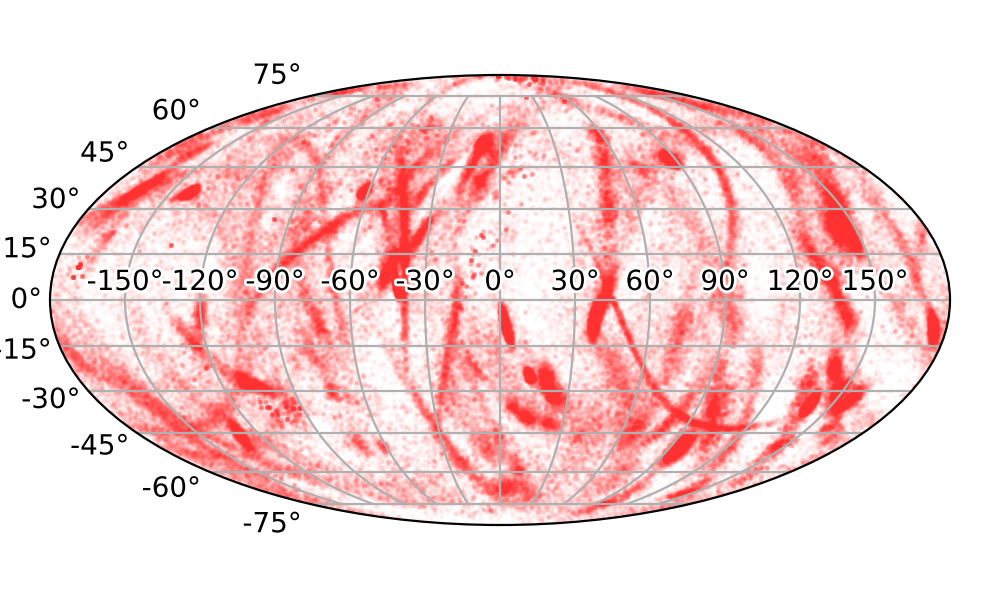}
    \caption{
        Superposition of localization estimates from all \result{63} events considered in this study.
        Each point is a fair draw from one event's posterior assuming an isotropic distribution over the sky and the mass, spin, and redshift distributions in Table~\ref{tab:fixed population models}.
        Darker shading corresponds to areas with many overlapping events or extremely well localized events, and roughly correspond to the hot-spots seen in the posterior means in Fig.~\ref{fig:GP maps}.
        See Table~\ref{tab:single-event parameters} for individual events' localizations.
    }
    \label{fig:overlaid event posteriors}
\end{figure}

Fig.~\ref{fig:overlaid event posteriors} shows the superposition of all the individual events' localizations.
Overdensities of points correspond to hot-spots in Fig.~\ref{fig:GP maps}, as expected based on the analysis in Appendix~\ref{sec:perturbation}.
Table~\ref{tab:single-event parameters} shows the medians and 90\% symmetric credible intervals for the component masses, spins, and redshifts of each of the \result{63} selected events assuming the distributions in Table~\ref{tab:fixed population models} and an isotropic distribution over the sky.

Single-event posterior samples for all events detected during the first half of O3 (O3a; GWTC-2~\cite{GWTC-2}) were taken from Ref.~\cite{GWTC-2-pe-samples}, with the exception of two events first published in GWTC-2.1~\cite{GWTC-2d1}: GW190725\_174728 and GW190805\_211137.
Posterior samples for these events are available in Ref.~\cite{GWTC-2d1-pe-samples}.
Samples for events from the second half of O3 (O3b; GWTC-3~\cite{GWTC-3}) are available within Ref.~\cite{GWTC-3-pe-samples}.

Although only used to benchmark our constraints from O3, posterior samples for events from GWTC-1 are available in Ref.~\cite{GWTC-1-pe-samples}.
Because these samples do not include all the Cartesian spin components and because spin inference largely decouples from localization, we do not include the prior for the spin within analyses of GWTC-1.

\newpage

\begin{table}
    \caption{
        Medians and 90\% symmetric credible regions for each of our \result{63} events assuming an isotropic source distribution and the mass, spin, and redshift distributions from Table~\ref{tab:fixed population models}.
        We also show the smallest area on the sky that contains 90\% of the posterior probability and a scatter plot of posterior samples.
        Brighter colors in scatter plots correspond to higher likelihoods, and each point in a fair draw from the posterior.
    }
    \label{tab:single-event parameters}
    {\renewcommand{\arraystretch}{1.0}
    \begin{tabular}{l c c c c c p{2.5cm}}
        \hline\hline
        \multicolumn{1}{c}{name} & $m_1\,[M_\odot]$ & $m_2\,[M_\odot]$ & $z$ & $D_L\,[\mathrm{Mpc}]$ & $\Delta \Omega_{90\%} [\mathrm{deg}^2]$ & \multicolumn{1}{c}{skymap} \\
        \hline\hline
        GW190408\_181802 & $22.89^{+4.14}_{-2.11}$ & $19.88^{+2.60}_{-3.20}$ & $0.29^{+0.07}_{-0.10}$ & $1540.36^{+427.70}_{-601.78}$ & $179.2$ & \includegraphics[hsmash=r, align=c, width=2.5cm]{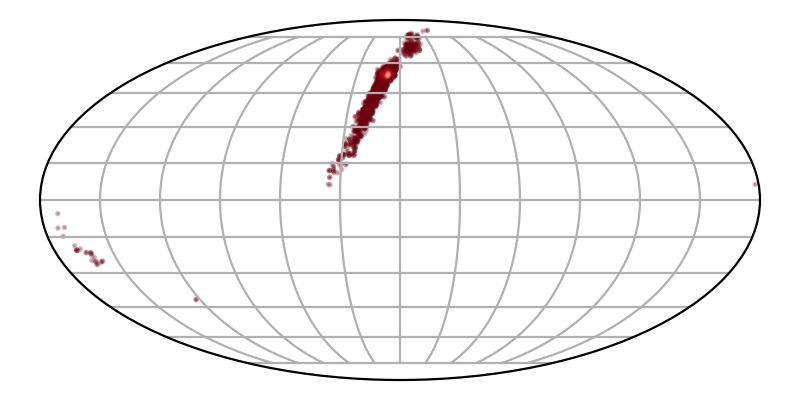} \\
        GW190412 & $26.10^{+5.91}_{-6.26}$ & $9.45^{+3.06}_{-1.53}$ & $0.14^{+0.03}_{-0.04}$ & $694.54^{+174.19}_{-217.11}$ & $83.3$ & \includegraphics[hsmash=r, align=c, width=2.5cm]{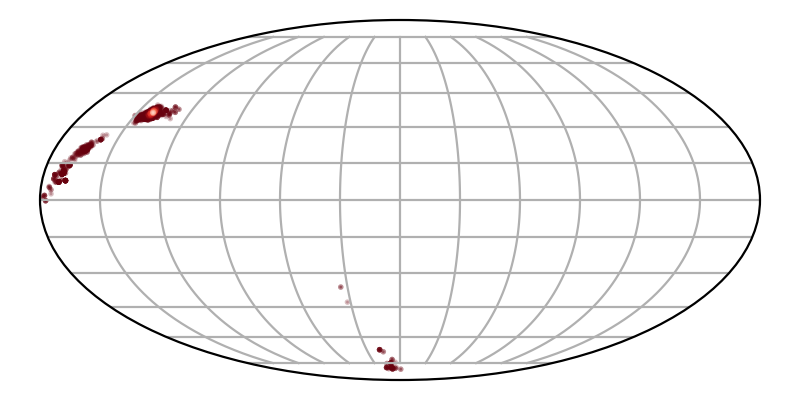} \\
        GW190413\_052954 & $29.96^{+8.25}_{-5.29}$ & $25.65^{+6.03}_{-5.34}$ & $0.60^{+0.26}_{-0.24}$ & $3633.56^{+2041.16}_{-1627.01}$ & $1425.1$ & \includegraphics[hsmash=r, align=c, width=2.5cm]{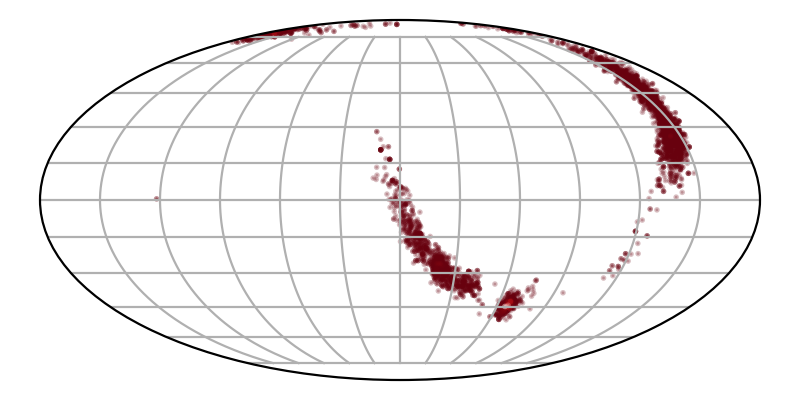} \\
        GW190413\_134308 & $41.19^{+10.54}_{-7.08}$ & $35.24^{+8.75}_{-7.96}$ & $0.74^{+0.30}_{-0.31}$ & $4734.55^{+2403.83}_{-2238.56}$ & $562.5$ & \includegraphics[hsmash=r, align=c, width=2.5cm]{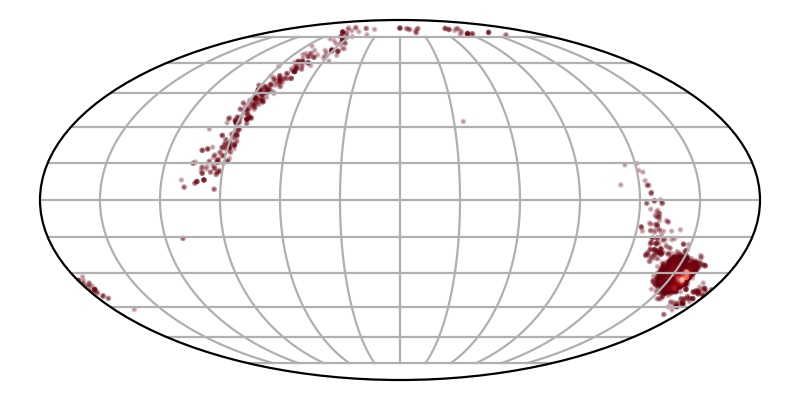} \\
        GW190421\_213856 & $38.18^{+7.54}_{-5.33}$ & $33.24^{+6.73}_{-5.93}$ & $0.50^{+0.19}_{-0.20}$ & $2936.10^{+1363.16}_{-1339.61}$ & $1033.4$ & \includegraphics[hsmash=r, align=c, width=2.5cm]{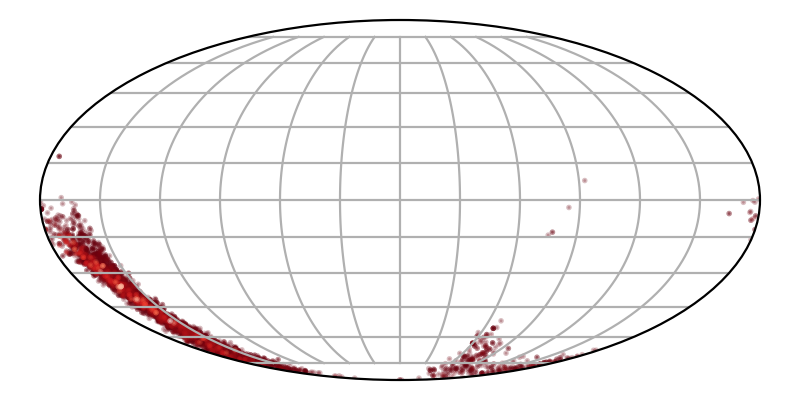} \\
        GW190424\_180648 & $37.30^{+7.63}_{-5.57}$ & $32.78^{+6.24}_{-5.79}$ & $0.41^{+0.21}_{-0.19}$ & $2297.44^{+1483.64}_{-1174.10}$ & $25972.7$ & \includegraphics[hsmash=r, align=c, width=2.5cm]{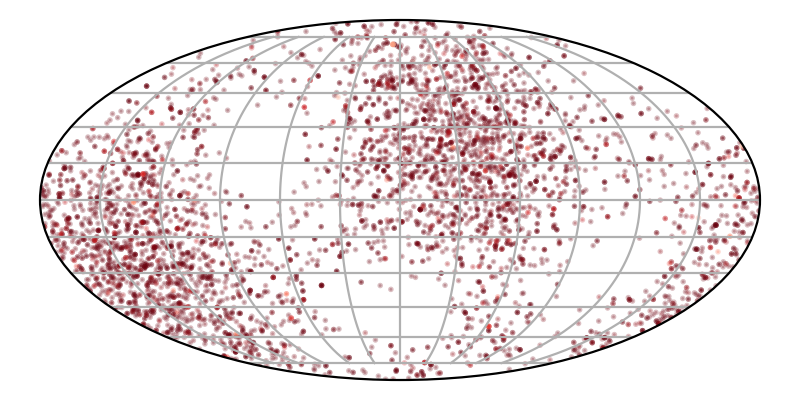} \\
        GW190425 & $1.90^{+0.45}_{-0.23}$ & $1.44^{+0.18}_{-0.25}$ & $0.04^{+0.01}_{-0.02}$ & $161.23^{+65.85}_{-72.41}$ & $8517.3$ & \includegraphics[hsmash=r, align=c, width=2.5cm]{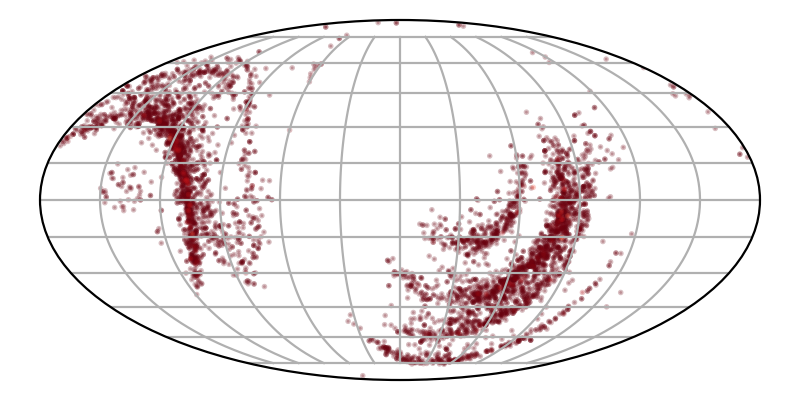} \\
        GW190503\_185404 & $38.26^{+7.94}_{-5.17}$ & $31.87^{+5.56}_{-7.00}$ & $0.29^{+0.11}_{-0.12}$ & $1519.47^{+692.77}_{-695.95}$ & $108.3$ & \includegraphics[hsmash=r, align=c, width=2.5cm]{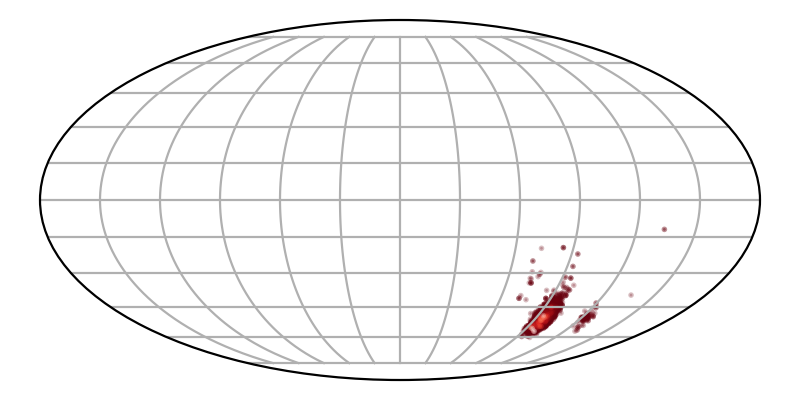} \\
        GW190512\_180714 & $18.86^{+5.65}_{-2.52}$ & $15.12^{+2.33}_{-3.04}$ & $0.28^{+0.09}_{-0.10}$ & $1470.52^{+577.61}_{-595.76}$ & $245.9$ & \includegraphics[hsmash=r, align=c, width=2.5cm]{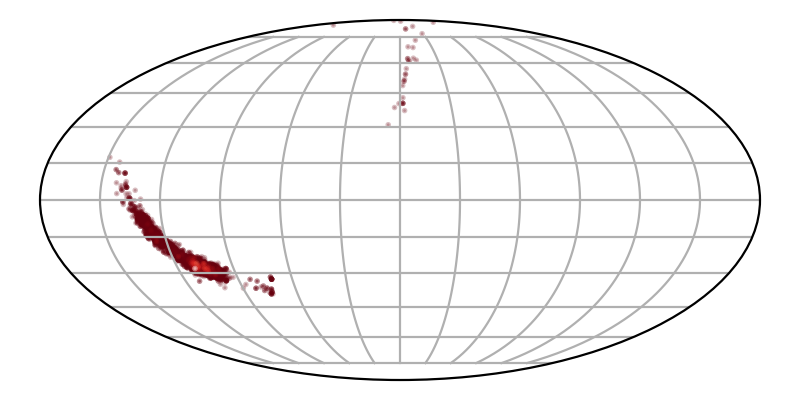} \\
        GW190513\_205428 & $28.33^{+7.97}_{-3.93}$ & $23.24^{+4.30}_{-5.91}$ & $0.38^{+0.13}_{-0.15}$ & $2096.74^{+900.07}_{-941.55}$ & $462.5$ & \includegraphics[hsmash=r, align=c, width=2.5cm]{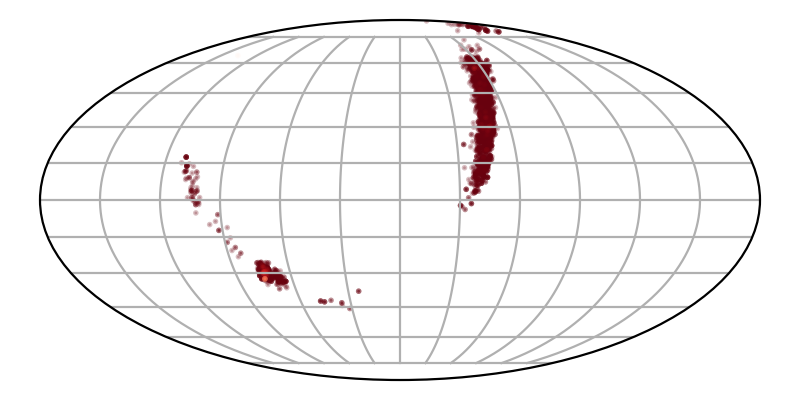} \\
        GW190517\_055101 & $33.08^{+7.38}_{-5.58}$ & $27.98^{+5.03}_{-5.61}$ & $0.36^{+0.23}_{-0.15}$ & $1998.60^{+1622.41}_{-940.58}$ & $429.2$ & \includegraphics[hsmash=r, align=c, width=2.5cm]{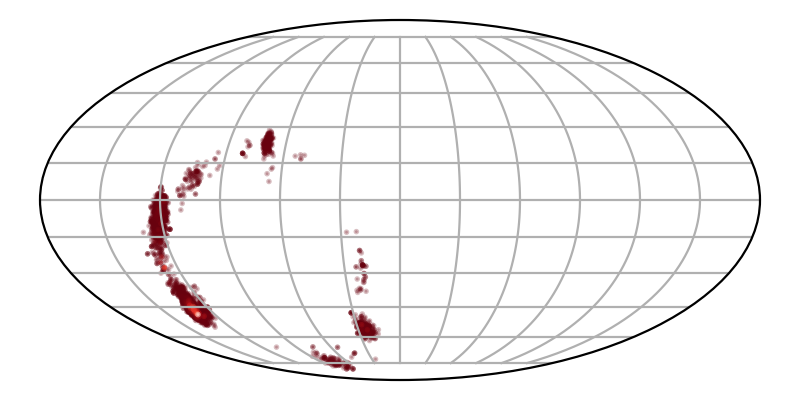} \\
        GW190519\_153544 & $59.04^{+11.70}_{-12.54}$ & $44.10^{+9.68}_{-10.02}$ & $0.49^{+0.30}_{-0.16}$ & $2835.40^{+2212.13}_{-1092.74}$ & $820.9$ & \includegraphics[hsmash=r, align=c, width=2.5cm]{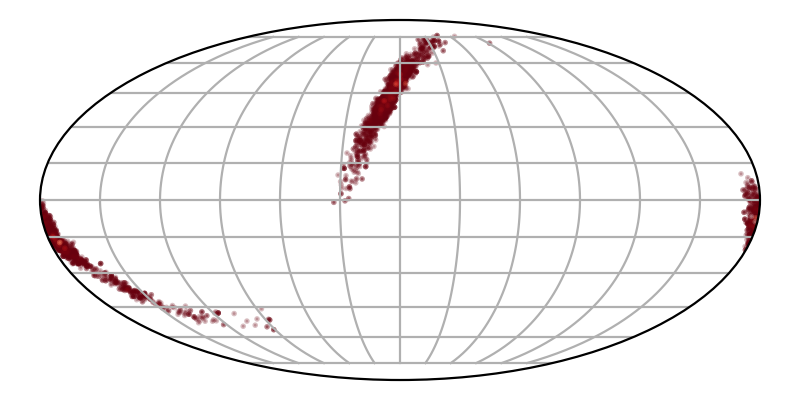} \\
        GW190521 & $82.90^{+19.55}_{-12.45}$ & $70.36^{+17.39}_{-13.88}$ & $0.72^{+0.29}_{-0.28}$ & $4514.92^{+2346.49}_{-2003.55}$ & $887.6$ & \includegraphics[hsmash=r, align=c, width=2.5cm]{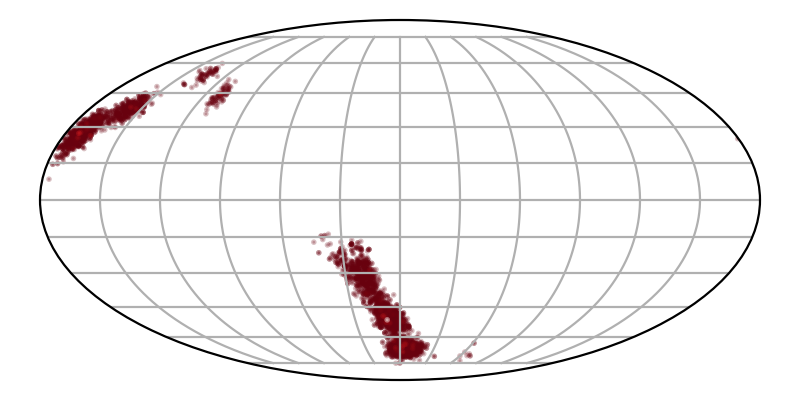} \\
        GW190521\_074359 & $40.25^{+5.25}_{-3.87}$ & $34.61^{+4.23}_{-5.26}$ & $0.24^{+0.07}_{-0.10}$ & $1252.81^{+405.74}_{-552.87}$ & $491.7$ & \includegraphics[hsmash=r, align=c, width=2.5cm]{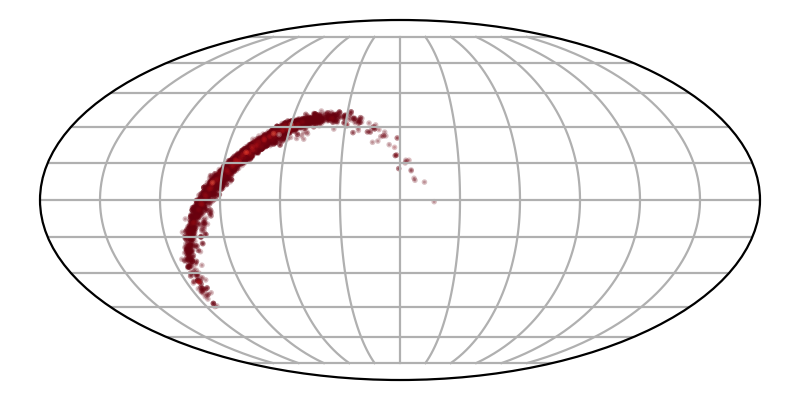} \\
        GW190527\_092055 & $30.70^{+8.05}_{-5.44}$ & $25.67^{+6.36}_{-6.12}$ & $0.44^{+0.27}_{-0.20}$ & $2506.95^{+1986.41}_{-1268.74}$ & $3329.4$ & \includegraphics[hsmash=r, align=c, width=2.5cm]{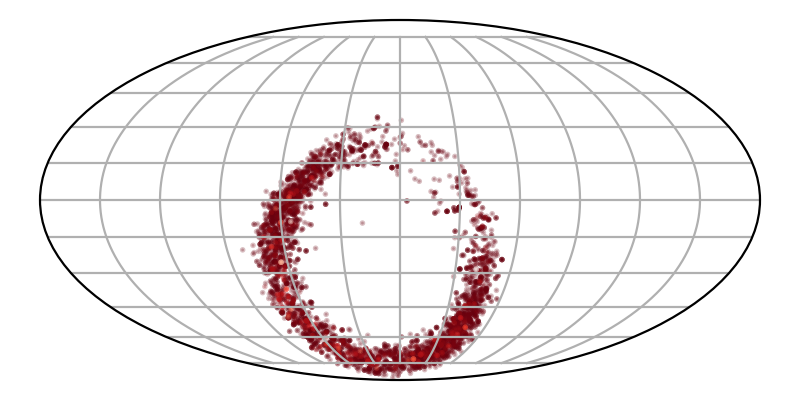} \\
        GW190602\_175927 & $61.27^{+12.96}_{-9.71}$ & $51.57^{+10.20}_{-11.95}$ & $0.51^{+0.25}_{-0.18}$ & $2982.27^{+1881.03}_{-1207.87}$ & $725.1$ & \includegraphics[hsmash=r, align=c, width=2.5cm]{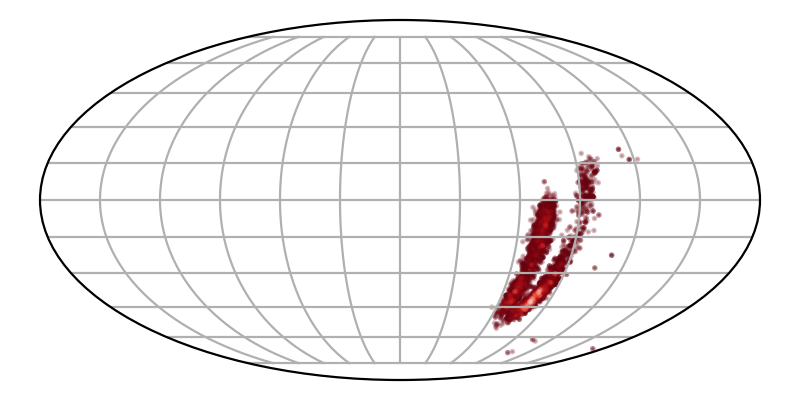} \\
        GW190620\_030421 & $48.09^{+12.52}_{-7.76}$ & $39.74^{+8.43}_{-9.34}$ & $0.54^{+0.21}_{-0.21}$ & $3185.49^{+1588.11}_{-1422.76}$ & $6158.8$ & \includegraphics[hsmash=r, align=c, width=2.5cm]{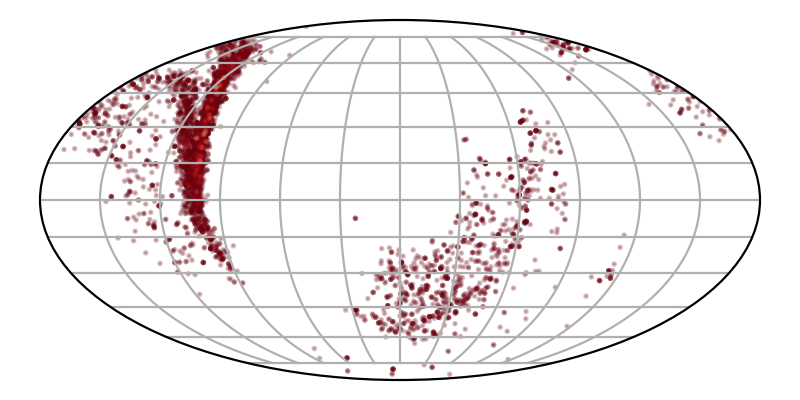} \\
        \hline
    \end{tabular}
    }
\end{table}

\begin{table}
    {\renewcommand{\arraystretch}{1.0}
    \begin{tabular}{l c c c c c p{2.5cm}}
        \hline\hline
        \multicolumn{1}{c}{name} & $m_1\,[M_\odot]$ & $m_2\,[M_\odot]$ & $z$ & $D_L\,[\mathrm{Mpc}]$ & $\Delta \Omega_{90\%} [\mathrm{deg}^2]$ & \multicolumn{1}{c}{skymap} \\
        \hline\hline
        GW190630\_185205 & $31.73^{+5.66}_{-3.57}$ & $26.65^{+3.24}_{-4.89}$ & $0.17^{+0.11}_{-0.06}$ & $837.61^{+619.59}_{-336.30}$ & $1558.4$ & \includegraphics[hsmash=r, align=c, width=2.5cm]{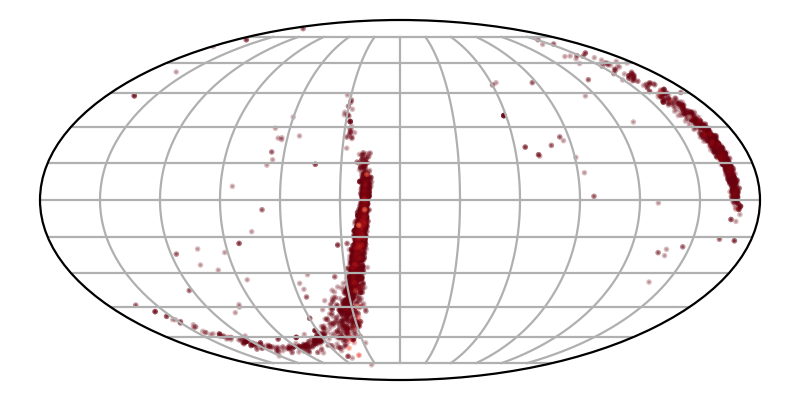} \\
        GW190701\_203306 & $49.52^{+8.74}_{-5.85}$ & $43.15^{+6.51}_{-8.13}$ & $0.38^{+0.11}_{-0.11}$ & $2096.04^{+738.49}_{-712.16}$ & $66.7$ & \includegraphics[hsmash=r, align=c, width=2.5cm]{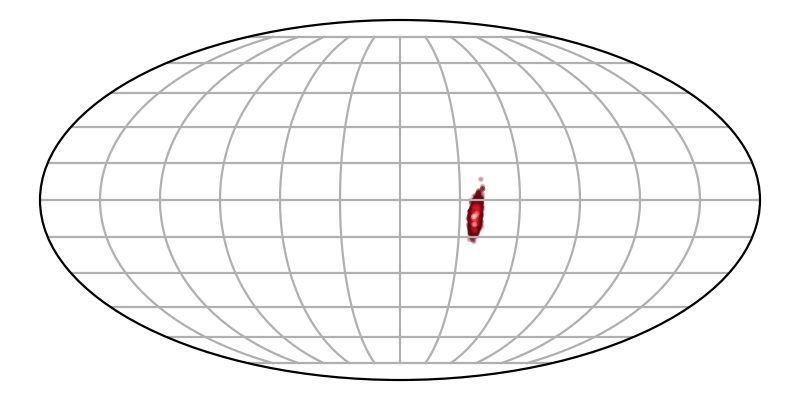} \\
        GW190706\_222641 & $55.49^{+15.88}_{-10.15}$ & $43.59^{+10.47}_{-10.37}$ & $0.82^{+0.29}_{-0.31}$ & $5335.31^{+2430.24}_{-2363.03}$ & $620.9$ & \includegraphics[hsmash=r, align=c, width=2.5cm]{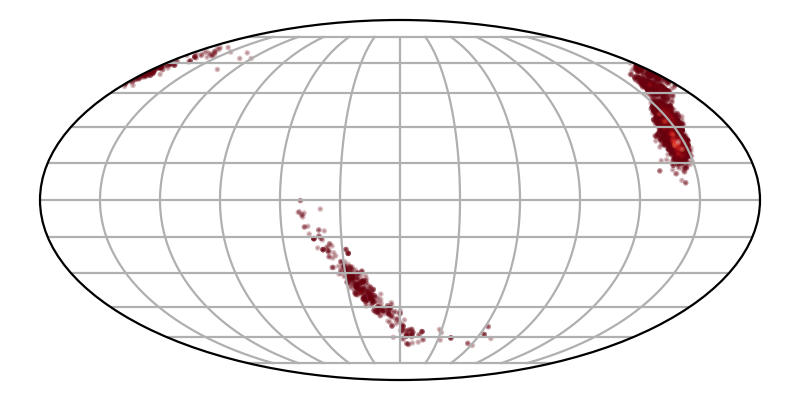} \\
        GW190707\_093326 & $10.58^{+1.95}_{-0.97}$ & $9.06^{+0.90}_{-1.32}$ & $0.17^{+0.06}_{-0.08}$ & $857.19^{+333.14}_{-427.25}$ & $1416.8$ & \includegraphics[hsmash=r, align=c, width=2.5cm]{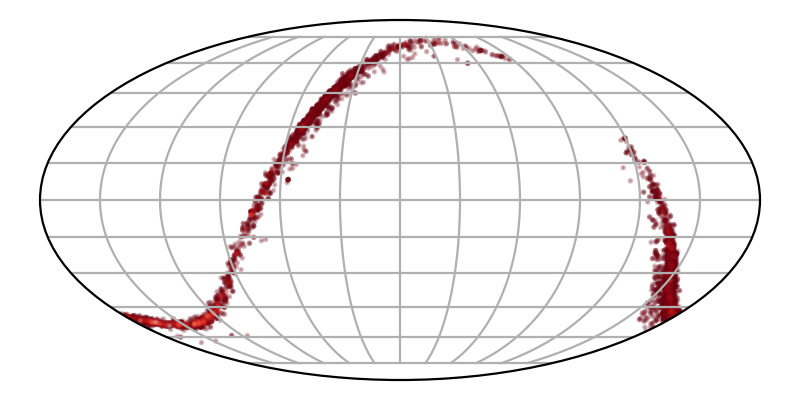} \\
        GW190708\_232457 & $16.31^{+2.76}_{-1.32}$ & $14.20^{+1.37}_{-2.16}$ & $0.17^{+0.06}_{-0.07}$ & $876.65^{+334.85}_{-383.05}$ & $10021.6$ & \includegraphics[hsmash=r, align=c, width=2.5cm]{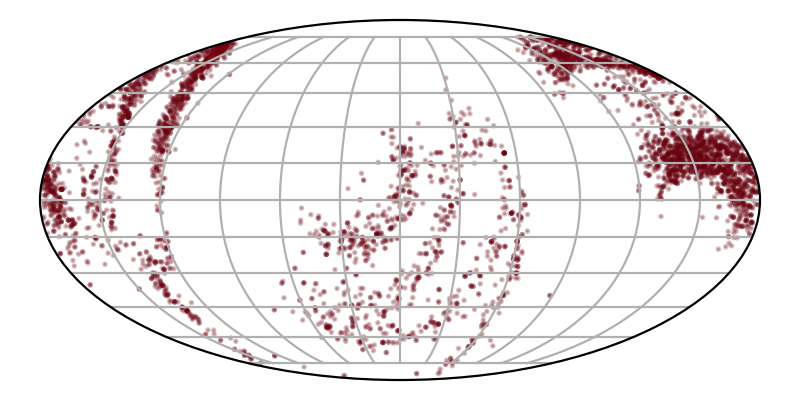} \\
        GW190719\_215514 & $28.97^{+9.29}_{-5.41}$ & $24.50^{+6.37}_{-5.60}$ & $0.64^{+0.31}_{-0.28}$ & $3942.61^{+2475.88}_{-1938.22}$ & $2579.4$ & \includegraphics[hsmash=r, align=c, width=2.5cm]{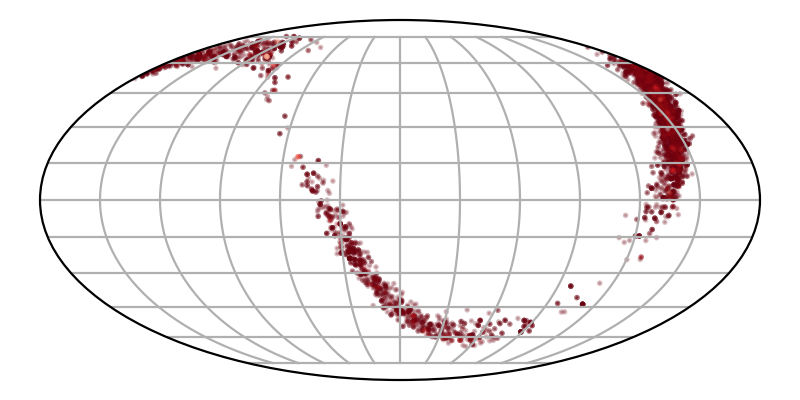} \\
        GW190720\_000836 & $10.94^{+2.69}_{-1.29}$ & $9.21^{+1.07}_{-1.71}$ & $0.18^{+0.11}_{-0.07}$ & $882.93^{+668.84}_{-372.22}$ & $616.7$ & \includegraphics[hsmash=r, align=c, width=2.5cm]{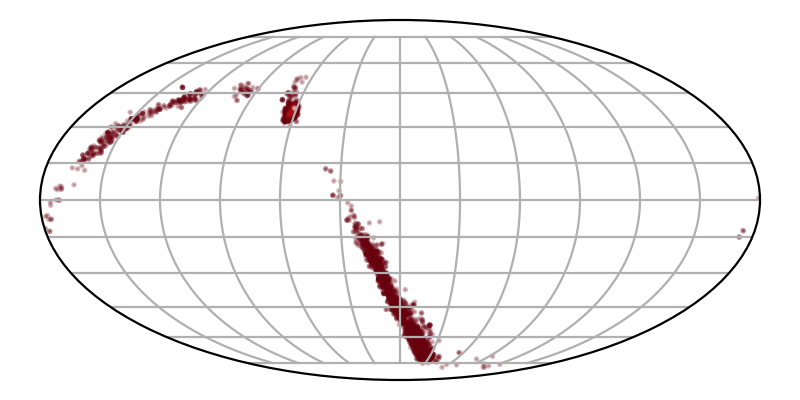} \\
        GW190725\_174728 & $9.32^{+2.69}_{-1.01}$ & $7.88^{+0.99}_{-1.64}$ & $0.20^{+0.10}_{-0.09}$ & $1034.40^{+598.20}_{-476.09}$ & $2162.7$ & \includegraphics[hsmash=r, align=c, width=2.5cm]{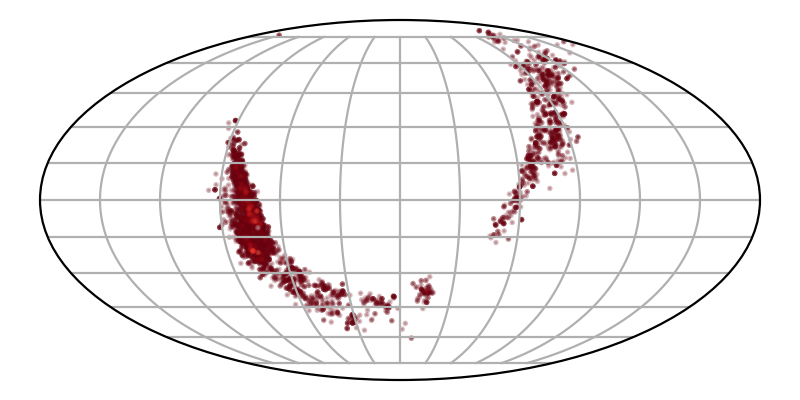} \\
        GW190727\_060333 & $35.37^{+6.95}_{-4.66}$ & $30.87^{+5.82}_{-5.50}$ & $0.56^{+0.20}_{-0.22}$ & $3367.62^{+1488.39}_{-1502.40}$ & $741.7$ & \includegraphics[hsmash=r, align=c, width=2.5cm]{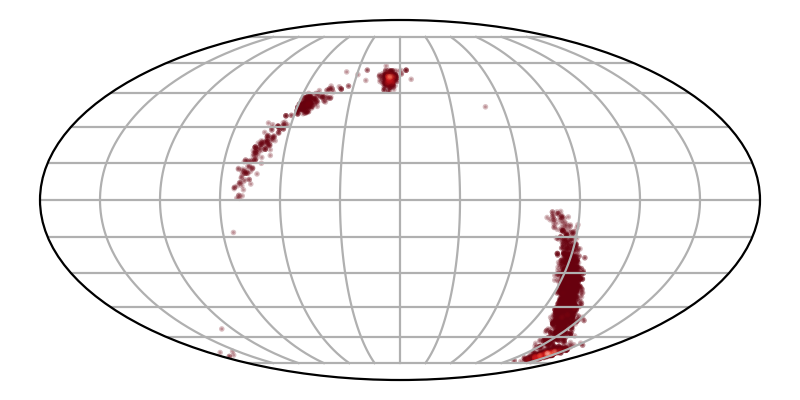} \\
        GW190728\_064510 & $10.74^{+2.26}_{-0.85}$ & $9.25^{+0.87}_{-1.49}$ & $0.17^{+0.05}_{-0.07}$ & $856.36^{+271.90}_{-355.00}$ & $325.0$ & \includegraphics[hsmash=r, align=c, width=2.5cm]{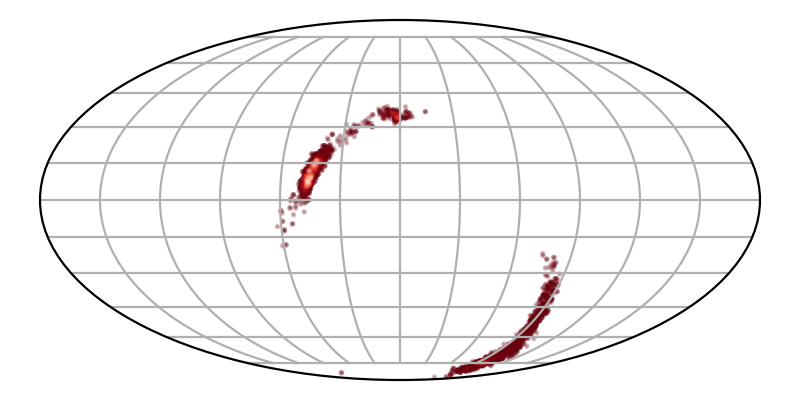} \\
        GW190731\_140936 & $36.95^{+9.29}_{-6.92}$ & $31.37^{+7.43}_{-7.46}$ & $0.58^{+0.31}_{-0.26}$ & $3535.48^{+2373.90}_{-1769.26}$ & $3091.9$ & \includegraphics[hsmash=r, align=c, width=2.5cm]{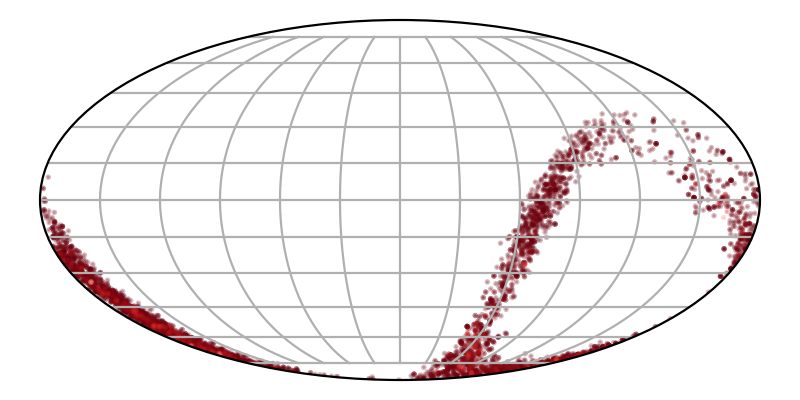} \\
        GW190803\_022701 & $33.81^{+7.51}_{-5.18}$ & $29.19^{+6.34}_{-5.74}$ & $0.57^{+0.24}_{-0.24}$ & $3412.79^{+1822.16}_{-1610.66}$ & $1458.4$ & \includegraphics[hsmash=r, align=c, width=2.5cm]{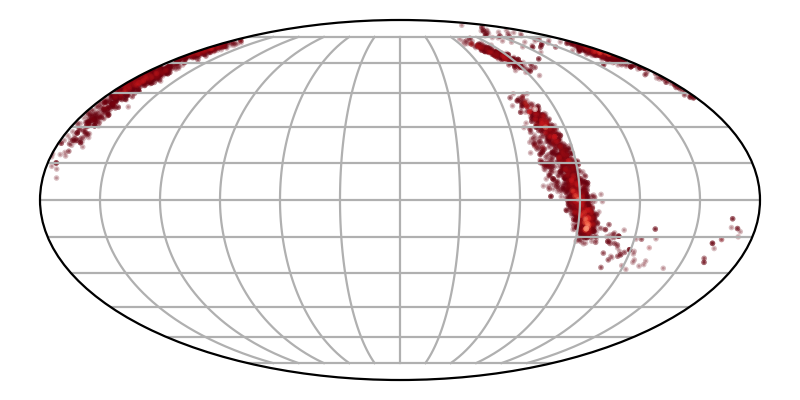} \\
        GW190805\_211137 & $41.34^{+11.95}_{-8.35}$ & $34.74^{+9.88}_{-8.26}$ & $0.89^{+0.44}_{-0.39}$ & $5914.41^{+3760.37}_{-2975.96}$ & $3533.6$ & \includegraphics[hsmash=r, align=c, width=2.5cm]{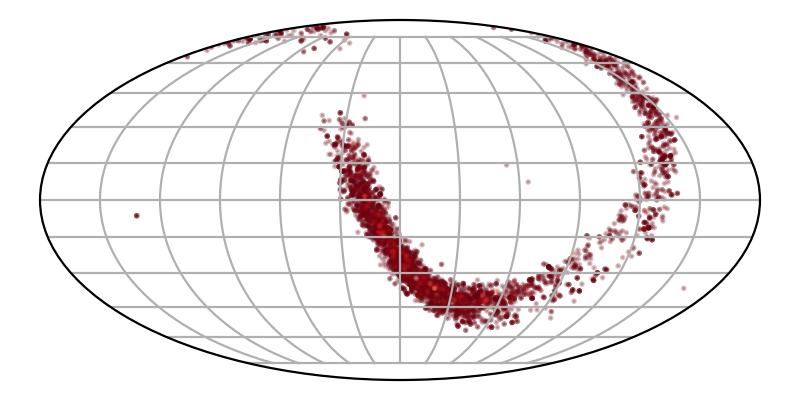} \\
        GW190814 & $23.41^{+1.12}_{-0.92}$ & $2.57^{+0.08}_{-0.09}$ & $0.05^{+0.01}_{-0.01}$ & $245.77^{+40.13}_{-43.26}$ & $29.2$ & \includegraphics[hsmash=r, align=c, width=2.5cm]{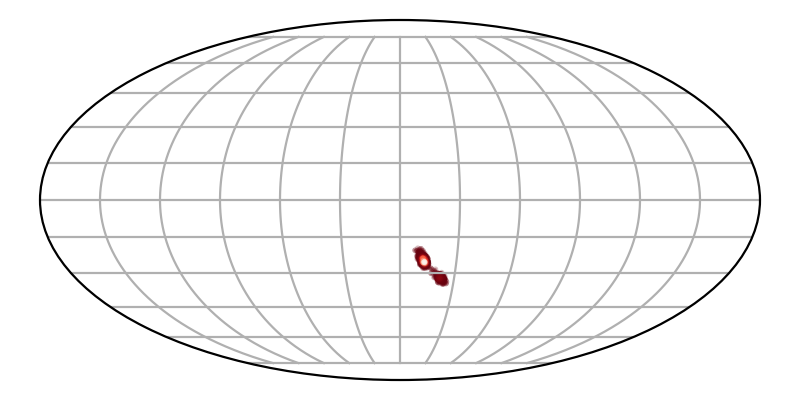} \\
        GW190828\_063405 & $30.67^{+4.91}_{-3.15}$ & $27.12^{+4.26}_{-3.70}$ & $0.38^{+0.10}_{-0.15}$ & $2128.25^{+668.61}_{-928.63}$ & $475.0$ & \includegraphics[hsmash=r, align=c, width=2.5cm]{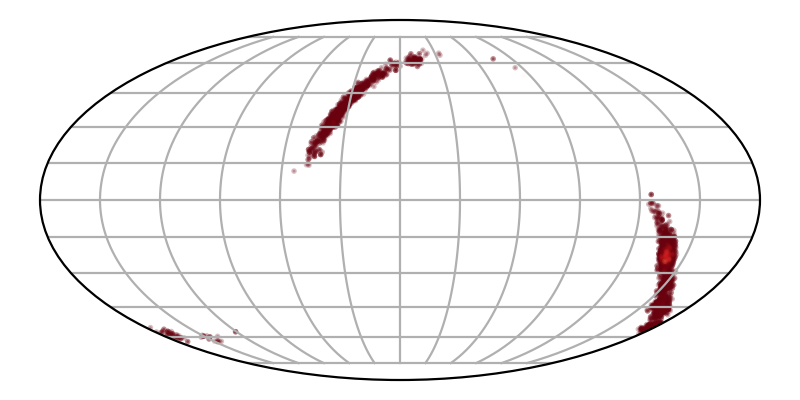} \\
        GW190828\_065509 & $17.39^{+6.81}_{-2.57}$ & $13.44^{+2.19}_{-3.22}$ & $0.32^{+0.10}_{-0.11}$ & $1714.51^{+658.82}_{-682.49}$ & $745.9$ & \includegraphics[hsmash=r, align=c, width=2.5cm]{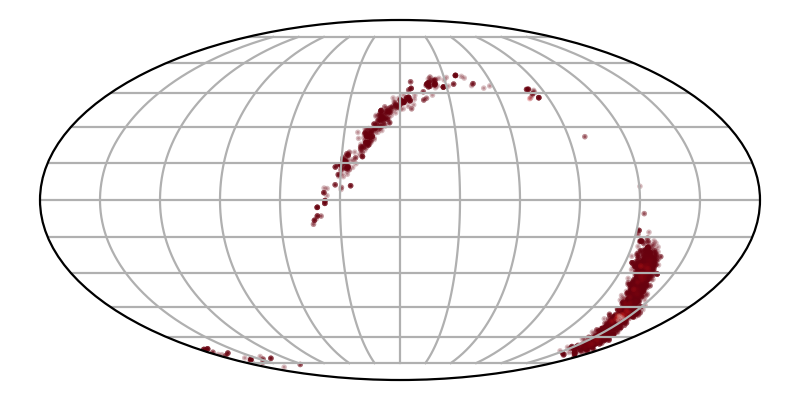} \\
        GW190910\_112807 & $41.62^{+6.16}_{-5.53}$ & $36.51^{+5.31}_{-5.82}$ & $0.29^{+0.16}_{-0.10}$ & $1535.74^{+1021.35}_{-604.01}$ & $9838.2$ & \includegraphics[hsmash=r, align=c, width=2.5cm]{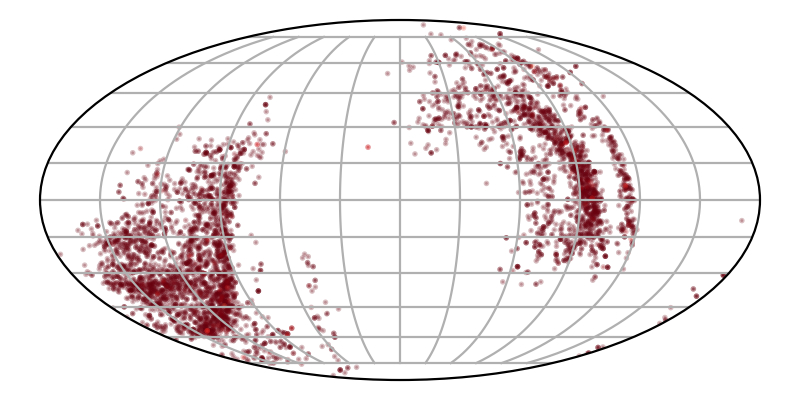} \\
        GW190915\_235702 & $30.84^{+5.22}_{-3.61}$ & $26.76^{+3.91}_{-4.28}$ & $0.31^{+0.10}_{-0.11}$ & $1699.15^{+678.11}_{-649.29}$ & $362.5$ & \includegraphics[hsmash=r, align=c, width=2.5cm]{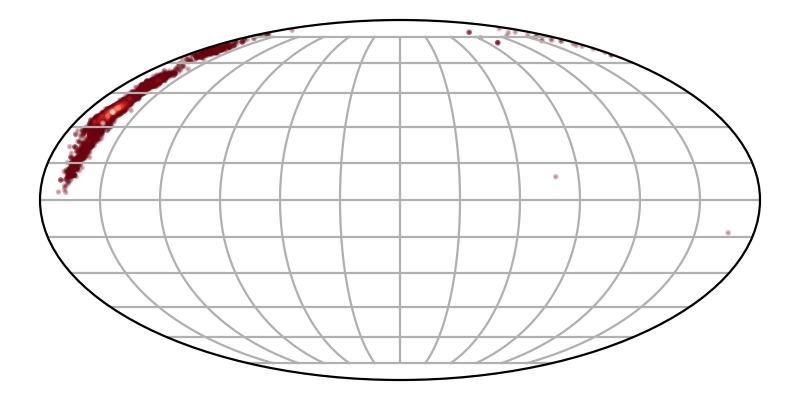} \\
        \hline
    \end{tabular}
    }
\end{table}

\begin{table}
    {\renewcommand{\arraystretch}{1.0}
    \begin{tabular}{l c c c c c p{2.5cm}}
        \hline\hline
        \multicolumn{1}{c}{name} & $m_1\,[M_\odot]$ & $m_2\,[M_\odot]$ & $z$ & $D_L\,[\mathrm{Mpc}]$ & $\Delta \Omega_{90\%} [\mathrm{deg}^2]$ & \multicolumn{1}{c}{skymap} \\
        \hline\hline
        GW190924\_021846 & $7.20^{+1.90}_{-0.59}$ & $6.06^{+0.56}_{-1.11}$ & $0.12^{+0.04}_{-0.05}$ & $572.77^{+215.26}_{-229.08}$ & $358.4$ & \includegraphics[hsmash=r, align=c, width=2.5cm]{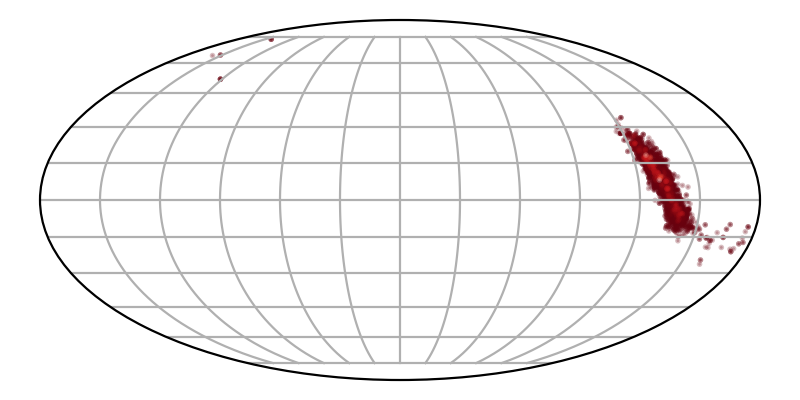} \\
        GW190929\_012149 & $50.07^{+17.25}_{-11.85}$ & $35.38^{+11.44}_{-11.85}$ & $0.71^{+0.35}_{-0.30}$ & $4440.24^{+2803.54}_{-2138.08}$ & $1954.3$ & \includegraphics[hsmash=r, align=c, width=2.5cm]{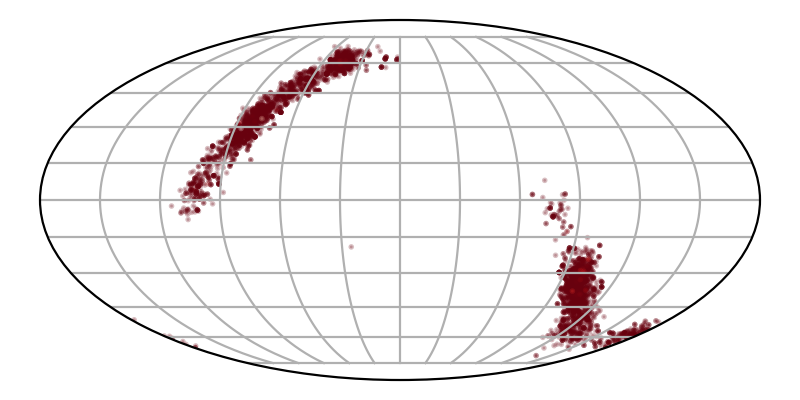} \\
        GW190930\_133541 & $10.58^{+2.24}_{-0.95}$ & $9.02^{+0.91}_{-1.48}$ & $0.15^{+0.06}_{-0.06}$ & $758.74^{+351.63}_{-328.50}$ & $1683.5$ & \includegraphics[hsmash=r, align=c, width=2.5cm]{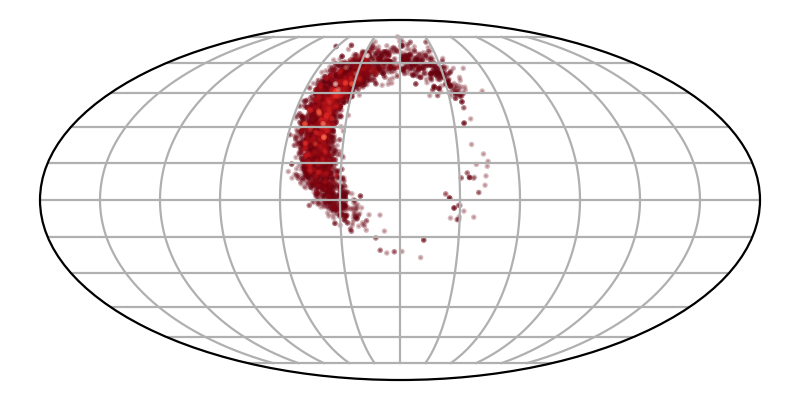} \\
        GW191103\_012549 & $10.33^{+2.24}_{-0.97}$ & $8.98^{+1.01}_{-1.61}$ & $0.19^{+0.08}_{-0.09}$ & $968.08^{+488.21}_{-468.99}$ & $2558.5$ & \includegraphics[hsmash=r, align=c, width=2.5cm]{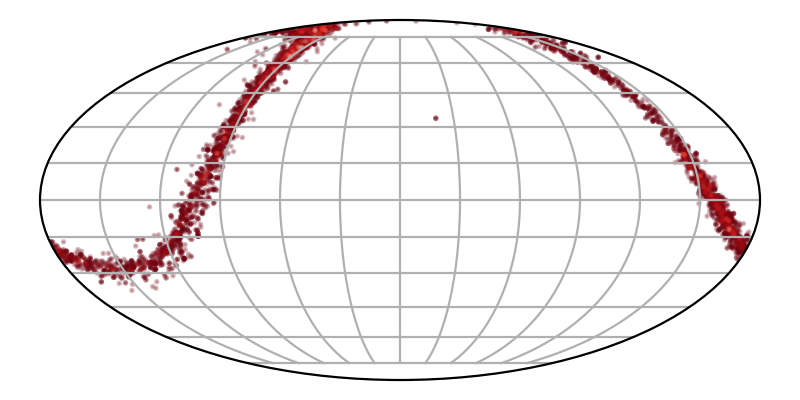} \\
        GW191105\_143521 & $9.69^{+1.96}_{-0.86}$ & $8.41^{+0.89}_{-1.42}$ & $0.22^{+0.07}_{-0.09}$ & $1147.37^{+406.73}_{-479.83}$ & $820.9$ & \includegraphics[hsmash=r, align=c, width=2.5cm]{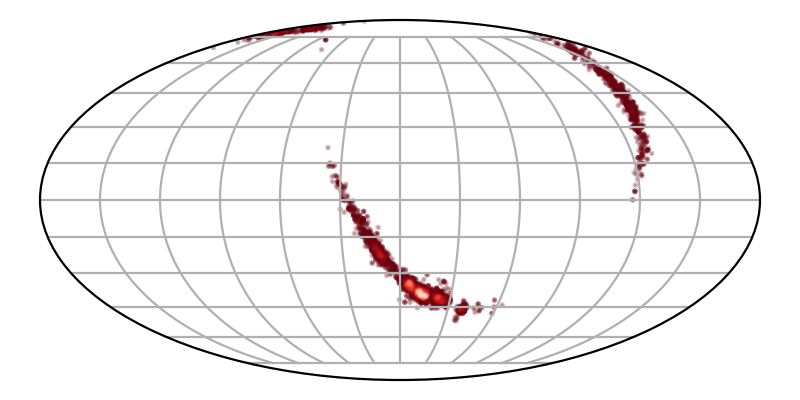} \\
        GW191109\_010717 & $60.18^{+10.44}_{-10.59}$ & $48.18^{+12.55}_{-11.35}$ & $0.27^{+0.22}_{-0.13}$ & $1409.53^{+1453.76}_{-760.26}$ & $1579.3$ & \includegraphics[hsmash=r, align=c, width=2.5cm]{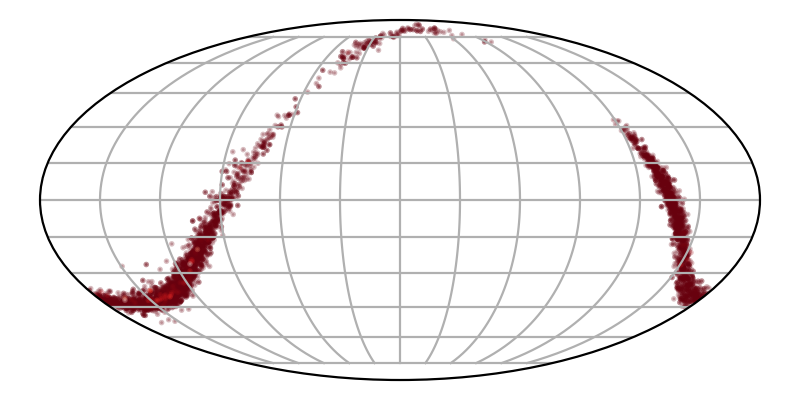} \\
        GW191127\_050227 & $32.60^{+13.87}_{-6.59}$ & $28.11^{+10.78}_{-7.04}$ & $0.52^{+0.42}_{-0.33}$ & $3041.05^{+3244.28}_{-2094.05}$ & $1029.2$ & \includegraphics[hsmash=r, align=c, width=2.5cm]{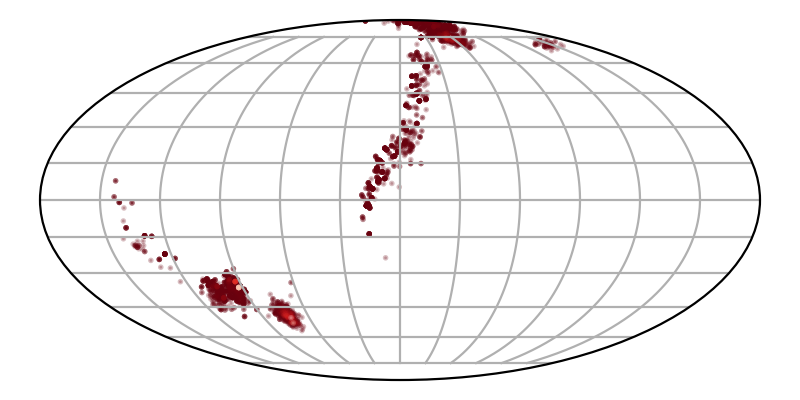} \\
        GW191129\_134029 & $9.18^{+2.47}_{-0.79}$ & $7.77^{+0.85}_{-1.58}$ & $0.15^{+0.05}_{-0.06}$ & $765.83^{+269.14}_{-328.52}$ & $1333.4$ & \includegraphics[hsmash=r, align=c, width=2.5cm]{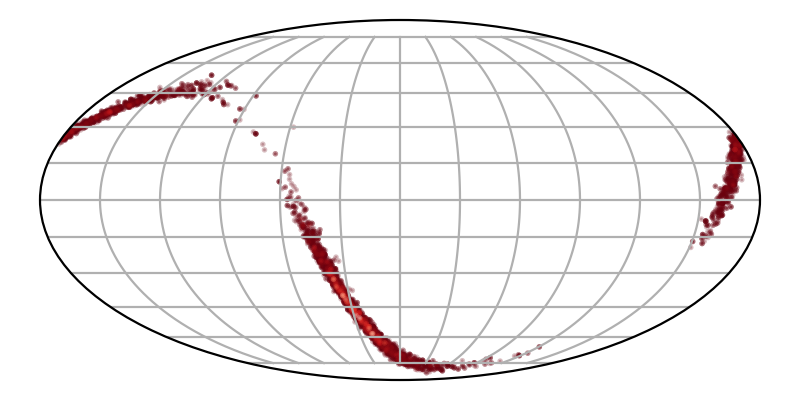} \\
        GW191204\_171526 & $10.82^{+2.41}_{-0.95}$ & $9.03^{+0.85}_{-1.54}$ & $0.13^{+0.04}_{-0.05}$ & $632.63^{+192.64}_{-234.48}$ & $329.2$ & \includegraphics[hsmash=r, align=c, width=2.5cm]{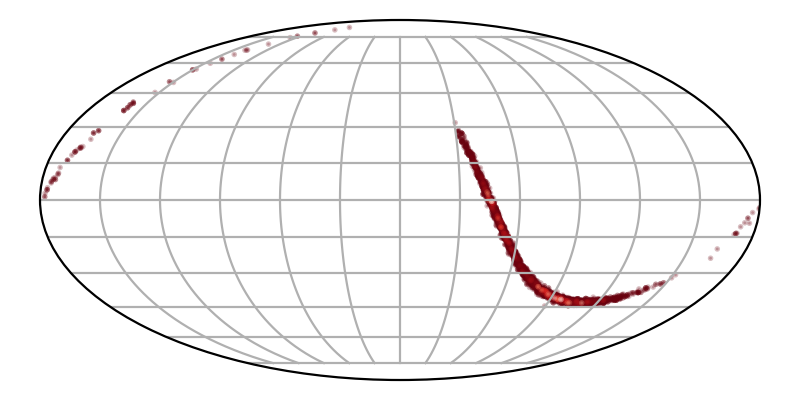} \\
        GW191215\_223052 & $22.85^{+4.53}_{-2.76}$ & $19.76^{+2.98}_{-3.37}$ & $0.34^{+0.14}_{-0.14}$ & $1879.79^{+940.85}_{-851.88}$ & $562.5$ & \includegraphics[hsmash=r, align=c, width=2.5cm]{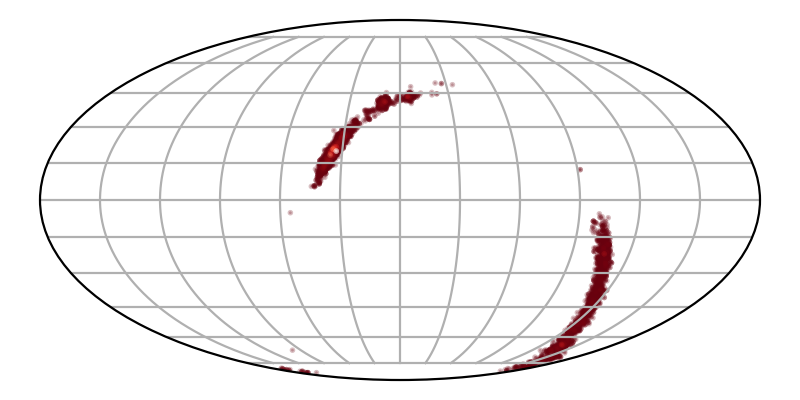} \\
        GW191216\_213338 & $10.40^{+2.76}_{-0.83}$ & $8.82^{+0.73}_{-1.70}$ & $0.07^{+0.02}_{-0.03}$ & $338.99^{+113.86}_{-133.57}$ & $241.7$ & \includegraphics[hsmash=r, align=c, width=2.5cm]{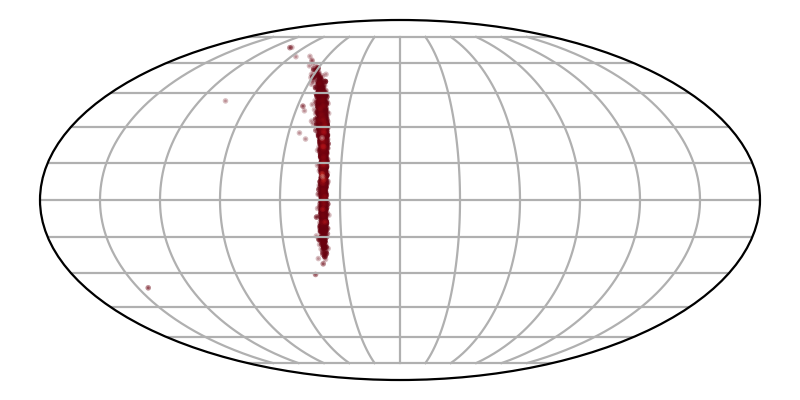} \\
        GW191222\_033537 & $41.82^{+8.59}_{-6.58}$ & $36.40^{+7.70}_{-7.21}$ & $0.50^{+0.24}_{-0.24}$ & $2903.37^{+1792.76}_{-1574.71}$ & $1991.8$ & \includegraphics[hsmash=r, align=c, width=2.5cm]{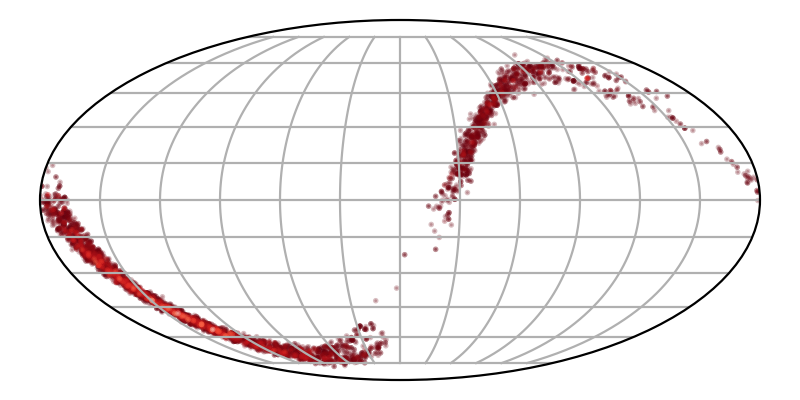} \\
        GW191230\_180458 & $44.21^{+9.88}_{-7.17}$ & $38.46^{+8.55}_{-7.67}$ & $0.72^{+0.25}_{-0.27}$ & $4529.72^{+2043.75}_{-1969.25}$ & $1100.1$ & \includegraphics[hsmash=r, align=c, width=2.5cm]{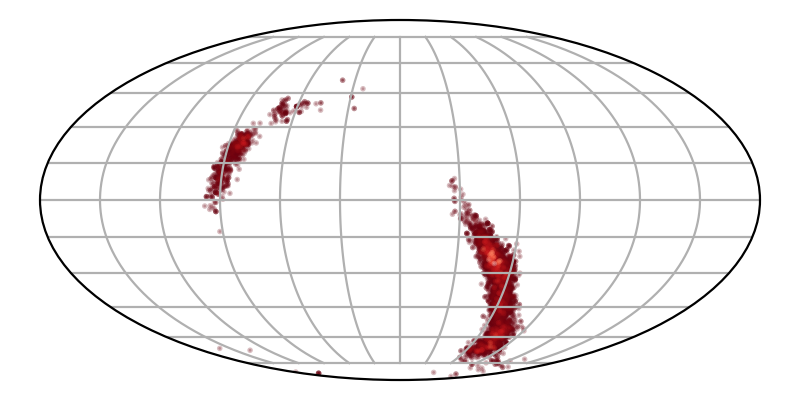} \\
        GW200105\_162426 & $9.13^{+2.40}_{-1.57}$ & $1.89^{+0.29}_{-0.31}$ & $0.06^{+0.02}_{-0.02}$ & $271.86^{+113.79}_{-112.24}$ & $7496.4$ & \includegraphics[hsmash=r, align=c, width=2.5cm]{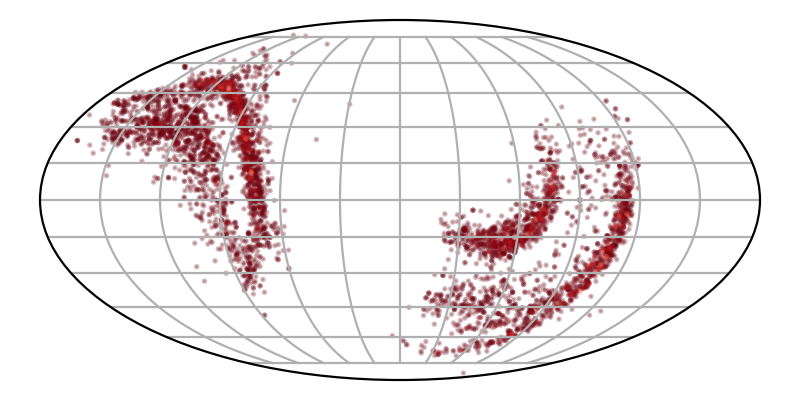} \\
        GW200112\_155838 & $33.67^{+5.33}_{-3.29}$ & $29.29^{+3.40}_{-4.90}$ & $0.24^{+0.07}_{-0.09}$ & $1276.29^{+431.75}_{-490.32}$ & $3204.4$ & \includegraphics[hsmash=r, align=c, width=2.5cm]{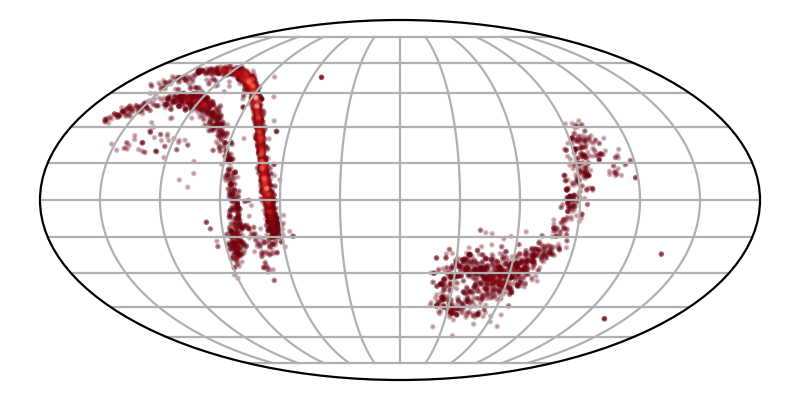} \\
        GW200115\_042309 & $6.32^{+1.19}_{-1.08}$ & $1.37^{+0.23}_{-0.16}$ & $0.06^{+0.03}_{-0.02}$ & $288.40^{+128.24}_{-93.89}$ & $366.7$ & \includegraphics[hsmash=r, align=c, width=2.5cm]{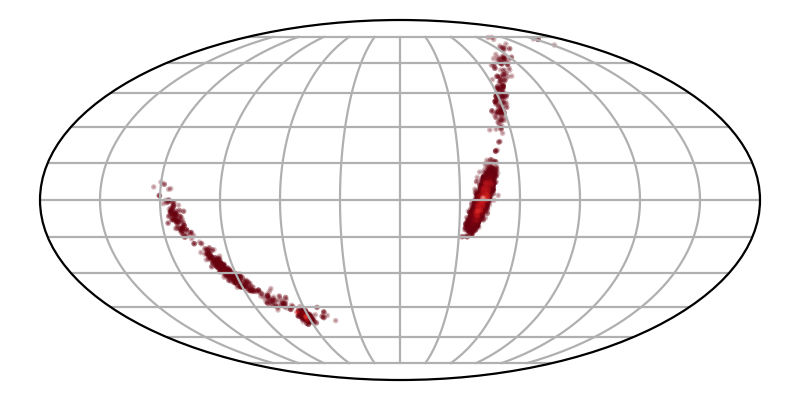} \\
        GW200128\_022011 & $38.41^{+9.03}_{-6.73}$ & $33.34^{+8.24}_{-6.95}$ & $0.57^{+0.28}_{-0.27}$ & $3399.78^{+2147.79}_{-1801.03}$ & $2466.8$ & \includegraphics[hsmash=r, align=c, width=2.5cm]{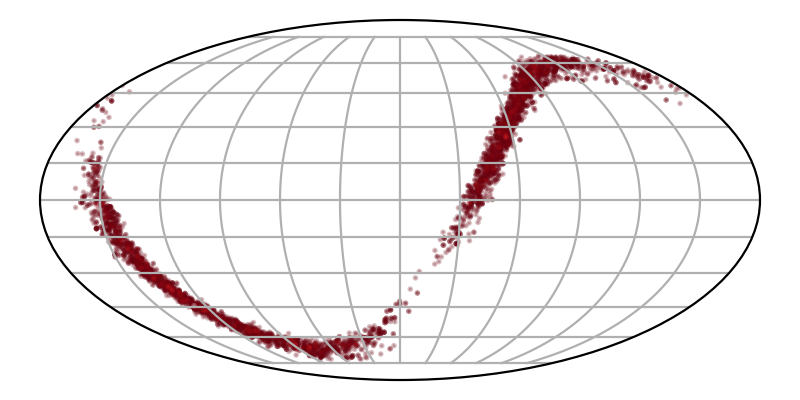} \\
        GW200129\_065458 & $33.66^{+5.33}_{-2.63}$ & $29.55^{+3.27}_{-5.16}$ & $0.19^{+0.04}_{-0.07}$ & $942.17^{+250.98}_{-387.45}$ & $45.8$ & \includegraphics[hsmash=r, align=c, width=2.5cm]{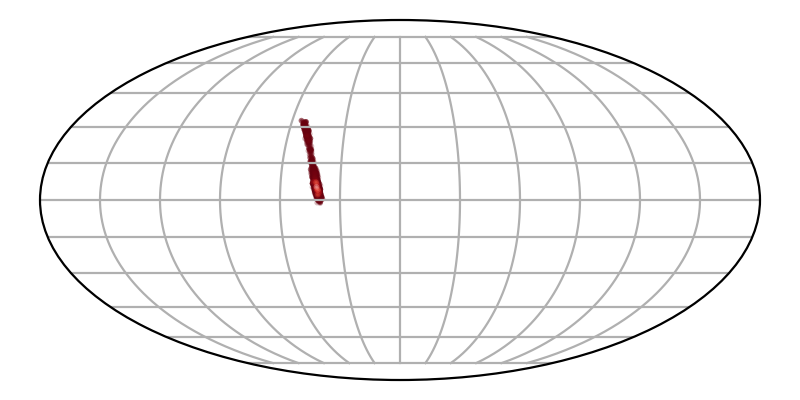} \\
        \hline
    \end{tabular}
    }
\end{table}

\begin{table}
    {\renewcommand{\arraystretch}{1.0}
    \begin{tabular}{l c c c c c p{2.5cm}}
        \hline\hline
        \multicolumn{1}{c}{name} & $m_1\,[M_\odot]$ & $m_2\,[M_\odot]$ & $z$ & $D_L\,[\mathrm{Mpc}]$ & $\Delta \Omega_{90\%} [\mathrm{deg}^2]$ & \multicolumn{1}{c}{skymap} \\
        \hline\hline
        GW200202\_154313 & $9.17^{+1.81}_{-0.60}$ & $8.09^{+0.59}_{-1.28}$ & $0.09^{+0.03}_{-0.04}$ & $411.93^{+140.20}_{-176.53}$ & $158.3$ & \includegraphics[hsmash=r, align=c, width=2.5cm]{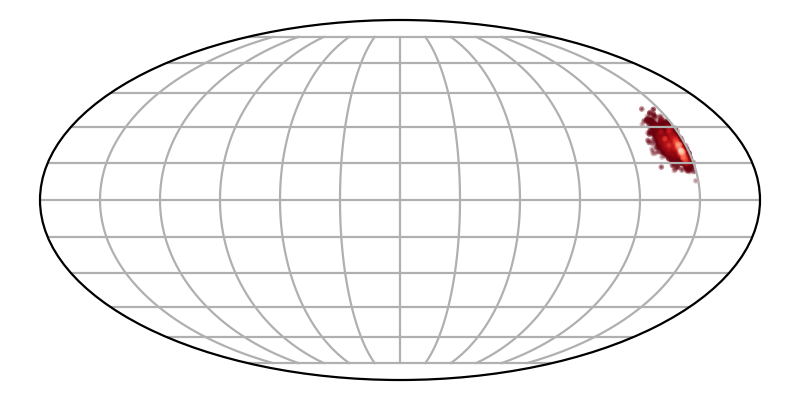} \\
        GW200208\_130117 & $34.53^{+6.52}_{-4.44}$ & $29.72^{+4.72}_{-5.62}$ & $0.40^{+0.14}_{-0.14}$ & $2244.01^{+992.89}_{-901.31}$ & $33.3$ & \includegraphics[hsmash=r, align=c, width=2.5cm]{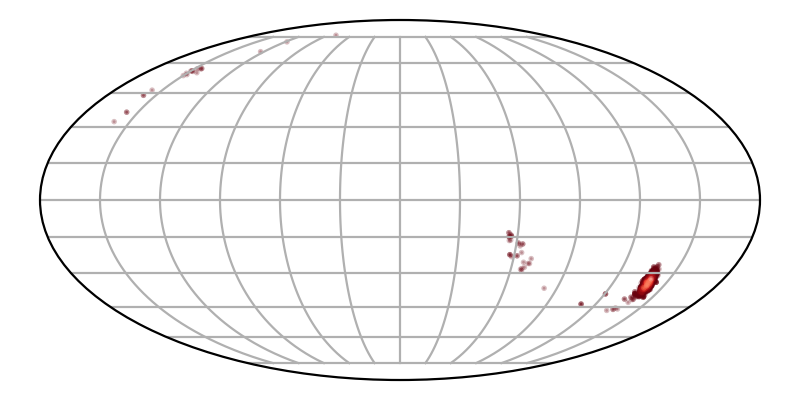} \\
        GW200209\_085452 & $32.47^{+7.19}_{-5.42}$ & $27.99^{+6.79}_{-5.49}$ & $0.56^{+0.24}_{-0.29}$ & $3352.15^{+1852.25}_{-1931.87}$ & $1025.1$ & \includegraphics[hsmash=r, align=c, width=2.5cm]{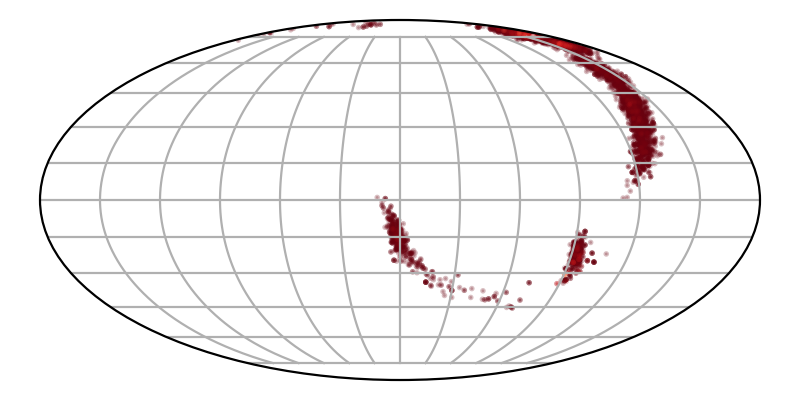} \\
        GW200216\_220804 & $41.65^{+11.55}_{-7.80}$ & $35.68^{+8.75}_{-9.53}$ & $0.68^{+0.35}_{-0.34}$ & $4272.53^{+2811.53}_{-2392.08}$ & $3104.4$ & \includegraphics[hsmash=r, align=c, width=2.5cm]{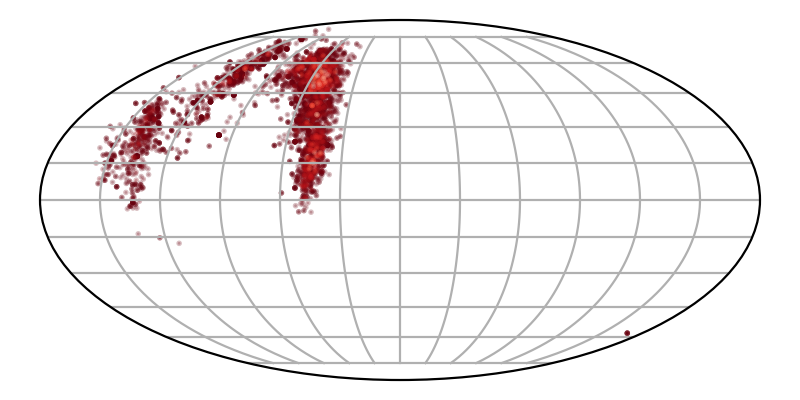} \\
        GW200219\_094415 & $34.11^{+7.64}_{-5.13}$ & $29.40^{+6.34}_{-5.71}$ & $0.58^{+0.22}_{-0.24}$ & $3522.91^{+1680.98}_{-1655.16}$ & $745.9$ & \includegraphics[hsmash=r, align=c, width=2.5cm]{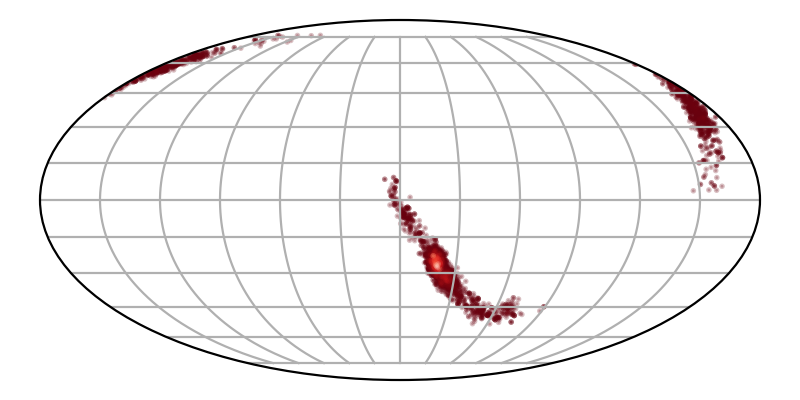} \\
        GW200224\_222234 & $38.35^{+5.64}_{-3.64}$ & $33.67^{+4.29}_{-5.22}$ & $0.32^{+0.08}_{-0.12}$ & $1725.50^{+488.84}_{-691.54}$ & $50.0$ & \includegraphics[hsmash=r, align=c, width=2.5cm]{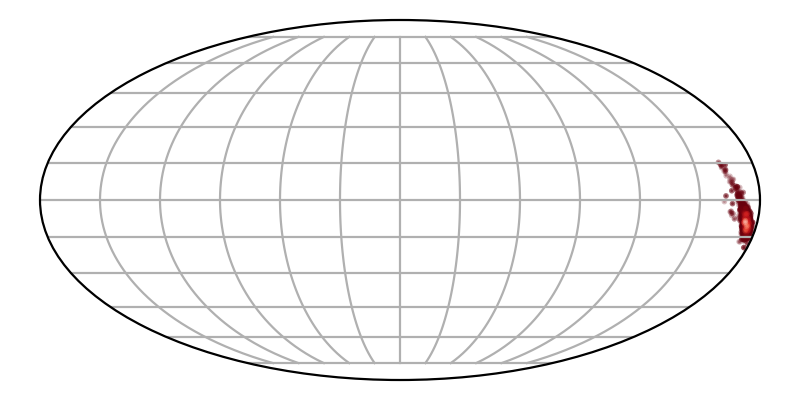} \\
        GW200225\_060421 & $17.91^{+3.22}_{-2.11}$ & $15.44^{+1.96}_{-2.93}$ & $0.22^{+0.09}_{-0.09}$ & $1104.17^{+526.97}_{-493.85}$ & $616.7$ & \includegraphics[hsmash=r, align=c, width=2.5cm]{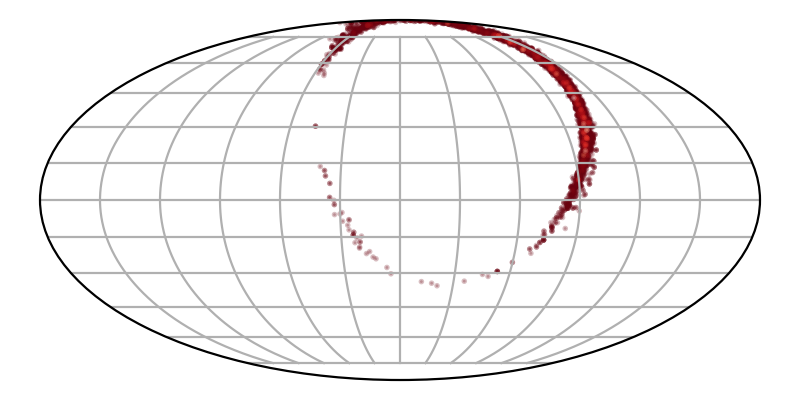} \\
        GW200302\_015811 & $31.47^{+8.84}_{-5.68}$ & $24.26^{+5.11}_{-6.64}$ & $0.30^{+0.16}_{-0.13}$ & $1608.82^{+1041.31}_{-780.03}$ & $8684.0$ & \includegraphics[hsmash=r, align=c, width=2.5cm]{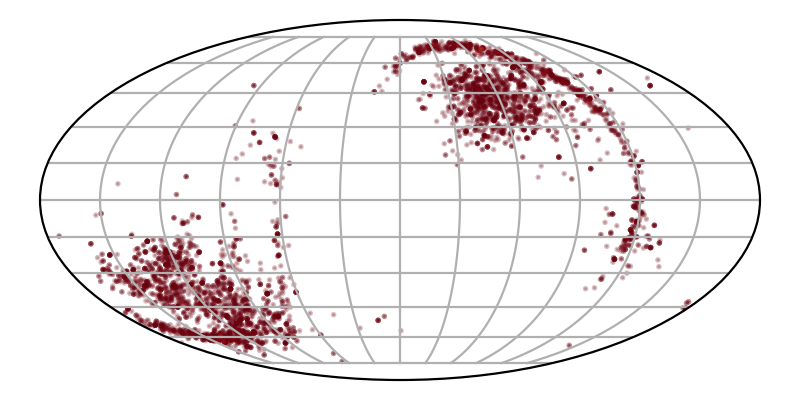} \\
        GW200311\_115853 & $32.60^{+4.69}_{-3.04}$ & $28.83^{+3.49}_{-4.53}$ & $0.23^{+0.05}_{-0.08}$ & $1169.68^{+284.52}_{-433.83}$ & $45.8$ & \includegraphics[hsmash=r, align=c, width=2.5cm]{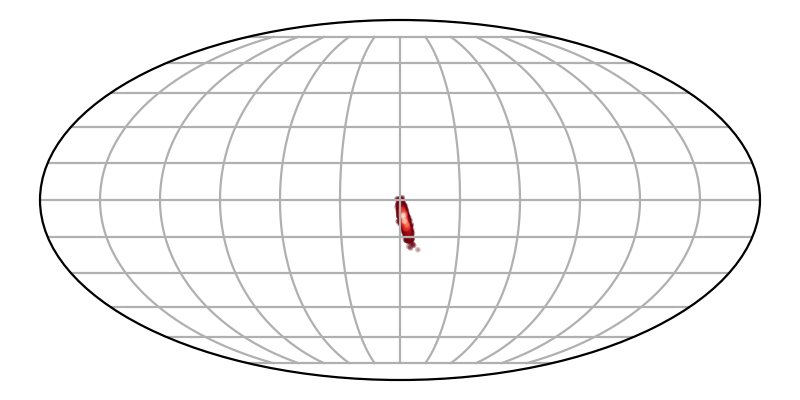} \\
        GW200316\_215756 & $11.00^{+3.10}_{-1.17}$ & $9.20^{+0.97}_{-1.77}$ & $0.22^{+0.08}_{-0.08}$ & $1120.21^{+448.67}_{-437.91}$ & $370.9$ & \includegraphics[hsmash=r, align=c, width=2.5cm]{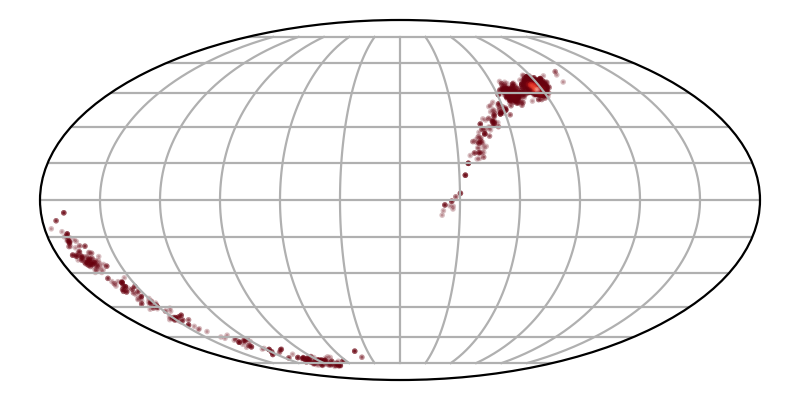} \\
\hline
\end{tabular}
}
\end{table}


\newpage
\newpage

\twocolumngrid

\end{document}